\documentclass[article,twocolumn,amsmath,amssymb,superscriptaddress,longbibliography]{revtex4-2}
\usepackage[english]{babel}
\usepackage{amssymb,amsmath,mathrsfs}       % 数学公式/符号（去重）
\usepackage{dcolumn,bm}                     % 表格列对齐/粗体符号（去重）
\usepackage{tabularx,multirow,makecell}     % 表格相关（去重）
\usepackage{graphicx,float}                 % 图片/浮动体（去重）
\graphicspath{{pict/}{fig2_2D_asymmetric/}} % 图片路径（补全/）
\usepackage[caption=false]{subfig}          % 子图
\usepackage{xcolor}                         % 颜色（去重）
\usepackage{txfonts}                        % 字体
\usepackage{verbatim,comment}               % 注释/逐字
\usepackage{epstopdf}                       % eps转pdf（去重）
\usepackage{hhline}                         % 表格线
\usepackage[normalem]{ulem}                 % 下划线（去重）
\usepackage[utf8]{inputenc}                 % 编码
\usepackage[raggedleft]{titlesec}           % 标题格式（保留基础包）
\usepackage[pdftex,colorlinks=true,linkcolor=red,citecolor=red,urlcolor=red]{hyperref}

\def\red{\textcolor{red}}

\def\LLH{\textcolor{cyan}}   % 保留你的原命令

\def\LZT{\textcolor{black}}
\def\MOD{\textcolor{black}}

\titleformat{\section}{\bfseries\large}{\thesection.\,}{0.24em}{}
\titlespacing{\section}{0cm}{0.5cm}{0cm}

\titleformat{\subsection}{\bfseries\normalsize}{\thesubsection.\,}{0.24em}{}
\titlespacing{\subsection}{0cm}{0.3cm}{0cm}

\titleformat{\subsubsection}{\bfseries\small}{\thesubsubsection.\,}{0.24em}{}
\titlespacing{\subsubsection}{0cm}{0.2cm}{0cm}

\begin{document}
\onecolumngrid
\affiliation{Quantum Science Center of Guangdong-Hong Kong-Macao Greater Bay Area (Guangdong), Shenzhen, China}
\affiliation{Guangdong Provincial Key Laboratory of Quantum Metrology and Sensing $\&$ School of Physics and Astronomy, Sun Yat-Sen University (Zhuhai Campus), Zhuhai 519082, China}

\title{Inter-species topological phases via a dynamical gauge field}
\author{Zhoutao Lei}
\affiliation{Guangdong Provincial Key Laboratory of Quantum Metrology and Sensing $\&$ School of Physics and Astronomy, Sun Yat-Sen University (Zhuhai Campus), Zhuhai 519082, China}
\author{Linhu Li}\email{lilinhu@quantumsc.cn}
\affiliation{Quantum Science Center of Guangdong-Hong Kong-Macao Greater Bay Area (Guangdong), Shenzhen, China}
\date{\today}

\begin{abstract}
We uncover a class of inter-species topological phases in a one-dimensional lattice, loaded with two species of non-identical particles interacting via a dynamical gauge field (DGF).
Two types of topological states are found to emerge from different inter-species topology activated by the DGF. 
Specifically, edge confined states with co-localization of both species arise from an extrinsic inter-species topology, which can be decomposed into  the single-particle topology for each species.
On the other hand, bulk bound states with extended distribution emerge from an intrinsic inter-species topology that cannot be understood from single-particle ones.
The two classes of inter-species topology are found to be independent of each other, characterized by different sets of inter-species topological invariants.
Thus, their topological states can coexist in certain parameter regimes and compete with each other, leading to distinguished dynamical signatures. 
We further propose a feasible cold-atom realization of our model to demonstrate experimental accessibility of inter-species topological phases.
Our work establishes inter-species topology as a new organizing principle of topological matter, revealing how correlations between distinct particle species can generate topological phenomena beyond single-particle paradigms.
\end{abstract}

\maketitle
\clearpage
\
\newpage
\
\newpage

\maketitle

\noindent
Topological phases of matter have generated great excitements 
%spanning from condensed matter physics to quantum physics over the past decade, 
spanning from condensed matter physics to quantum simulations,
with wide-range applications in quantum computing, spintronics, and dissipationless transports owing to their
robust boundary phenomena guaranteed by global topological natures of the systems~\cite{RevModPhys.82.3045,RevModPhys.83.1057}.
At single-particle level, these phases feature boundary states protected by bulk band topology, which is characterized by different topological invariants depending on the system's symmetries and dimensionality.
In many-body physics, 
ubiquitous particle interactions dramatically alter the single-particle band topology~\cite{PhysRevLett.95.226801,PhysRevLett.96.106401,PhysRevB.73.045322,fidkowski2010effects,fidkowski2011topological,hassan2017many,Liu_2023} and greatly enrich the formalisms of topological phases, %beyond the topology band theory, 
inducing novel topological phenomena such as topological Mott insulators~\cite{Pesin2010,PhysRevX.6.011034,Chen2021,Wagner2023}, fractional topological phases~\cite{PhysRevLett.106.236802,PhysRevLett.106.236803,PhysRevLett.106.236804,Zeng2023,Redekop2024,PhysRevLett.133.186602}, topological spin liquids~\cite{Balents2010,Yan2021,semeghini2021probing}, and many-body topological pumping~\cite{PhysRevA.95.063630,Walter2023}.

%Recently, significant progress has been made in effective non-Hermitian Hamiltonians of open quantum systems~\cite{}, which hosts a variaty of exotic phenomena of many-body static~\cite{} and dynamic behaviors~\cite{}.

Recently, significant progress has been made in quantum simulations of dynamical gauge fields (DGF), a novel type of interaction where local density fields exert an inverse influence on the gauge fields~\cite{PhysRevLett.109.175302,PhysRevLett.111.110504,RevModPhys.93.025001,martinez2016real,PhysRevLett.118.045302,barbiero2019coupling,gorg2019realization,PhysRevX.10.021041,zhou2022thermalization,PhysRevLett.130.171901,PhysRevLett.129.180401,PhysRevLett.132.023401,PhysRevA.111.013319}.
In particular, it has been shown~\cite{PhysRevLett.129.180401} that
DGF can induce a non-Hermitian point-gap topology characterized by a winding number and massive edge localization of eigenstates, representing a many-body extension of the celebrated non-Hermitian skin effect (NHSE)~\cite{PhysRevB.97.121401,PhysRevLett.121.086803,PhysRevX.8.031079,PhysRevLett.124.056802,PhysRevLett.124.086801,PhysRevLett.125.126402}.
However, can DGFs mediate topological phases between distinct particle species, creating topology that is neither single-particle nor purely many-body in nature?

%It is naturally to ask whether DGF can induce even more exotic many-body topological phases without a single-particle counterpart.

\begin{figure}
    \centering
    \includegraphics[width=0.9\linewidth]{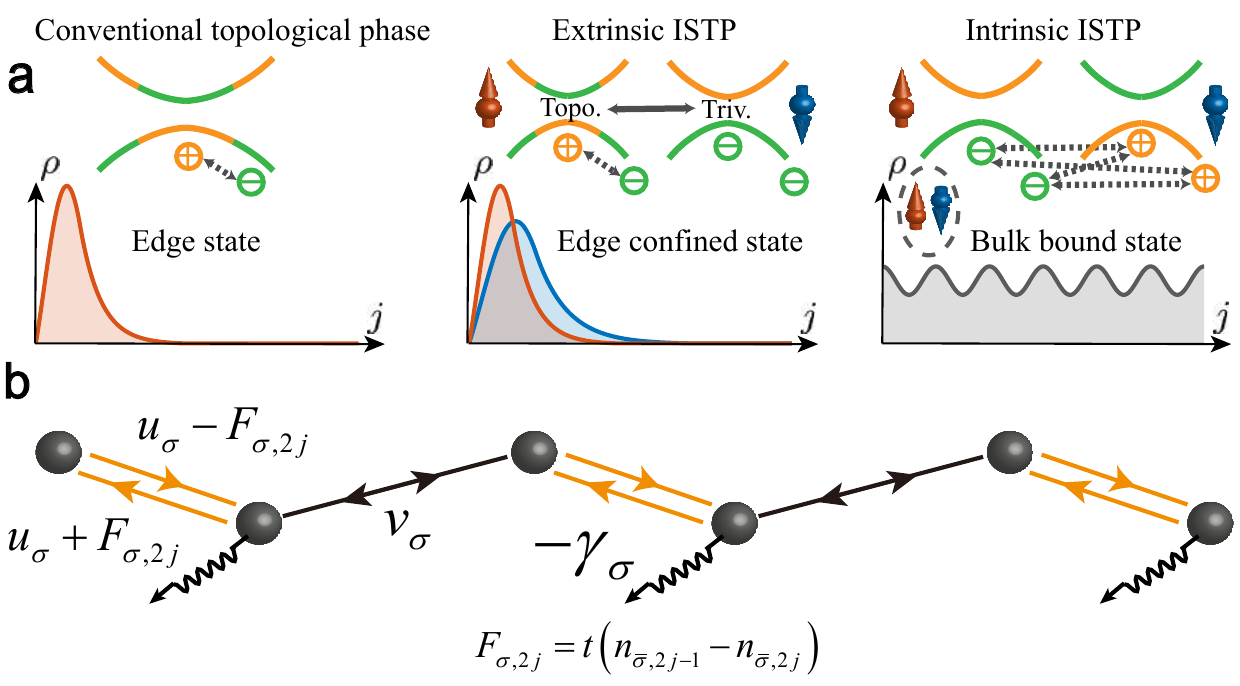}
   \caption{\textbf{Inter-species topological phases and our model.} \textbf{a} Topological band structures of conventional and inter-species topological phases and their topological states. Orange (green) circles with plus (minus) sign indicate the symmetry indicators at high-symmetric momenta, and topological band inversion is marked by dashed lines with arrows. 
Band inversion of a single species gives rise to the conventional topological phases.
Extrinsic ISTPs emerge from the correlation between two species with and without single-species band inversion.
Intrinsic ISTPs are induced by inter-species band inversion, regardless whether single-species band inversion occurs.
%As shown here, extrinsic ISTPs form when one particle species is trivial and the other nontrivial, and intrinsic ISTPs are induced when inter-species band inversion occurs. 
\textbf{b} A sketch of the model of Eq.\eqref{eq:DGF_H}. 
$\bar{\sigma},\sigma\in\{\uparrow,\downarrow\}$ with $\bar{\sigma}\neq\sigma$ denote the two species interacting with each other via a DGF ($F_{\sigma,2j}$). 
%Here, the symbol $\bar{\sigma}$ denotes the species of particles other then $\sigma$ (i.e., $\bar{\sigma}\neq{\sigma}$). As implied here, the two species of particles, influence each other only via a DGF.
}
\label{figsketch}
\end{figure}

In this work, we uncover a class of inter-species topological phases (ISTPs) emerging from a non-Hermitian DGF between two non-identical species of particles.
Specifically, via the DGF, the topological localization of one species induces a non-reciprocal pumping to the other, resulting in edge confined or anti-confined states depending on the direction of the non-reciprocity. 
These states constitute an extrinsic ISTP, as their associated topological characterization can be decomposed into single-particle ones defined for each species.
More intriguingly, we also identify an intrinsic ISTP with extended two-particle bound states, arising from correlations protected by an inter-species band inversion that cannot be reduced to single-particle topological features.
Notably, as sketched in Fig. \ref{figsketch}\textbf{a}, these ISTPs arise from the inter-species topological band \LZT{structure} of both particles, fundamentally different from conventional topological phases rooted in single-particle band topology.
Furthermore, both types of ISTPs exhibit imaginary energy shifts controlled respectively by species-dependent particle loss and the non-Hermitian DGF, leading to distinct dynamical signatures accessible in experiments. 
%Finally, we also propose an implementation of our model with Floquet engineering in cold-atom systems.
Finally, we propose a cold-atom realization of our model via Floquet engineering, offering a potential platform to explore these many-body non-Hermitian topological phases.

\vspace{1em} 
\noindent
{\large{\textbf{Results}}}.\\
\noindent
\textbf{Model and Hamiltonian.}
As an explicit example of ISTPs, we consider a minimal model with two spinless non-identical particles (denoted as pseudospin-up and -down species, $\sigma=\uparrow,\downarrow$) interacted with each other through a density-dependent DGF, loaded in a one-dimensional Su-Schrieffer-Heeger (SSH) model~\cite{PhysRevLett.42.1698,PhysRevB.22.2099} with $L$ lattice sites.
%We consider two non-identical particles (denoted as pseudospin-up and -down species, $\sigma=\uparrow,\downarrow$) interacted with each other through a density-dependent DGF, loaded in a Su-Schrieffer-Heeger (SSH) model~\cite{PhysRevLett.42.1698,PhysRevB.22.2099} with $L$ lattice sites.
%We consider a two-species Su-Schrieffer-Heeger (SSH) model with $N$ lattice sites, suffered from a DGF between different particles.
Its Hamiltonian is given by
\begin{eqnarray}
H&=&\sum_{\sigma=\uparrow,\downarrow}\left(H_{\sigma}-i\sum_{j=1}^{L/2}\gamma_\sigma n_{\sigma,2j}\right)+H_{\rm DGF},\label{eq:DGF_H}\\
H_{\sigma}&=&\sum_{j=1}^{L/2}\left(u_\sigma a^\dagger_{\sigma,2j-1}a_{\sigma,2j}+v_\sigma a^\dagger_{\sigma,2j}a_{\sigma,2j+1}\right)+h.c.,\nonumber\\
H_{\rm DGF}&=&\sum_{\sigma\neq\bar{\sigma}}\sum_{j=1}^{L/2}\left[t\left(n_{\bar{\sigma},2j-1}-n_{\bar{\sigma},2j}\right)a^\dagger_{\sigma,2j-1}a_{\sigma,2j}\right]-h.c.,\nonumber
\end{eqnarray}
where $a^\dagger_{\sigma, j}$ creates a $\sigma$-particle at site $j$, $n_{\sigma,j}=a^\dagger_{\sigma, j}a_{\sigma, j}$ is the corresponding density operator,
$u_\sigma$ and $v_\sigma$ are pseudospin-depdendent staggered hopping amplitudes,
$t$ describes the density-dependent DGF, 
and $\gamma_\sigma$  represents species-dependent particle loss added on even lattice sites,
as sketched in Fig. \ref{figsketch}\textbf{b}.
%Species-dependent particle loss $\gamma_\sigma$ is introduced only on even lattice sites,
Without loss of generality, $\gamma_\downarrow=0$ and $t>0$ are chosen in the rest of this paper.
This model satisfies the parity-time ($\mathcal{PT}$) symmetry upon a uniform shift of eigenenergies along imaginary axis,
with the spinless time-reversal symmetry represented by a complex-conjugate operation.
Namely, the shifted Hamiltonian 
%$H+\sum_{j} \sum_\sigma i\gamma_\sigma n_{\sigma,j}/2$ 
$H+ \sum_\sigma i\gamma_\sigma N_\sigma/2$ 
remains unchanged after transformation of $a_{\sigma,j}\rightarrow a_{\sigma,L+1-j}$ and $i\rightarrow -i$, with $N_\uparrow=N_\downarrow=1$ the numbers of pseudospin-up and -down particles.
\LZT{
In addition, we note that the for single-particle model (without $H_{\rm DGF}$), each species is in its PT-broken (-unbroken) phase when $\gamma_\sigma^2 > 4(u_\sigma^2 + v_\sigma^2 - 2|u_\sigma v_\sigma|)$ ($\gamma_\sigma^2 <4(u_\sigma^2 + v_\sigma^2 - 2|u_\sigma v_\sigma|)$), and exceptional points emerge when the inequality become equality~\cite{Ding2022}.
%For the dissipation-hopping condition [$\gamma_\sigma^2 > 4(u_\sigma^2 + v_\sigma^2 - 2|u_\sigma v_\sigma|)$], single-component bulk states are $\mathcal{PT}$-broken; for [$\gamma_\sigma^2 < 4(u_\sigma^2 + v_\sigma^2 - 2|u_\sigma v_\sigma|)$], they are $\mathcal{PT}$-unbroken, with exceptional points emerging at the boundary~\cite{Ding2022}. 
Here we focus primarily on the regime where single-component bulk states are $\mathcal{PT}$-unbroken, 
so that the $\mathcal{PT}$-broken inter-species topological states can be more clearly identified, as demonstrated below.
%yet the system still supports $\mathcal{PT}$-broken nontrivial confined/bound states due to the ISTPs, as discussed below.
}
In the following discussion, $2t=1$ is fixed as the energy unit for a clearer demonstration of different ISTPs.
\\

%This Hamiltonian satisfies the parity-time ($\mathcal{PT}$) symmetry upon a uniform shift of eigenenergies along imaginary axis, namely, the shifted Hamiltonian $H+\sum_{j} \sum_\sigma i\gamma_\sigma n_{\sigma,j}/2$ remains unchanged after transformation of $a_{\sigma,j}\rightarrow a_{\sigma,N-j}$ and $i\rightarrow -i$.
%Consequently, the bulk states hold the same imaginary energy ($-i\sum_\sigma \gamma_\sigma/2$) in the $\mathcal{PT}$-unbroken phase, while the topologically-correlated bound states can be made $\mathcal{PT}$-broken to acquire extra imaginary energies, allowing them to dominate the time-evolution and manifest in a dynamical process.

%\begin{figure}
%    \centering
%    \includegraphics[width=1.0\linewidth]{fig1_sketch_2.pdf}
% \caption{\textbf{Sketch of model and main results} (a) Schematic of the model described in Eq.~\eqref{SSHDGF}. (b) and (c) The illustrations of the edge and bulk bound states respectively. The edge confined state is formed by the cooperation of the topological edge states and the DGF. The bulk bound states are induced by the interplay between the DGF and inter-band inversion, which is the relative bulk topology between two particles.}
%    \label{fig1}
%\end{figure}
%\\

\begin{figure*}
    \centering
    \includegraphics[width=0.8\linewidth]{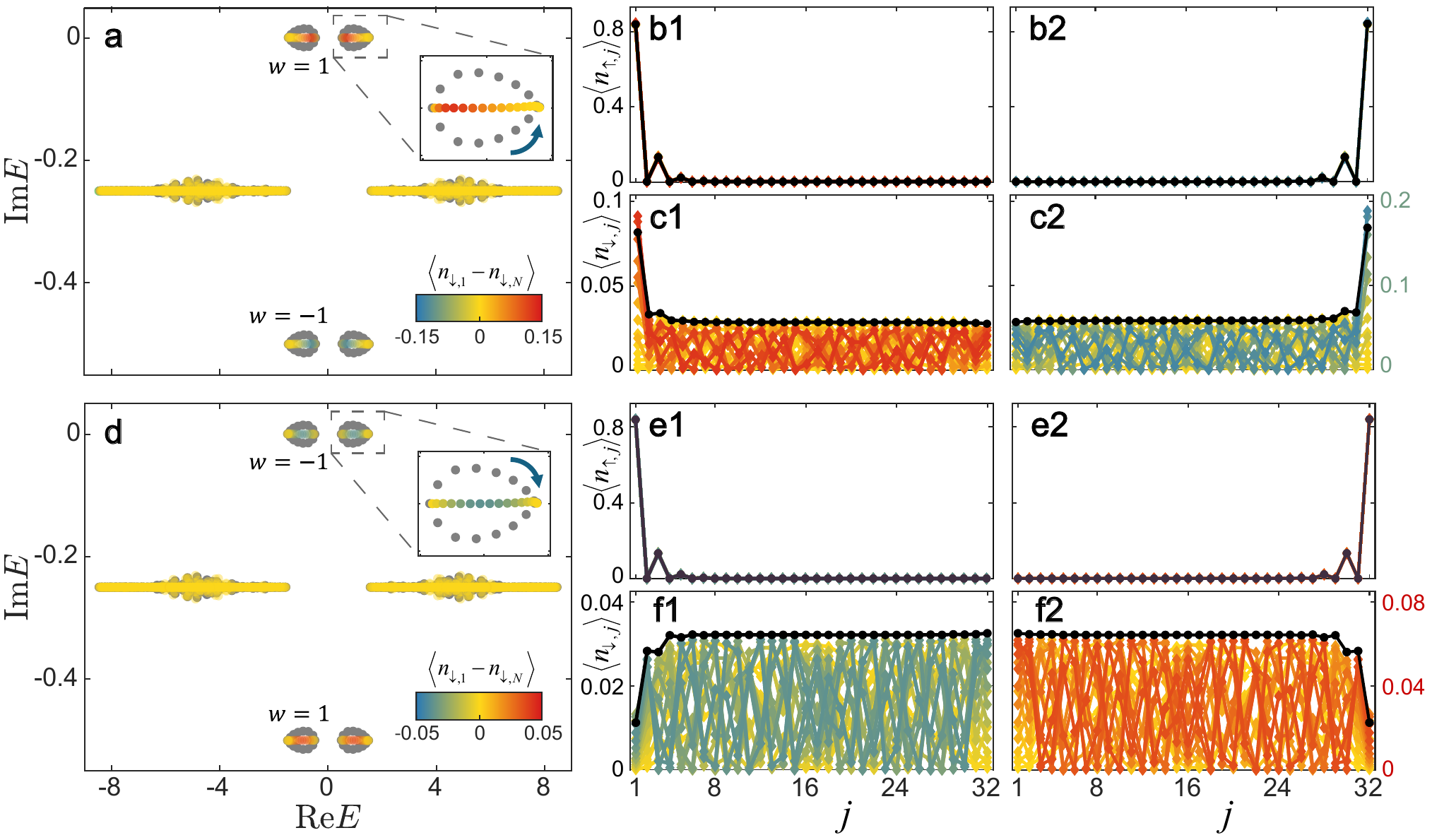}
   \caption{\textbf{edge confined and anti-confined states.} \textbf{a} OBC spectrum of the Hamiltonian in Eq.~\eqref{eq:DGF_H}, marked by the edge-density imbalance of the pseudospin-down particle. 
   Gray dots are the eigenenergies with PBCs taken only for the pseudospin-down particle.
   %Here, spin-$\uparrow$ particle is always under OBCs, and spin-$\downarrow$ particle takes OBCs (PBCs) for colored and black (gray) circles. 
  \textbf{b\LZT{1}} and \textbf{c\LZT{2}} distributions of pseudospin-up and -down particles, respectively, for OBC eigenstates in the enlarged insets in \textbf{a} marked by the same colors, \LZT{while \textbf{b2} and \textbf{c2} show these for the associated states with eigenvalues ${\rm Im}E\approx-\gamma_\uparrow$.} Their average distributions are shown in black in each panel.
   %(b) and (c) distributions of pseudospin-up and -down particles, respectively, averaged over eigenstates with ${\rm Im}E\approx 0$.
   edge confined states are characterized by the co-localization at the edge for both particles.
   %Here, distributions of single eigenstates are also demonstrated, marked by the same colors as in (a).
   Parameters in \textbf{a} to \textbf{c\LZT{2}} are $v_{\uparrow}=5$, $v_{\downarrow}=0.5$, $\gamma_{\uparrow}=0.5$, $u_{\uparrow}=2$, and $u_{\downarrow}=1$.
   \textbf{d} to \textbf{f\LZT{2}} the same as \textbf{a} to \textbf{c\LZT{2}}, but with $u_{\uparrow}=-2$ and $u_{\downarrow}=-1$. 
Edge anti-confined states are characterized by the pseudospin-up localization and the drop of pseudospin-down density at the same edge.
In \textbf{c\LZT{1}}-\LZT{\textbf{c2}} and \textbf{f\LZT{1}}-\LZT{\textbf{f2}}, distributions of single eigenstates and their average correspond to different y-axis coordinates on the right and left sides of the figure, respectively.
The chosen system size is $L=32$, with $L^2=1024$ the Hamiltonian dimension.
}
\label{fig1}
\end{figure*}

\noindent\textbf{Extrinsic ISTP and edge confined states.}
At the single-particle level, 
$H_\sigma$ describes the standard SSH model that is topologically nontrivial when $|u_\sigma|<|v_{\sigma}|$.
%topological edge localization of $\sigma$-particle emerges when $|u_\sigma|<|v_{\sigma}|$, where the corresponding SSH model is topologically nontrivial.
%Two-particle edge states naturally arise when both particles are topologically nontrivial, even in the absence of DGF.
In the presence of the DGF, we find that an ISTP with edge confined states \LZT{emerges} when the two particles fall in topologically different phases. 
%when the two particles fall in topologically different phases, where the interplay between DGF and non-trivial topology of one particle leads to an ISTP with edge confined states.
An example hosting such states is shown in Fig. \ref{fig1}, with pseudospin-up and -down particles chosen to be topologically nontrivial and trivial ($|u_\uparrow|<|v_{\uparrow}|$ and $|u_\downarrow|>|v_{\downarrow}|$), respectively. 
Fig. \ref{fig1}\textbf{a} \LZT{displays} the eigenenergies marked by the edge-density imbalance of the pseudospin-down particle for each eigenstate,
\begin{eqnarray}
\Delta \langle n_{\downarrow,{\rm edge}}\rangle=
\langle n_{\downarrow,1}-n_{\downarrow,L}\rangle,
\end{eqnarray}
with $\langle O\rangle$ the expectation value of an operator $O$ on a right normalized eigenstate $|\psi_m\rangle$.

%Due to the pseudospin-up particle loss $\gamma_\uparrow$,
Several branches of $\mathcal{PT}$-broken eigenstates under open boundary conditions (OBCs) are found to emerge and separate from the $\mathcal{PT}$-symmetric imaginary energy ${\rm Im}E=\gamma_\uparrow /2$, with the pseudospin-up particle localized at edges by its nontrivial topology.
In particular, 
left (right) localization corresponds to imaginary energies ${\rm Im}E\approx 0$ (${\rm Im}E\approx -\gamma_\uparrow$),
as illustrated in Fig. \ref{fig1}\textbf{b\LZT{1}} \LZT{(Fig. \ref{fig1}\textbf{b{2}})}. \LZT{Simultaneously}, even under a trivial topology, the pseudospin-down particle is found to show the same localization tendency as the pseudospin-up particle, indicated by nonzero $\Delta \langle n_{\downarrow,{\rm edge}}\rangle$ in Fig. \ref{fig1}\textbf{a}.
This is in sharp contrast to the behaviors of the standard single-particle SSH model,
%that describes the pseudospin-down paricle at single-particle level, 
and can only be attributed to the inter-species coupling from the DGF.
However, despite the strong edge localization, these edge confined states show uniform nonzero pseudospin-down distributions in the bulk [Figs. \ref{fig1}\textbf{c\LZT{1}} \LZT{and \textbf{c2}}], which greatly differ from the exponential decay of most topological or other forms of eigenstate localization. 

The origin of these edge confined states can be understood by taking a mean-field approximation $n_{\uparrow,j}\rightarrow \langle n_{\uparrow,j}\rangle$ in Eq.~\eqref{eq:DGF_H} (see {Supplementary Note 1}). Through the DGF, the edge localization of pseudospin-up particle generates a non-reciprocal hopping that decays exponential from the edge for the pseudospin-down particle, leading to its strong non-reciprocal edge localization and extended bulk distribution.
\LZT{The opposite topological localization of the pseudospin-up particle for different states (Figs. \ref{fig1}\textbf{b1} and \textbf{b2}) further results in a perfectly symmetric pattern of the pseudospin-down particle (Figs. \ref{fig1}\textbf{c1} and \textbf{c2}).}
This non-reciprocal localization also resembles an inhomogeneous NHSE~\cite{PhysRevB.97.121401,PhysRevLett.121.086803}. Indeed, by taking periodic boundary conditions (PBCs) only for the pseudospin-down particle, energy spectrum of these states forms some loops enclosing the full-OBC eigenenergies in Fig. \ref{fig1}\textbf{a}, resembling the nontrivial point-gap topology of NHSE~\cite{PhysRevX.8.031079,PhysRevLett.124.056802,PhysRevLett.124.086801,PhysRevLett.125.126402}. 

In addition, note that when $u_\downarrow<0$, the effective non-reciprocal direction reverses, resulting in some edge anti-confined states with the pseudospin-down particle distributing less at the edge, as shown in Figs. \ref{fig1}\textbf{d} to \ref{fig1}\textbf{f}.
\LZT{Thus, while extrinsic ISTPs exhibit similar point gap behavior to NHSE systems~\cite{PhysRevLett.124.086801,PhysRevLett.125.126402,PhysRevB.103.L140201,Wang2025-uo}: when the left hopping strength of pseudospin-down particles is greater than (less than) the right hopping strength, the corresponding spectrum shows a point gap with a winding number of $+1$ ($-1$), the correspondence between the winding number and localization phenomena is non-unique here. For example, a winding number of $+1$ can correspond to either enhanced leftward hopping or weakened rightward hopping. Both cases lead to left-localized systems with NHSE, but here they respectively result in an increase in pseudospin-down particle density at the left boundary (see Fig. \ref{fig1}\textbf{c1} for edge confined states) and a decrease at the right boundary (see Fig. \ref{fig1}\textbf{f2} for edge anti-confined states).}
Notably, with an origin similar to that of NHSE, 
the edge confined and anti-confined distributions rely crucially on the $\mathcal{PT}$-symmetry breaking, see {Supplementary Note 1 B} \LZT{, and the choice of the right eigenbasis instead of the biorthogonal one~\cite{kunst2018biorthogonal} (also see Supplementary Note 6).}

Finally, when both particles are in topologically nontrivial regimes, 
%the edge and bulk lattices are effectively decoupled by the single-particle topology for each species. In this case, 
the inter-species topological edge states disappear as the bulk and edge are decoupled for each species by single-particle topology,
and some other edge confined and anti-confined states may coincidentally appear from the direct product of single-particle eigenstates (see {Supplementary Note 2}).
Therefore, the DGF-mediated edge confined and anti-confined states represent a type of extrinsic ISTP, in the sense that their emergence requires certain topological conditions for both species, which can be decomposed into single-particle topology of each of them.
% (trivial and nontrivial, respectively). But the inter-species topology here is extrinsic, since it can be decomposed into a combination of single-particle topology. 
An explicit definition of corresponding topological invariants will be given in the following discussion of the intrinsic ISTP.
\\

\begin{figure}
    \centering
    \includegraphics[width=1\linewidth]{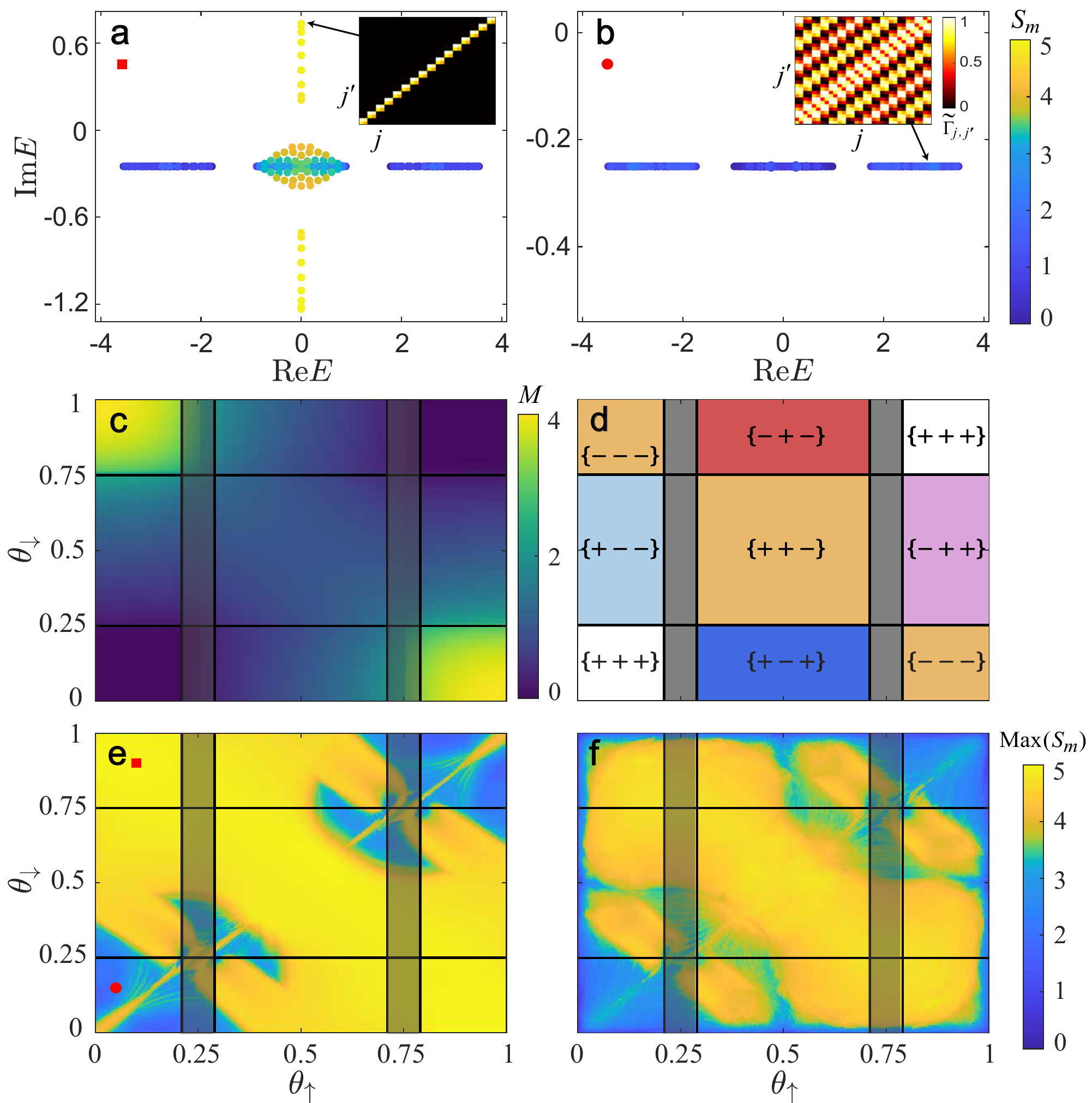}
   \caption{\textbf{Bulk bound states induced by inter-species band inversion and DGF.} \textbf{a} and \textbf{b} The PBC spectra of the Hamiltonian in Eq.~\eqref{eq:DGF_H} with different parameters. The normalized two-particle correction of bulk bound states [Eq.~\eqref{Twocorr}] with the highest entanglement
entropy (marked by arrows) is shown in the insets. 
   Parameters are \textbf{a} $\theta_{\uparrow}=0.1\pi$ and $\theta_{\downarrow}=0.9\pi$, and \textbf{b} $\theta_{\uparrow}=0.05\pi$ and $\theta_{\downarrow}=0.15\pi$, with $\theta_{\sigma}={\rm arg}(u_{\sigma}+iv_{\sigma})$, $|u_{\sigma}+iv_{\sigma}|=\sqrt{2}$, and $\gamma_\uparrow=0.5$.
   \textbf{c} Magnitude of the DGF term $M$ defined in Eq.~\eqref{eq:DGF_matrix}.  
   \textbf{d} A phase diagram spanned by $\theta_{\uparrow}$-$\theta_{\downarrow}$ based on $\{I_{00},I_{\pi\pi},I_{0\pi}\}$ defined in Eq.~\eqref{ISbi}.
   Nontrivial regions with at least one $I_{kk'}=-1$ are marked by different colors, and shaded regions represent the single-particle $\mathcal{PT}$-broken phase of pseudospin-up particle (see {Supplementary Note 4 C}), where the topological invariants are ill-defined.   
   Trivial regions marked by $\{+++\}$ have vanishing DGF ($M\approx 0$) in \textbf{c}.
  Additionally, dark and light blue (red and pink) regions possess edge confined (anti-confined) states induced by single-particle topology of the pseudospin-up and -down particles, respectively.
    \textbf{e} The maximal entanglement entropy $S_m$ of all eigenstates. Red square and diamond mark the parameters of \textbf{a} and \textbf{b} with the same symbols.
    \textbf{f} The same as \textbf{e}, but with an extra disorder term $\lambda\sum_{\sigma}\sum_j\varepsilon_{\sigma,j}n_{\sigma,j}$ with $\lambda=0.2$ and $\varepsilon_{\sigma,j}$ randomly drawn from a uniform distribution $[-1,1]$.
    $L=32$ is set for all panels.
    }
\label{fig2}
\end{figure}

\noindent\textbf{Intrinsic ISTP and Bulk bound states.}
Other than the edge confined states, we identify another class of two-particle bound states in this model, which can be characterized by an inter-species topological band inversion.
In Figs. \ref{fig2}\textbf{a} and \ref{fig2}\textbf{b}, we demonstrate the PBC energy spectrum marked by inter-species entanglement entropy of each eigenstate $|\psi_m\rangle$,
\begin{eqnarray}
S_m=-{\rm Tr}(\rho_{m,\downarrow}\log_2\rho_{m,\downarrow}),~~~~
\rho_{m,\downarrow}={\rm Tr}_{\uparrow}(|\psi_m\rangle\langle\psi_m|).
\label{Tro}
\end{eqnarray} 
In Fig. \ref{fig2}\textbf{a},  a class of $\mathcal{PT}$-broken states with ${\rm Re}E\approx 0$ and ${\rm Im}E$ diverged from $-\gamma_\uparrow/2$
are found to emerge, with much larger $S_m$ that indicates a strong correlation between the two particles.
Explicitly, we display in the insets of Figs. \ref{fig2}\textbf{a} and \ref{fig2}\textbf{b} the normalized two-particle correlation function% \blue{and its normalized form},
%\blue{the local two-particle correlation $\Gamma_0\equiv \sum_j\Gamma_{jj}$, with the correlation function defined as}
\begin{eqnarray}
\widetilde{\Gamma}_{j,j'}=\Gamma_{j,j'}/{\rm Max}(\Gamma_{j,j'}),~~~\Gamma_{j,j'}=\langle n_{\uparrow,j}n_{\downarrow,j'}\rangle,
\label{Twocorr}
\end{eqnarray}
for the eigenstate with maximal $S_m$, which also corresponds to maximal ${\rm Im}E$ in Fig. \ref{fig2}\textbf{a}.
%for the eigenstates with maximal ${\rm Im}E$ in the insets of Fig. \ref{fig2}(***).
$\Gamma_{j,j'}$ is seen to take nonzero values only near the diagonal line $j=j'$, indicating a two-particle bound state with extended distributions in the bulk, \LZT{referred to as bulk bound states}.
\LZT{Such distribution
features allow these bulk bound states to persist also
in the biorthogonal basis (also see Supplementary Note 6).}

The emergence of these states can be attributed to the \LZT{correlation} of DGF enabled by inter-species topological properties, i.e., the occurrence of inter-species band inversion.
To see this, we first provide a description of topological bands in our model.
Note that at single-particle level, due to the protection of inversion symmetry, topological phases of the SSH model ($H_\sigma$) can be characterized by the band inversion~\cite{PhysRevB.76.045302}
 at high-symmetric momenta $k=0$ and $k=\pi$. Explicitly, the topologically trivial (nontrivial) phase is characterized by $I_{\rm sin}=I_0I_\pi=1$ ($-1$)~\cite{Slager2013,PhysRevB.89.155114,PhysRevX.7.041069,PhysRevB.103.024205,doi:10.1126/sciadv.aao4748}, with 
$I_k=\langle\sigma_x\rangle$
the expectation value of the Pauli's matrix $\sigma_x$ on the lower-band eigenstate at momentum $k$.
\LZT{Here $\sigma_x$ represents the inversion symmetry operator ($a_{\sigma,j} \rightarrow a_{\sigma,L+1-j}$),
%that exchanges the two sublattices. 
which commutes with $H_\sigma$ at high-symmetric momenta.
Thus, $I_{\rm sin}$ marks the lower band and changes sign when the two bands topologically inverse in energy at these momenta.}
The DGF obeys the inversion symmetry, while the imaginary on-site potential $\gamma_{\sigma}$ breaks it but keeps topological properties~\cite{li2019geometric}.
%As both symmetries remains in our model with DGF \LZT{(The dissipation breaks inversion and chiral symmetry)}, 
Therefore, we define topological invariants as
\begin{eqnarray}
I_{k_{\uparrow}k_{\downarrow}}=
\langle\psi_{--}(k_{\uparrow},k_{\downarrow})|
\sigma_x\otimes\sigma_x |\psi_{--}(k_{\uparrow},k_{\downarrow})\rangle
%\\I_{k_{\uparrow}k_{\downarrow}}&=&[\langle\mu^{(\uparrow)}_{-}(k_{\uparrow})|\otimes\langle\mu^{(\downarrow)}_{-}(k_{\downarrow})|]\sigma_x\otimes\sigma_x[|\mu^{(\uparrow)}_{-}(k_{\uparrow})\rangle\otimes|\mu^{(\downarrow)}_{-}(k_{\downarrow})\rangle]\nonumber\\
\label{ISbi}
\end{eqnarray} 
on the high-symmetric points of $k_\uparrow,k_\downarrow=0$ or $\pi$,
with $|\psi_{\alpha\beta}(k_{\uparrow},k_{\downarrow})\rangle=|\varphi^{(\uparrow)}_{\alpha}(k_{\uparrow})\rangle\otimes|\varphi^{(\downarrow)}_{\beta}(k_{\downarrow})\rangle$, $|\varphi^{(\sigma)}_{\alpha}(k_{\sigma})\rangle$ the Bloch state of $\alpha$ band of $H_{\sigma}$, and $\alpha,\beta=\pm$ the single-particle band index.
$I_{k_{\uparrow}k_{\downarrow}}=1$ ($-1$) indicates the absence (presence) of inter-species band inversion at the corresponding momenta.
The conservation of the total momentum $K=k_\uparrow+k_\downarrow$~\cite{PhysRevA.95.063630,PhysRevA.101.023620,PhysRevB.107.125161,PhysRevLett.133.140202}
(also see {Supplemental Note 3 A})
decouples subspaces with $K=0$ and $\pi$ we concern.
Thus the system can be characterized by four mutually dependent topological invariants with $k_\uparrow,k_\downarrow\in\{0,\pi\}$.
%and only three of them are independent from each other. 
Without loss of generality, we use the three invariants $\{I_{00}, I_{\pi\pi}, I_{0\pi}\}$
to characterize our system, with the forth one given by $I_{00}I_{\pi\pi}=I_{0\pi}I_{\pi0}$.

With further derivation~(see {Supplementary Notes 3 A and 3 B}), we find that the DGF effectively vanishes
in these subspaces ($K=0$ or $\pi$) unless at least one invariant takes a nontrivial value of $-1$.
%in the subspace $K=0$ ($K=\pi$) unless when $I_{00}$ and/or $I_{\pi\pi}$ ($I_{0\pi}$ and/or $I_{\pi0}$) takes a nontrivial value $-1$. 
To see this, we expand the DGF on the basis of $|\psi_{\alpha\beta}(k_\uparrow, k_\downarrow)\rangle$, and define its magnitude as the sum of absolute values of all matrix elements in the concerned subspaces,
\begin{eqnarray}
M=\sum%_{p,p',K}
\big|\langle\psi_{\alpha\beta}(k, K-k)|
H_{\rm DGF} |\psi_{\alpha'\beta'}(k',K-k')\rangle\big|^2.
\label{eq:DGF_matrix}
\end{eqnarray} 
%with $p=[\alpha,k]$ represents a set of band and momentum indices. 
The summation runs over (for indices with and without apostrophes) $\alpha=\pm$, $k\in[0,2\pi)$, and $K\in\{0,\pi\}$; 
and $\beta=-\alpha$ 
{so that $H_\sigma$ gives single-particle eigenenergies with opposite signs for the two species, contributing to the centra energy cluster in the spectrum that hosts bulk bound states in Fig. \ref{fig2}\textbf{a}.}
%as $H_\sigma$ would otherwise give nonzero real energies, in contrast to the observed bulk bound states with ${\rm Re}E=0$ in Fig. \ref{fig2}~(see \blue{Supplementary Note 3 A} \red{[It is a simple description even in SuppNote. Can we move it to the main text?]}).
In Figs. \ref{fig2}\textbf{c} and \ref{fig2}\textbf{d}, we display phase diagrams regarding the DGF magnitude $M$ and the topological invariants,
in the parameter space of $\theta_{\sigma}={\rm arg}(u_{\sigma}+iv_{\sigma})$ with fixed $|u_\sigma+iv_\sigma|$
for a better demonstration.
It is seen that $M\approx 0$ when and only when all invariants are trivial [labeled as $\{+++\}$ in Fig. \ref{fig2}\textbf{d}].

In the absence of the DGF, the system is simply the direct sum of two single-particle Hamiltonians, which may only accidentally host some unstable two-particle bound states at ${\rm Im} E=-\gamma_{\uparrow}/2$.
Therefore, non-accidental two-particle bulk bound states with ${\rm Im}E$ diverged from $-\gamma_\uparrow/2$ may emerge only when inter-species band inversion occurs at least at one set of single-particle high-symmetric points (with the corresponding $I_{kk'}=-1$).
In Fig. \ref{fig2}\textbf{e}, we display the maximal $S_m$ of all eigenstates in the same parameter region as Fig. \ref{fig2}\textbf{c} and \textbf{d}.
We find that $S_m$ takes a relatively large value mainly in the regimes when $M\neq 0$ with at least one of $\{I_{00}, I_{\pi\pi}, I_{0\pi}\}$ being $-1$.
In particular, $S_m$ approaches its maximal value ${\rm log}_2 L$ ($S_m\approx5$ for $L=32$) in the bright yellow regions,
indicaiting a strong entanglement between different species of particles, which persists under different system sizes (no shown).
Among different parameter regimes, the $\{---\}$ phase has inter-species band inversion occuring between arbitrary of high-symmetric momenta of the two species [see Fig. \ref{figsketch}\textbf{a}], and hosts the bulk bound states in Fig. \ref{fig2}\textbf{a}.
The other phases with mixture of ``$+$" and ``$-$" indices
also support similar bulk bound states with large $S_m$ and ${\rm Im}E$, as presented in the {Supplementary Figure S3}.
Notably, both the $\{---\}$ and $\{+++\}$ phases have each species being topologically trivial at the single-particle level (with $I_0I_\pi=1$),
but only the former hosts the bulk bound states.
Thus, the inter-species band inversion here represents an intrinsic ISTP that cannot be understood from single-particle topology.
We also note that in Fig. \ref{fig2}\textbf{e}, when $\theta_\uparrow=\theta_\downarrow$, accidental bulk bound states with large $S_m$ emerge in the regime with $\{+++\}$ due to the significant degeneracy of $|\psi_m\rangle$~{(see Supplementary Note 3 C)}, which are unstable against disorder [Fig. \ref{fig2}\textbf{f}].

Finally, we note that the inter-species topological invariants can also be applied to described the extrinsic ISTP discussed previously. Namely,
topological edge confined states appear (disappear) when an inter-species invariant $I_{\rm ext}=I_{00}I_{\pi\pi}$ takes the value of $-1$ ($1$). However, this invariant can be decomposed into the product of single-particle topological invariants, $I_{\rm ext}=I^{\uparrow}_{\rm sin}I^{\downarrow}_{\rm sin}$ (extra superscripts label the two species), reflecting the extrinsic nature of its inter-species topology.
\\

\begin{figure}
    \centering
    \includegraphics[width=1.0\linewidth]{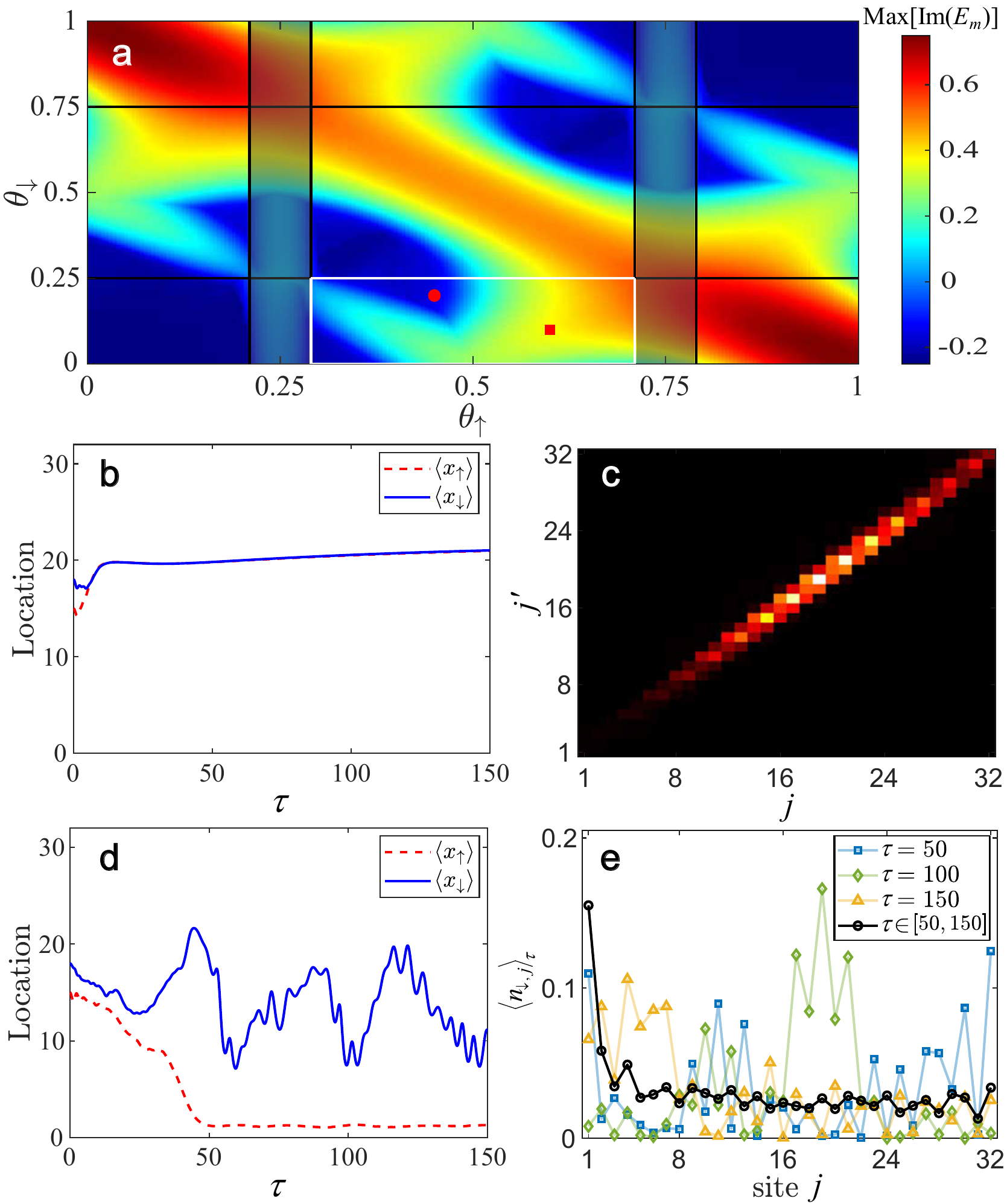}
   \caption{\textbf{The dynamical competition between two types of bound state.} \textbf{a} The largest imaginary energies for all PBC eigenstates, corresponding to that of bulk bound states in the inter-species nontrivial regimes. Black lines indicate the same phase boundaries as in Fig. \ref{fig2}.
   White box marks the region with $\mathcal{PT}$-broken edge confined states as in Fig. \ref{fig1}\textbf{a} to \textbf{c}.
   %It can reveal the largest imaginary energies of bulk bound states in the nontrivial regions. Here, the regions with $\mathcal{PT}$ broken phases of pseudospin-$\uparrow$ particle are covered by gray color. And edge and bulk bound states coexist in the region surrounded by white lines. 
   \textbf{b} The mean location of pseudospin-up and -down particles during the evolution process for the initial state $|\psi(0)\rangle=a^\dagger_{\uparrow,15}a^\dagger_{\downarrow,18}|0\rangle$.
   \textbf{c} The normalized two-particle correction $\widetilde{\Gamma}_{j,j'}$ of the evolved state at the end time $\tau=150$. 
   Parameters in \textbf{b} and \textbf{c} are $\theta_{\uparrow}=0.6\pi$ and $\theta_{\downarrow}=0.1\pi$, marked by the red square in \textbf{a}.
   \textbf{d} The same as \textbf{b}, but with $\theta_{\uparrow}=0.45\pi$ and $\theta_{\downarrow}=0.2\pi$, marked by the red circle in \textbf{a}.
   \textbf{e} Distribution of pseudospin-down particle at several different time $\tau$, and its average during the time interval $\tau\in[50,150]$, with the same parameters as in \textbf{d}.
   %In (b) and (c), $\theta_{\uparrow}=0.6\pi$ and $\theta_{\downarrow}=0.1\pi$, represented by the red square in (a), are set. (d) Similar to (b) but  with $\theta_{\uparrow}=0.45\pi$ and $\theta_{\downarrow}=0.2\pi$, marked by the red circle in (a). (e) The distribution of pseudospin-$\downarrow$ particle at several moments and the accumulation during the interval $t\in[50,150]$. 
   In all panels, $|u_{\sigma}+iv_{\sigma}|=\sqrt{2}$, $\gamma_{\uparrow}=0.5$, and $L=32$ are set.}
\label{fig3}
\end{figure}

\noindent
\textbf{Dynamical signatures.}
Since the edge confined and bulk bound states discussed above originate from different extrinsic and intrinsic inter-species topology, they can co-exist in certain parameter regimes, as shown in Fig. \ref{fig2}\textbf{a}. 
In such regimes, they are expected to compete with each other to dominate the system's dynamics, depending on which possesses the largest imaginary energies among all eigenstates.
In particular, the edge confined states in Fig. \ref{fig1}\textbf{a} have the pseudospin-up particle polarized at odd sites, which always leads to $\mathcal{PT}$-breaking and ${\rm Im}E\approx 0$, due to the on-site loss $\gamma_{\uparrow}$ on even sites~({see Supplementary Note 4 A}).
%because they have the pseudospin-up particle mostly localized at odd sites and are thus much less affected by the particle loss $\gamma_{\uparrow}$.
In contrast, $\gamma_{\uparrow}$ only assigns a $\mathcal{PT}$-symmetric imaginary energy $-\gamma_{\uparrow}/2$ to the bulk bound states with uniform distribution.
Alternatively, non-Hermitian DGF may also lead to their $\mathcal{PT}$-breaking and ${\rm Im}E> 0$, \LZT{associated with a transition at an exceptional point\red{[ON HOLD]}}~({see Supplementary Note 4 B}).
%whose $\mathcal{PT}$-breaking is induced by the non-Hermitian DGF, which can lead to ${\rm Im}E> 0$, as shown in Fig. \ref{fig2}(a)~\cite{suppmat}.
%thus $\gamma_{\uparrow}$ can only assign the same negative imaginary energy to them as to the rest bulk states. 
%However, they may also acquire positive imaginary energies from the non-Hermitian DGF, as shown in Fig. \ref{fig2}(a).
In Fig.~\ref{fig3}\textbf{a}, we display the largest imaginary energy under PBCs, ${\rm Max}[{\rm Im}(E_m)]$, 
whose positive values can only be attributed to bulk bound states (as edge confined states require OBCs).

To demonstrate the dynamics of different inter-species topological states, 
we consider two sets of of parameters in a parameter regime supporting both of them, marked by red square and circle in Fig.~\ref{fig3}\textbf{a}, with ${\rm Max}[{\rm Im}(E_m)]>0$ and $<0$, respectively.
We then choose an initial state $|\psi(0)\rangle$ with the two particles placed near the center of the lattice.
As shown by the single-species mean locations
$\langle x_{\sigma} \rangle$ in Fig. \ref{fig3}\textbf{b}, 
when ${\rm Max}[{\rm Im}(E_m)]>0$,
the two spatially separated particles soon evolve to the same position for the evolved state $|\psi(\tau)\rangle=e^{-i H \tau}|\psi(0)\rangle$ under OBCs, which describes the probability amplitude for the particles to remain in the system at time $\tau$.
The normalized two-particle correlation $\widetilde{\Gamma}_{j,j'}$ at $\tau=150$ is demonstrated in Fig.~\ref{fig3}\textbf{c}, which is exactly the form of bulk bound states.
On the other hand, in the other case with ${\rm Max}[{\rm Im}(E_m)]<0$,
the mean location of the pseudospin-up particle reaches the edge of the lattice at time $\tau\approx50$, while pseudospin-down particle oscillates persistently over time, as shown in Fig. \ref{fig3}\textbf{d}.
To unveil the pseudospin-down edge-localization dynamically,
we display in Fig. \ref{fig3}\textbf{e} the pseudospin-down distribution 
$\langle  n_{\downarrow,j} \rangle_{\tau}$ for the normalized evolved state $|\tilde{\psi}(\tau)\rangle=|\psi(\tau)\rangle/\sqrt{\langle \psi(\tau)|\psi(\tau)\rangle}$
at several different times, and its average during the time interval $\tau\in[50,150]$.
Although at each time the pseudospin-down particle is distributed across the lattice, clear edge localization can still be seen from its average position over time, reflecting the edge confined states of the extrinsic ISTP.

\noindent
\textbf{Experimental implementation.}
Our model may be implemented via Floquet engineering in cold-atoms loaded in optical lattices.
Here we provide a summary of our proposal and leave the explict implementation in Supplementary Note \LZT{7}, as it is based on several established studies.
Firstly, staggered hopping amplitudes can be realized by dimerized optical lattices~\cite{Atala2013}, and their pseudospin-dependency can be induced by ``tune-out" wavelengths with suitable polarization for Bose-Einstein condensates systems~\cite{Wen:21,Meng2023}.
The pseudospin-dependent loss can be realized by applying a \LZT{resonant} optical beam~\cite{Li2019,PhysRevLett.129.070401,Zhao2025}.
Finally, the DGF and the explicit form of the Hamiltonian in Eq. \eqref{eq:DGF_H} can be realized by introducing a two-time-scale Floquet modulation%~\cite{PhysRevA.68.013820,PhysRevX.4.031027,Eckardt_2015} 
of both hopping parameters and the pseudospin-dependent loss,
with a modulated inter-species interaction introduced through Feshbach resonance,%~\cite{RevModPhys.82.1225}, 
and a periodically-modulated intra-cell hopping with a phase difference $\pm\pi/2$ with respect to the natural tunneling implemented via Raman-assisted tunneling~\cite{PhysRevLett.107.255301,PhysRevLett.111.185301,PhysRevLett.111.185302,Jotzu2014,Aidelsburger2015,Kennedy2015,doi:10.1126/science.1259052} (see Supplementary Note \LZT{7}).

\vspace{1em} 
\noindent
{\large{\textbf{Discussions}}}.\\
\noindent
%In this work, we have studied a pseudospin-dependent SSH model with DGF. Under this exotic form of interaction, we discovered two types of topologically-correlated bound states: edge confined states that cluster around the edge and are characterized by intra-particle topology, and bulk bound states that display extended spread in the bulk and characterized by inter-particle topology. These states coexist and compete in the dynamical process. We also elaborate on how to realize this model using cold atoms in pseudospin-dependent optical lattices with DGF induced by Floquet protocols. Our work not only deepens our understanding of the role of DGF in topological states but also paves the way for further exploration of complex interactions in quantum systems, thereby contributing to the development of topological physics.
Through a minimal model with pseudospin-dependent parameters and an non-Hermitian DGF, we reveal two types of extrinsic and intrinsic ISTPs characterized by inter-species topology with corresponding topological states.
Notably, the DGF acts differently for the two ISTPs. For extrinsic ISTP, topological localization of one species induces non-reciprocal pumping to the other via DGF, 
leading to edge confined or anti-confined states depending on the non-reciprocal direction. 
%while the corresponding edge confined states are shifted in their imaingary energies due to an extra particle loss.
For intrinsic ISTP, DGF is activated by inter-species topology, inducing correlation that binds the two species of particles in the bulk.
%, and also shifting the imaginary energy due to its anti-Hermiticity.
Moreover, 
$\mathcal{PT}$-symmetry ensures real energies for most non-topological states, 
thus the inter-species topological states can be energetically identified and produce clear dynamical signatures when they acquire large imaginary energies from $\mathcal{PT}$-breaking, as further discussed in Supplemental Materials~{Supplementary Note 4}.
\LZT{On the other hand, since the functionality of DGF hinges on inter-species band inversion, ISTPs are not only expected to exist in $\mathcal{PT}$-symmetric non-Hermitian systems but also potentially realizable in other setups, including Hermitian ones.}
Note that non-Hermitian systems may suffer from numerical instability due to the non-normality of the Hamiltonian matrices \cite{xu2025numerical,1bvp-p2cz},
and the correctness of our numerical results is justified by an analysis of the non-normality of our model Hamiltonian in {Supplementary Note 5}.

In literature, topological properties are usually defined for a single species of particles.
Thus our results open an area of inter-species topological properties for non-identical particles, induced by strong correlations such as DGF.
Note that here we have focused on the case with a single particle of each species for a clearer demonstration. 
Nevertheless, similar inter-species topological states can also emerge with more particles loaded in the system, for different scenarios with the two species of particles being bosons, fermions, or their mixtures~(see {Supplementary Note \LZT{8}}).
The concept of inter-species topology suggests a new layer of topological organization based on correlation channels, opening prospects for classifying and engineering topology in hybrid quantum systems.

%\section*{Methods}
%\noindent
\vspace{1em} 
\noindent
{\large{\textbf{Data availability}}}.\\
\noindent
Raw numerical data from the plots presented are available from the authors upon reasonable request.
\\

\noindent
{\large{\textbf{Code availability}}}.\\
\noindent
Though not essential to the central conclusions of this work, computer codes for generating our figures are available from the authors upon reasonable request.
\\

%\bibliography{refs_2}
%\bibliographystyle{apsrev4-2}

\noindent
{\large{\textbf{References}}}.\\

\vspace{1em} 
\noindent
{\large\textbf{Acknowledgements}}.\\
\noindent
We acknowledge helpful discussion with  Zhihao Xu, Haiping Hu, Sen Mu, Hui-Qiang Liang, Jizhou Wu, Xiang-Chan Cheng and Zong-Yao Wang.
This work is supported by 
the National Natural Science Foundation of China (Grant No. 12474159).
\\

\noindent
{\large{\textbf{Author contributions}}}.\\
\noindent
Z. Lei initiated this project and performed the calculations with supervision from L. Li.
Both authors contribute significantly to the writing.
\\

\noindent
{\large{\textbf{Competing interests}}}.\\
\noindent
The authors declare no competing interests.
\\

\noindent
{\large{\textbf{Additional information}}}.\\
\noindent
Supplementary information is available for this paper at...
\\

\clearpage

\onecolumngrid

\begin{center}
\textbf{\large Supplementary information for ``Inter-species topological phases via a dynamical gauge field"}
\end{center}

\tableofcontents
\setcounter{secnumdepth}{2}

		%%%%%%%%%% Prefix a "S" to all equations, figures, tables and reset the counter %%%%%%%%%%
	\setcounter{equation}{0} \setcounter{figure}{0} \setcounter{table}{0} %
	\renewcommand{\theequation}{S\arabic{equation}} \renewcommand{\thefigure}{S%
		\arabic{figure}} \renewcommand{\bibnumfmt}[1]{[S#1]} 
	%\renewcommand{\citenumfont}[1]{S#1}
	%%%%%%%%%% Prefix a "S" to all equations, figures, tables and reset the counter %%%%%%%%%%

\section{Supplementary Note 1: detailed analysis of the edge (anti-)confined states}\label{appB}

\subsection{Effective single-species Hamiltonian for edge (anti-)confined states}\label{appBEFFH}

In this section, we derive the effective single-species Hamiltonian for edge (anti-)\MOD{confined} states \MOD{and analysis their $\mathcal{PT}$ symmetry breaking properties, aiming} to gain a thorough understanding of their underlying mechanism. 
For the states depicted in Fig.2 of the main text, the pseudospin-up particle is localized at the left edge of the lattice, which resembles the edge states of the Hermitian Su-Schrieffer-Heeger (SSH) model ($H_{\uparrow}$ in Eq.(1) of the main text).
%which can be obtained using the transfer matrix approach \cite{MacKinnon1983,PhysRevLett.70.982,RevModPhys.69.731,Liu_2023}. 
In the thermodynamic limit (where the number of sites $L\rightarrow\infty$), the left edge state has zero eigenenergy and 
can be obtained as followed by solving the Schr\"odinger equation:
\begin{eqnarray}
|\psi_{\rm left}\rangle\propto%\frac{1}{\sqrt{\mathcal{N}}}
|1\rangle+\kappa|3\rangle+\cdots+\kappa^{j - 1}|2j - 1\rangle+\cdots.
\label{rightEd}
\end{eqnarray}
Here, $|j\rangle$ represents the state of a pseudospin-up particle on the $j$-th site, and $\kappa = u_{\uparrow}/v_{\uparrow}$ is the decay index.%, and $\mathcal{N}=1/(1-\kappa^{2})$ is the normalization factor. 

Explicitly, the left edge state distributes only on odd sites, ensured by a chiral symmetry $\sigma_z H_{\uparrow}\sigma_z=-H_{\uparrow}$.
Therefore, this state remains a zero-energy eigenstate of the Hamiltonian $H_{\uparrow}-i\sum_{j = 1}^{L/2}\gamma_{\uparrow}n_{\uparrow,2j}$ when dissipation terms (added on to even sites) are included.

\begin{figure}
    \centering
    \includegraphics[width=1\linewidth]{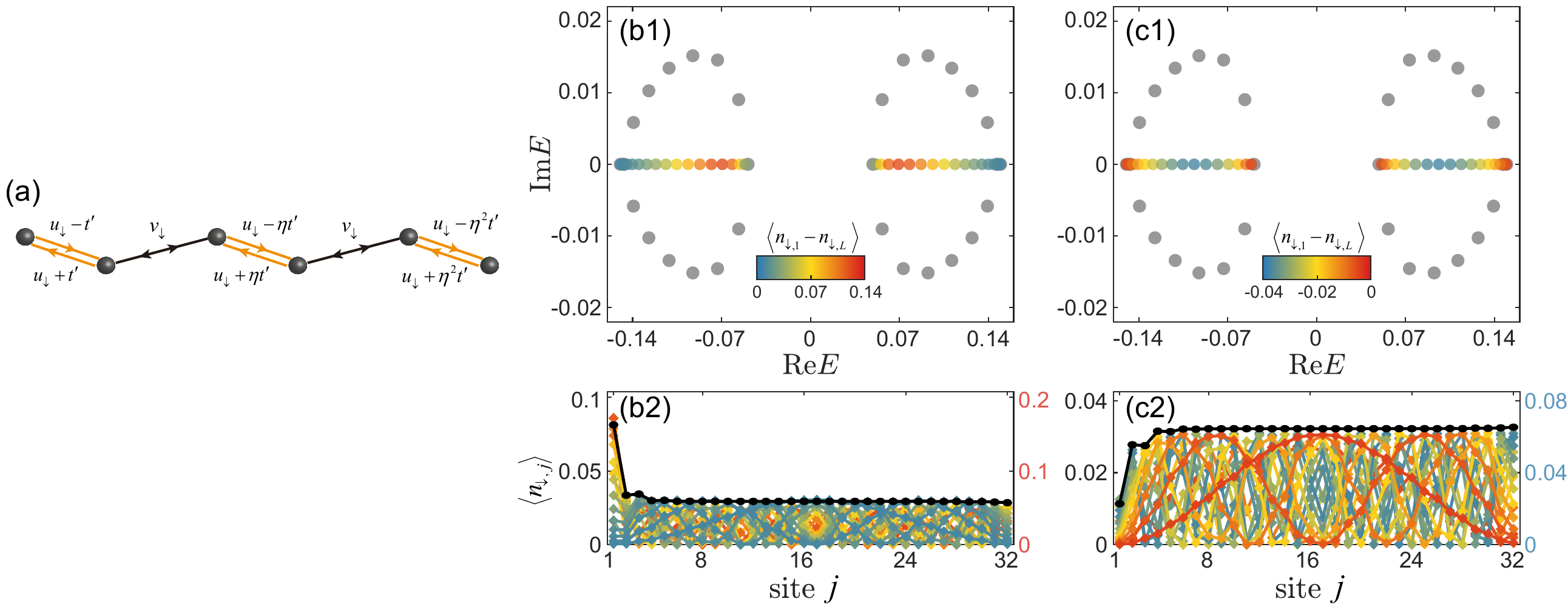}
 \caption{\textbf{A sketch and eigensolutions of the effective model $H_{\downarrow,{\rm eff}}$.} 
    (a) Lattice structure of the Hamiltonian $H_{\downarrow,{\rm eff}}$ in Eq.~\eqref{meanDO}. Here $\eta=\kappa^2=u^2_{\uparrow}/v^2_{\uparrow}$ and $t'=t(1-\eta)$. (b1) The PBC (gray dots) and OBC (colored) spectra of$H_{\downarrow,{\rm eff}}$. The OBC spectrum is marked by colors according to the edge-density imbalance. (b2) Distributions of OBC eigenstates, marked by the same colors as in (b1). Their average distribution are shown by the black line and dots. Parameters in (b1-b2) are $v_{\uparrow}=5$, $v_{\downarrow}=0.5$, $\gamma_{\uparrow}=t = 0.5$, $u_{\uparrow}=2$, and $u_{\downarrow}=1$, i.e., same as those for Figs. 2(a-c) of the main text. (c1-c2) The same as (b1-b2), but with $u_{\uparrow}=-2$ and $u_{\downarrow}=-1$, which are the same as those for Figs. 2(d-f) of the main text.
   The features presented in (b1-b2) resemble the edge \MOD{confined} states, while those in (c1-c2) resemble the edge anti-\MOD{confined} states.}
\label{figS2}
\end{figure}

For the edge (anti)-\MOD{confined} states in the multi-particle scenario, we assume that the DGF does not affect distribution of the pseudospin-up particle, and treat it as a mean field acting on the pseudospin-down particle. Thus, an effective Hamiltonian for the pseudospin-down particle can be obtained from Eq. (1) of the main text, given by 
\begin{eqnarray}
H_{\downarrow,{\rm eff}}&=&H_{\downarrow}+H_{\downarrow,{\rm DGF}},\nonumber\\
H_{\downarrow,{\rm DGF}}&=&\sum_{j = 1}^{L/2}\left[t\left(\langle n_{\uparrow,2j - 1}\rangle-\langle n_{\uparrow,2j}\rangle\right)a^\dagger_{\downarrow,2j - 1}a_{\downarrow,2j}\right]-h.c.,
\label{meanDO}
\end{eqnarray}
where $\langle n_{\uparrow,j}\rangle\equiv\langle \psi_{\rm left}|a^\dagger_{\uparrow,j}a_{\uparrow,j}|\psi_{\rm left}\rangle$.
Substituting the the left edge state in Eq.~\eqref{rightEd} into Eq.~\eqref{meanDO}, 
the effective Hamiltonian can be expressed as
\begin{eqnarray}
H_{\downarrow,{\rm eff}}=H_{\downarrow}+\sum_{j=1}^{L/2}\left[t_{j}^{\rm nr}a^\dagger_{\downarrow,2j-1}a_{\downarrow,2j}-h.c.\right],
\end{eqnarray}
 with $\eta=\kappa^2$, anti-Hermitian hopping amplitudes $t^{\rm nr}_j=t'\eta^{j-1}$, and $t'=t(1-\eta)$.
%the formula of the effective Hamiltonian in Eq.(3) of the main text can be obtained. 
A sketch of the lattice structure of $H_{\downarrow,{\rm eff}}$ is given by Fig.~\ref{figS2}(a).
We can see that non-reciprocity arises from the modification of  $t^{\rm nr}_j$ on intra-cell hopping,
and its strength decays exponentially from the left edge to the right, in contrast to the non-Hermitian SSH model with a uniform non-reciprocal strength in Refs.~\cite{PhysRevLett.121.086803,PhysRevLett.121.026808}.

%Compared with the non-Hermitian SSH model with intra-cell non-reciprocal hopping studied in Refs.~\cite{PhysRevLett.121.086803,PhysRevLett.121.026808}, the strength of non-reciprocal hopping in this effective model decays exponentially towards the bulk.

We plot the spectra and particle distributions in Figs.~\ref{figS2}(b1-b2) and (c1-c2) for $H_{\downarrow,{\rm eff}}$ with the same parameters as those in Fig. 2 of the main text.  
As demonstrated here, the distribution of pseudospin-down particle under OBC shows a bulge (dip) at the left edge in Fig.~\ref{figS2}(b2) [Fig.~\ref{figS2}(c2)], capturing the essence of the edge \MOD{confined} (anti-\MOD{confined}) states. 
These results demonstrate that the edge \MOD{confined} (anti-\MOD{confined}) states are induced by the interplay between DGF and the edge localization induced by non-trivial single-particle topology. 
In addition, we note that the PBC spectrum holds nontrivial spectral winding (i.e., loop-like PBC spectrum) and becomes pure real under OBCs [see Figs.~\ref{figS2}(b1) and (c1)], 
analogous to the spectral features of conventional non-Hermitian skin effect (NHSE) \cite{PhysRevLett.124.056802,PhysRevLett.124.086801,PhysRevLett.125.126402}.
%which originates from the non-reciprocal hopping and similar to the one studied in Refs.~\cite{PhysRevLett.121.086803,PhysRevLett.121.026808}.
However, as the non-Hermitian non-reciprocity decays exponentially from the left edge, the Hamiltonian remains roughly Hermitian in rest of the system, resulting in an evenly distribution away from the left edge. Finally, on the left edge of the lattice, we may have edge \MOD{confined} states with a large pseudospin-down distribution, or edge anti-\MOD{confined} states with a drop of pseudospin-down distribution, depending on the direction of non-reciprocity [Figs. \ref{figS2}(b2) and (c2)].
These observations are in sharp contrast to the conventional NHSE in non-Hermitian systems with a uniform non-reciprocity along the lattice.

\subsection{$\mathcal{PT}$ symmetry and edge (anti-)confined states}\label{appBPT}
\MOD{Here, we analyze the relationship between $\mathcal{PT}$ symmetry and edge (anti-)confined states. 
In our study, we have considered two species of particles without an actual spin degree of freedom, 
and thus the time-reverisal symmetry (TRS) for spinless system represented by a complex conjugation, $i\rightarrow -i$.
When both a system and its eigenstates satisfy the spinless $\mathcal{PT}$ symmetry (i.e., in a $\mathcal{PT}$-unbroken phase), the spectrum must be purely real, which rules out the occurrence of spectral winding and the NHSE. However, if some eigenstates of the system break $\mathcal{PT}$ symmetry, their corresponding energies can be complex even when the entire system is $\mathcal{PT}$-symmetric. Thus, in the $\mathcal{PT}$-broken phase, the NHSE can be induced by appropriately designing the system such that its complex eigenenergies possess nontrivial winding~\cite{lei2024activating}.}

\MOD{In our model, the explicit origin of skin-like localization under $\mathcal{PT}$ symmetry can be understood from two aspects: the $\mathcal{PT}$ symmetry of the overall system and the single-particle $\mathcal{PT}$ symmetry of each species when DGF is absent. The details are as follows.
When DGF is absent, our system can be divided into two single-particle models corresponding to the two species, whose Bloch Hamiltonians are given by
\begin{eqnarray}\label{sigbloch} 
H_{\sigma,\mathcal{PT}}(k)=(u_{\sigma}+v_{\sigma})\tau_x+(v_{\sigma}\sin k)\tau_y+\frac{i\gamma_{\sigma}}{2}(\tau_z-\tau_0),
\end{eqnarray}
where $\tau_{x,y,z,0}$ denote the Pauli matrices and identity matrix acting on the sublattice degree of freedom. After a global energy shift (i.e., by removing the $\tau_{0}$ term), these Hamiltonians satisfy the single-particle $\mathcal{PT}$ symmetry for spin-less systems, $\tau_xH_{\sigma,\mathcal{PT}}(k)\tau_x=H^*_{\sigma,\mathcal{PT}}(k)$. In real space, the symmetry operations involve $a_{\sigma,j}\rightarrow a_{\sigma,L+1-j}$ (for a single species of pseudospin) and $i\rightarrow-i$.
}

\MOD{Moving forward, when $|u_{\sigma}|<|v_{\sigma}|$, the associated single-particle Hamiltonian enters a topologically nontrivial phase. In our setup, we assign the pseudospin-up Hamiltonian to this nontrivial topological phase. As elaborated in our main text, the single-particle topological edge states, mediated by the DGF, give rise to non-reciprocal hopping for the pseudospin-down particle—an effect that disrupts its single-particle $\mathcal{PT}$ symmetry. This is corroborated by results in Subsec.~\ref{appBEFFH}, where the single-particle mean-field Hamiltonian for the pseudospin-down particle is shown to violate the aforementioned single-particle $\mathcal{PT}$ symmetry and support both a loop-like PBC spectrum and skin-like OBC eigenstates.}

\MOD{Turning to the overall multi-particle Hamiltonian, it breaks the individual single-particle $\mathcal{PT}$ symmetries yet preserves their combined $\mathcal{PT}$ symmetry. Specifically, the Hamiltonian remains invariant under the symmetry operations $a_{\sigma,j}\rightarrow a_{\sigma,L+1-j}$, $i\rightarrow-i$ applied to both pseudospin-up and pseudospin-down particles (after a global energy shift). 
Explicitly, 
\begin{align}
\mathcal{PT} H_{\sigma}(\mathcal{PT})^{-1}&=\sum_{j=1}^{L/2}\left(u_\sigma a^\dagger_{\sigma,L+2-2j}a_{\sigma,L+1-2j} +v_\sigma a^\dagger_{\sigma,L+1-2j}a_{\sigma,L-2j}\right)+h.c.\nonumber\\
&=\sum_{j^\prime=1}^{L/2}\left(u_\sigma a^\dagger_{\sigma,2j^\prime}a_{\sigma,2j^\prime-1} +v_\sigma a^\dagger_{\sigma,2j^\prime-1}a_{\sigma,2j^\prime-2}\right)+h.c.=H_\sigma,\\
\mathcal{PT} H_{\rm DGF}(\mathcal{PT})^{-1}&=\sum_{\sigma\neq\bar{\sigma}}\sum_{j=1}^{L/2}\bigg[t\left(n_{\bar{\sigma},L+2-2j}-n_{\bar{\sigma},L+1-2j}\right)a_{\sigma,L+2-2j}^\dagger a_{\sigma,L+1-2j}\bigg]-h.c.\nonumber\\
&=\sum_{\sigma\neq\bar{\sigma}}\sum_{j^\prime=1}^{L/2}\bigg[t\left(n_{\bar{\sigma},2j^\prime}-n_{\bar{\sigma},2j^\prime-1}\right)a_{\sigma,2j^\prime}^\dagger a_{\sigma,2j^\prime-1}\bigg]-h.c.
=H_{\rm DGF},
\end{align}
with ${2j\to L+2-2j^\prime}$.For the total Hamiltonian $H$ with species-dependent particle loss, we consider an extra global energy shift,
\begin{align}
H_{\rm s}&=H+i\sum_\sigma \gamma_\sigma N_\sigma/2\nonumber\\
&=\left[\sum_{\sigma=\uparrow,\downarrow}\left(H_\sigma - i\sum_j^{L/2}\gamma_\sigma n_{\sigma,2j}\right)+H_{\rm DGF}\right]+i\sum_{\sigma=\uparrow,\downarrow} \sum_{j=1}^{L/2}\left(n_{\sigma,2j-1}+n_{\sigma,2j} \right)\gamma_\sigma /2\nonumber\\
&=\sum_{\sigma=\uparrow,\downarrow}\left(H_\sigma + i\sum_j^{L/2}\left(n_{\sigma,2j-1}-n_{\sigma,2j} \right)\gamma_\sigma /2\right)+H_{\rm DGF}.
\end{align}
This shifted Hamiltonian is $\mathcal{PT}$-symmetric, as it satisfies
\begin{align}
\mathcal{PT} H_{\rm s} (\mathcal{PT})^{-1}&=\sum_{\sigma=\uparrow,\downarrow}\left(H_\sigma - i\sum_j^{L/2}\left(n_{\sigma,L+2-2j}-n_{\sigma,L+1-2j} \right)\gamma_\sigma /2\right)+H_{\rm DGF}\nonumber\\
&=\sum_{\sigma=\uparrow,\downarrow}\left(H_\sigma - i\sum_{j^\prime}^{L/2}\left(n_{\sigma,2j^\prime}-n_{\sigma,2j^\prime-1} \right)\gamma_\sigma /2\right)+H_{\rm DGF}=H_{\rm s}.
\end{align}
This global $\mathcal{PT}$ symmetry imposes a requirement: any complex eigenenergies must have their complex conjugate counterparts, a feature that should manifest in the loop-like PBC spectrum discussed earlier in the single-particle context. Our model confirms this behavior: additional eigenenergies form a loop-like spectrum near ${\rm Im}E=-\gamma_{\uparrow}$ as shown in Fig. 2(a) and (d) of the main text. After a global energy shift, these serve as the complex conjugates of those near ${\rm Im}E=0$, as illustrated in Fig. 2(b,c) and (e,f) of the main text.}

\MOD{Finally, the magnitudes of these imaginary energy components can also be interpreted within the single-particle framework. Specifically, the single-particle topological edge states of pseudospin-up break the corresponding single-particle $\mathcal{PT}$ symmetry. This results in a complex conjugate pair of eigenenergies at ${\rm Im}E=-\gamma_{\uparrow}$ and ${\rm Im}E=0$ after a global energy shift.
For the overall multi-particle system, these imaginary-energetic features are carried by the DGF to the pseudospin-down particle, which is PT-symmetric at single-particle level (without DGF); while the non-Hermitian DGF itself further induces extra complex energy loops centered at ${\rm Im}E=-\gamma_{\uparrow}$ and ${\rm Im}E=0$ under PBCs.
In other words, the complex eigenenergies of the $\mathcal{PT}$-broken inter-species edge (anti-)confined states come from two sources: (i) imaginary energies inherited from the single-particle $\mathcal{PT}$-broken features of pseudospin-up topological edge states; and (ii) $\mathcal{PT}$-asymmetric non-reciprocity mediated by the DGF, giving rise to the loop-like PBC spectrum and skin-like OBC states.}

\section{Supplementary Note 2: coincidental edge confined and anti-confined states}\label{appA}
In the main text, we have shown the emergence of edge \MOD{confined} and anti-\MOD{confined} states in extrinsic inter-species topological phases (ISTPs) where only one species of particle is topologically nontrivial.
%For the edge anti-bound states, one species of particle exhibits topological localization, while another shows a decay at the associated edge.
In this section, we show that eigenstates with similar distribution also emerge when two particles are both topologically nontrivial at the single-particle level.
However, these states arise solely from single-particle topology, and is trivial in the sense of \MOD{dynamical gauge field (DGF)}-mediated inter-species topology.

\begin{figure}
    \centering
    \includegraphics[width=1\linewidth]{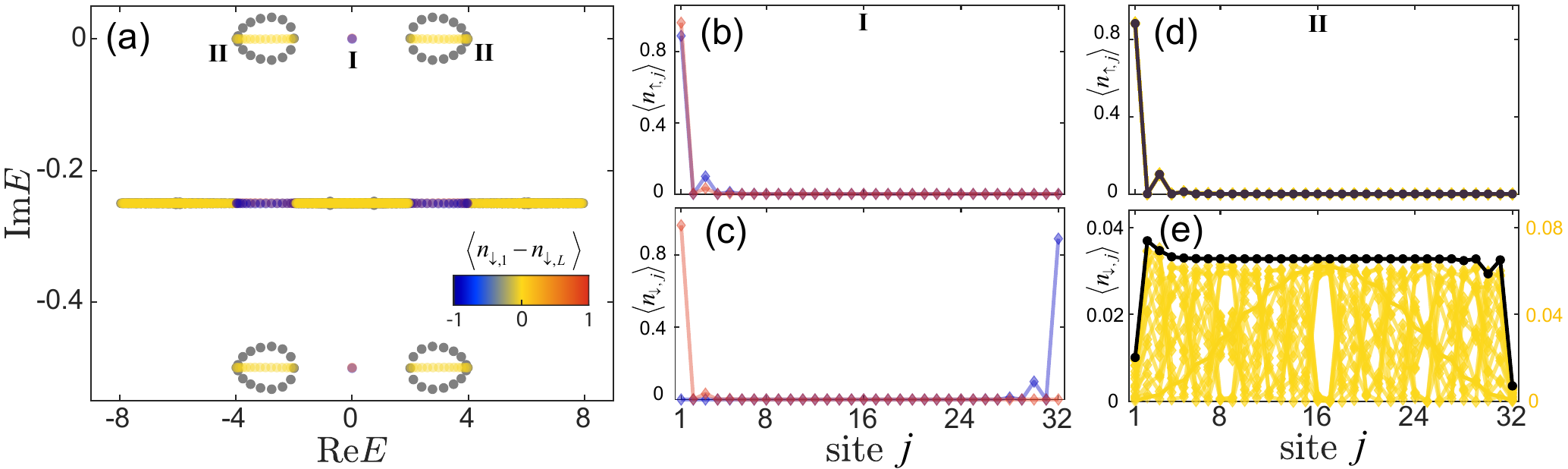}
 \caption{\textbf{Trivial edge anti-\MOD{confined} states.} 
    (a) The full-OBC spectrum of the Hamiltonian in Eq.(1) of the main text, marked by the edge-density imbalance of the pseudospin-down particle. Gray dots represent the eigenenergies with PBCs taken only for the pseudospin-down particle.
Parameters are chosen to have both species of particles being topologically nontrivial at single-particle level ($u_{\uparrow}=u_{\downarrow}=1$, $v_{\uparrow}=v_{\downarrow}=3$, and $\gamma_{\uparrow}=t=0.5$), in contrast to Fig. 2 of the main text where only the pseudospin-$\uparrow$ particle is nontrivial.
%In contrast to Fig.1 of the main text where only the spin-$\uparrow$ particle is topologically nontrivial, both species of particles are nontrivial here ($u_{\uparrow}=u_{\downarrow}=1$, $v_{\uparrow}=v_{\downarrow}=3$, and $\gamma_{\uparrow}=t=0.5$).
    (b) and (c) the distributions of pseudospin-up and pseudospin-down particles for the zero-energy eigenstates, respectively. These states are indexed by ${\rm I}$ and marked with the same colors as in (a).
    (d) to (e) the same as (b) to (c), but for the edge anti-\MOD{confined} states with ${\rm Im}(E) \approx 0$ and ${\rm Re}(E) \neq 0$, which are indexed by ${\rm II}$ in (a). The average distribution over eigenstates is also presented (black lines and dots). \MOD{The system size in this figure is $L=32$,
which gives a Hamiltonian matrix of dimension $L^2=1024$.}}
\label{figS1}
\end{figure}

Specifically, in Fig.~\ref{figS1}(a), we present the energy spectrum of the two particles under open boundary conditions (OBCs),
as well as that of the pseudospin-up particle under OBCs and the pseudospin-down particle periodic boundary conditions (PBCs).
%for both spin-up and spin-down particles under open boundary conditions (OBCs), as well as for spin-down particle under periodic boundary conditions (PBCs) with spin-up particle retained under OBCs.
Eigenenergies under full-OBCs are colored according to the edge-density imbalance of the spseudopin-down particle, $\langle n_{\downarrow,1}-n_{\downarrow,L}\rangle$.
It is seen that two pairs of eigenstates with zero energies $E=0$ and pure imaginary energies $E=-i\gamma_{\uparrow}$, respectively, are separated from the other states in energy and occur only under full-OBCs.
Explicitly, these states arises from the nontrivial single-particle topology, which drives each particle to localize at the edges even in the absence of DGF.
%correspond to topological-topological states, where each species of particles  exhibits topological localization. 
As shown in Figs. \ref{figS1}(b) and (c), the two zero-energy eigenstates have the pseudospin-up particle localized near the left edge, and the pseudospin-down particle localized at both edges,
Note that the other two eigenstates with $E=-i\gamma_{\uparrow}$ exhibit the same distribution for the pseudospin-down particle, 
but right-localization for the pseudospin-up particle (not shown), which suffers from the on-site loss term $\gamma_{\uparrow}$ on the right-most lattice site for our chosen parameters.
%In detail, the spin-up particle is localized only near the left boundary, while the spin-down particle is localized at both boundaries. 

Next, we focus on another set of eigenstates with ${\rm Im}(E) \approx 0$ and ${\rm Re}(E) \neq 0$.
As shown in Fig. \ref{figS1}(e), the pseudospin-up particle of these states also localize at the left, reflecting its single-particle topological localization.
On the other hand, as shown by Fig. \ref{figS1}(f), the pseudospin-down particle distributes almost evenly in the bulk, but much less on the two edges. 
Although these states resembles the edge anti-\MOD{confined} states discussed in the main text, they arise also from the nontrivial topology combining with orthogonality between different eigenstates.
Namely, the pseudospin-down particle is described by a Hermitian Hamiltonian $H_{\downarrow}$ at the single-particle level, thus, different eigenstates shall have pseudospin-down distributions nearly orthogonal to each other. 
Consequently, 
when $H_{\downarrow}$ is topologically nontrivial, its bulk states have vanishing distributions on edges, which, combined with the topological localization of the pseudospin-up particle, give rise to the edge anti-\MOD{confined} states in Fig. \ref{figS1}.
Similarly, the other set of eigenstates  with ${\rm Im}(E)  \approx -i\gamma_{\uparrow}$ and ${\rm Re}(E) \neq 0$ shall have the same distribution for the pseudospin-down particle, but right-localization for the pseudospin-up particle (no shown), due to the chosen on-site loss term $\gamma_{\uparrow}$.

Finally, we would like to emphasize that the formation of these edge \MOD{confined} and anti-\MOD{confined} states are coincidental at multi-particle level, as they origin from the direct product of single-particle eigenstates of the two species of particles, without relying on any sort of inter-particle interaction such as the \MOD{DGF}.
On the other hand, it is noteworthy that the DGF-mediated topological correlation between different species still functions here, as shown by the small density imbalance between the two edges in Fig.~\ref{figS1}(e). 
Explicitly, the density on the left is slightly larger than that on the right, as the DGF-generated non-reciprocal hopping [see Eq. (2) of the main text] is stronger toward the left for the chosen parameters.

\section{Supplementary Note 3: the effects and representation of dynamical gauge field}\label{appC}
 
In this section, we investigate the role of DGF in engineering bulk bound states of the intrinsic ISTPs, and derive their eigenenergies in certain symmetric parameter regimes, \MOD{as well as discuss the $\mathcal{PT}$ transition boundaries}.
%under symmetric parameter configurations, which can provide deeper insights into the topological origin of bulk bound states.
Without loss of generality, we only consider the scenario with $v_{\sigma}\geq0$.
This is because for the Hamiltonian $H$ in Eq.(1) of the main text, the scenario with $v_{\sigma}\leq0$ can be mapped to that with $v_{\sigma}\geq0$
%and the same other parameters 
through a gauge transformation $a_{\sigma,2j-1}\to(-1)^{j-1}a^\dagger_{\sigma,2j-1}$ and $a_{\sigma,2j}\to(-1)^{j-1}a^\dagger_{\sigma,2j}$, without affecting other parameters.
In addition, this transformation also transforms the momentum state defined below as $|k_{\sigma}\rangle\to|k_{\sigma}+\pi\rangle$.

\begin{figure}
    \centering
    \includegraphics[width=0.9\linewidth]{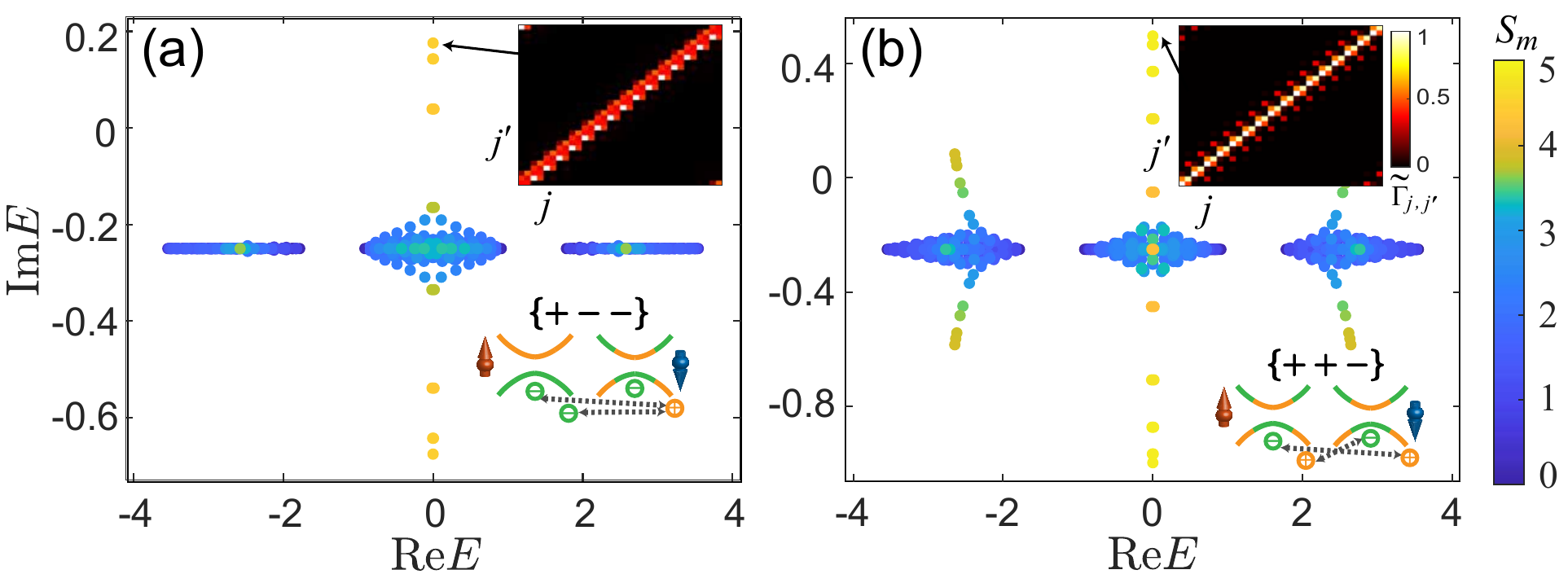}
   \caption{\textbf{Bulk bound states in different topological phases.}  (a) The PBC spectra of the Hamiltonian in Eq.(1) of the main text with $\theta_{\uparrow}=0.1\pi$ and $\theta_{\downarrow}=0.4\pi$ with $\theta_{\sigma}={\rm arg}(u_{\sigma}+iv_{\sigma})$, $|u_{\sigma}+iv_{\sigma}|=\sqrt{2}$, and $\gamma_\uparrow=t=0.5$. %[i.e., same as parameters used in Fig.~\ref{figS4}(a1)-(a2)]. 
The normalized two-particle correction of bulk bound states [Eq.(4) of the main text] with the highest entanglement entropy (marked by arrows), along with the associated topological invariants $\{I_{00},I_{\pi\pi},I_{0\pi}\}$ and symmetry indicators at high-symmetric momenta, are shown in the insets. Here, the occurrences of inter-species band inversion are emphasized by dashed lines with arrows. (b) same as (a), but with parameters $\theta_{\uparrow}=0.4\pi$ and $\theta_{\downarrow}=0.6\pi$. %[i.e., same as parameters used in Fig.~\ref{figS4}(b1)-(b2)].
These two cases show different inter-species band inversion compared to Fig. 1(a) in the main text, and also support bulk bound states with large $S_m$ and ${\rm Im}E$. \MOD{The system size in this figure is $L=32$, which gives a Hamiltonian matrix of dimension $L^2=1024$.} 
}
\label{figSBBS}
\end{figure}

Before moving on to our detailed analysis, we note that we have only shown numerical results of bulk bound states with inter-species topological invariants $\{I_{00},I_{\pi\pi},I_{0\pi}\}=\{---\}$.  In Fig. \ref{figSBBS}, we present the spectra and the appearance of bulk bound states 
in two other phases with nontrivial inter-species topology, 
verifying our conclusion that bulk bound states arise when at least one inter-species topological invariant takes the nontrivial value of $-1$.

\subsection{The representation of $H_{\rm DGF}$}\label{appC1}
In this subsection, we unveil that the DGF vanishes when and only when all the inter-species topological indices are trivial.
Here we will focus mostly on the high-symmetric single-particle quasi-momenta $k_{\sigma}\in\{0,\pi\}$ ($\sigma\in\{\uparrow,\downarrow\}$ for the two species), where the inter-species topological indices are defined on and the single-particle energies affect less the overall system (so that the DGF plays a more important role), as we will elaborate later.
In addition, note that the dissipation $\gamma_\sigma$ can be separated into a staggered imaginary potential and an overall imaginary-energy shift, namely, $\tau_z$ and $\tau_0$ terms in Eq.~\eqref{sigbloch}, respectively.
The former does not significantly affect the bulk bound states which distribute evenly on even and odd lattices; and the latter merely shifts all eigenenergies with the same amount of imaginary values.
Thus, we will omit the dissipation $\gamma_\sigma$ in the rest of this section.

\subsubsection{A brief analysis on the high-symmetric momenta}
The relation between the DGF and inter-species topology can be captured by analysing the single-particle eigenstates at the high-symmetric momenta $k_{\sigma}\in\{0,\pi\}$, where the single-particle Hamiltonian fo spin-$\sigma$ becomes $H_{\sigma}=(u_\sigma+v_\sigma)\tau_x$. 
The periodic part of the eigenstates at these momenta are either odd or even superpositions over the two sites in the unit cell: $|p\rangle_{\sigma}=(|2j-1\rangle_{\sigma}^{\rm cell}+p|2j\rangle_{\sigma}^{\rm cell})/\sqrt{2}$ with $p=\pm$, and there eigenenergies are given by $E_{\sigma,p} =p (u_\sigma+v_\sigma)$.
For a given set of high-symmetric momenta, the multi-particle eigenstates of $\sum_\sigma H_\sigma$ are given by $|p\rangle_{\uparrow} \otimes |p'\rangle_{\downarrow}$ with four different combinations of $p,p'=\pm$,
and we consider the two of them with ${\rm sgn}[E_{\uparrow,p}]=-{\rm sgn}[E_{\downarrow,p'}]$, denoted as 
$|p\rangle_{\uparrow} \otimes |p'\rangle_{\downarrow}$ and $|\overline{p}\rangle_{\uparrow} \otimes |\overline{p'}\rangle_{\downarrow}$.
This is because these two eigenstates have eigenenergies closest to $E=0$, corresponding to the central energy cluster in Fig. \ref{figSBBS} and Fig. 3(a) of the main text that gives the bulk bound states with imaginary eigenenergies.
%That is, the periodic part of the eigenstates for the spin-$\sigma$ Hamiltonian $H_{\sigma}$ at these momenta are either odd or even superpositions over the two sites in the unit cell: $|p\rangle_{\sigma}=(|2j-1\rangle_{\sigma}^{\rm cell}+p|2j\rangle_{\sigma}^{\rm cell})/\sqrt{2}$ with $p=\pm$.
%When both particle species are at high-symmetry points, two nearly degenerate eigenstates of $\sum_{\sigma} H_{\sigma}$, with the particles residing in opposite bands, can be expressed as $|p\rangle_{\uparrow} \otimes |p'\rangle_{\downarrow}$ and $|\overline{p}\rangle_{\uparrow} \otimes |\overline{p'}\rangle_{\downarrow}$.
Additionally, the DGF Hamiltonian can be decomposed as:
\begin{eqnarray}\label{DGFadd}
H_{\mathrm{DGF}} &=& H_{\mathrm{DGF},\uparrow \rightarrow \downarrow} + H_{\mathrm{DGF},\downarrow \rightarrow \uparrow}, \nonumber \\
H_{\mathrm{DGF},\bar{\sigma} \rightarrow \sigma} &=& \sum_{j=1}^{L/2} \left[ t \left(n_{\bar{\sigma},2j-1} - n_{\bar{\sigma},2j}\right) a^\dagger_{\sigma,2j-1} a_{\sigma,2j} - \mathrm{h.c.} \right].
\end{eqnarray}
Its action on the product state $|p\rangle_\sigma \otimes |p'\rangle_{\bar{\sigma}}$ is given by:
\begin{eqnarray}\label{DGFhier1}
H_{\mathrm{DGF},\bar{\sigma} \rightarrow \sigma} \, |p\rangle_{\sigma} \otimes |p'\rangle_{\bar{\sigma}} &=& p t \, |\bar{p}\rangle_{\sigma} \otimes |\bar{p}'\rangle_{\bar{\sigma}},
\end{eqnarray}
which leads to:
\begin{eqnarray}\label{DGFhier2}
H_{\mathrm{DGF}} \, |p\rangle_{\sigma} \otimes |p'\rangle_{\bar{\sigma}} &=& (p + p') t \, |\bar{p}\rangle_{\uparrow} \otimes |\bar{p}'\rangle_{\downarrow}.
\end{eqnarray}
Thereby, in the subspace spanned by these two nearly degenerate eigenstates, the matrix elements of the DGF Hamiltonian are evaluated as:
\begin{eqnarray}\label{DGFhier3}
{}_{\uparrow}\langle p| \otimes {}_{\downarrow}\langle p'| \, H_{\mathrm{DGF}} \, |p\rangle_{\uparrow} \otimes |p'\rangle_{\downarrow} =0,
{}_{\uparrow}\langle \overline{p}| \otimes {}_{\downarrow}\langle \overline{p}'| \, H_{\mathrm{DGF}} \, |p\rangle_{\uparrow} \otimes |p'\rangle_{\downarrow} = (p + p') t,
\end{eqnarray}
which reveals that the DGF term has non-vanishing contribution only when $p + p' \neq 0$.
Note that we have chosen the multi-particle eigenstates with ${\rm sgn}[E_{\uparrow,p}]=-{\rm sgn}[E_{\downarrow,p'}]$, namely,
the upper single-particle band of one species and the lower one of the other.
Thus, the above condition indicates that these two bands have the same single-particle topological indices,
corresponding to the occurrence of an inter-species band inversion (which is defined for the lower single-particle bands of the two species).

\subsubsection{A more comprehensive analysis of the DGF}
More specifically, we express the DGF in terms of the eigenbasis of $\sum_\sigma H_\sigma$, 
\begin{eqnarray}\label{Blcoh} 
|\psi_{\alpha\beta}(k_{\uparrow},k_{\downarrow})\rangle=|\varphi^{(\uparrow)}_{\alpha}(k_{\uparrow})\rangle\otimes|\varphi^{(\downarrow)}_{\beta}(k_{\downarrow})\rangle,\nonumber\\
|\varphi^{(\sigma)}_{\alpha}(k_{\sigma})\rangle=|\mu^{(\sigma)}_{\alpha}(k_{\sigma})\rangle\otimes|k_{\sigma}\rangle,
%|\varphi_{\beta}(k_{\downarrow})\rangle=|\mu^{(\downarrow)}_{\beta}(k_{\downarrow})\rangle\otimes|k_{\downarrow}\rangle,
\end{eqnarray}
where 
$|k_{\sigma}\rangle=\sqrt{2/L}\sum_j^{L/2}e^{ik_{\sigma}j}|j\rangle_{\sigma}^{\rm cell}$ is the momentum state with $k_{\sigma}\in\{-L\delta_k/4,...,\delta_k,2\delta_k,...,(L-1)\delta_k/4\}$, $\delta_k=4\pi/L$, and $|j\rangle_\sigma^{\rm cell}$ the basis of the $j$th unit cell; and 
$$|\mu^{(\sigma)}_{\alpha}(k_{\sigma})\rangle=[e^{i\phi_{\sigma}(k_{\sigma})/2},\alpha e^{-i\phi_{\sigma}(k_{\sigma})/2}]^T/\sqrt{2}=\left(e^{i\phi_{\sigma}(k_{\sigma})/2}|2j-1\rangle_\sigma+\alpha e^{-i\phi_{\sigma}(k_{\sigma})/2}|2j\rangle_\sigma\right)/\sqrt{2}$$ 
denotes the periodic part of the Bloch state of the $\alpha$-th band with $\phi_{\sigma}(k_{\sigma})={\rm arg}(u_{\sigma}+v_{\sigma}e^{-ik_{\sigma}})$.
In this basis, the SSH-eigenenergies $E_{\alpha\beta}(k_{\uparrow},k_{\downarrow})$ act as diagonal elements while $H_{\rm DGF}$ becomes a real skew-symmetric matrix $[M]$, which are given by
\begin{align}
E_{\alpha\beta}(k_{\uparrow},k_{\downarrow})=&\alpha\big|u_{\uparrow}+v_{\uparrow}e^{ik_{\uparrow}}\big|+\beta\big|u_{\downarrow}+v_{\downarrow}^{ik_{\downarrow}}\big|,\nonumber\\
[M]^{(\alpha'\beta'),(\alpha\beta)}_{(k'_{\uparrow},k'_{\downarrow}),(k_{\uparrow},k_{\downarrow})}=&\langle\psi_{\alpha'\beta'}(k'_{\uparrow},k'_{\downarrow})|H_{\rm DGF}|\psi_{\alpha\beta}(k_{\uparrow},k_{\downarrow})\rangle.
\label{SMDGF1}
\end{align}

\begin{figure}
    \centering
    \includegraphics[width=0.8\linewidth]{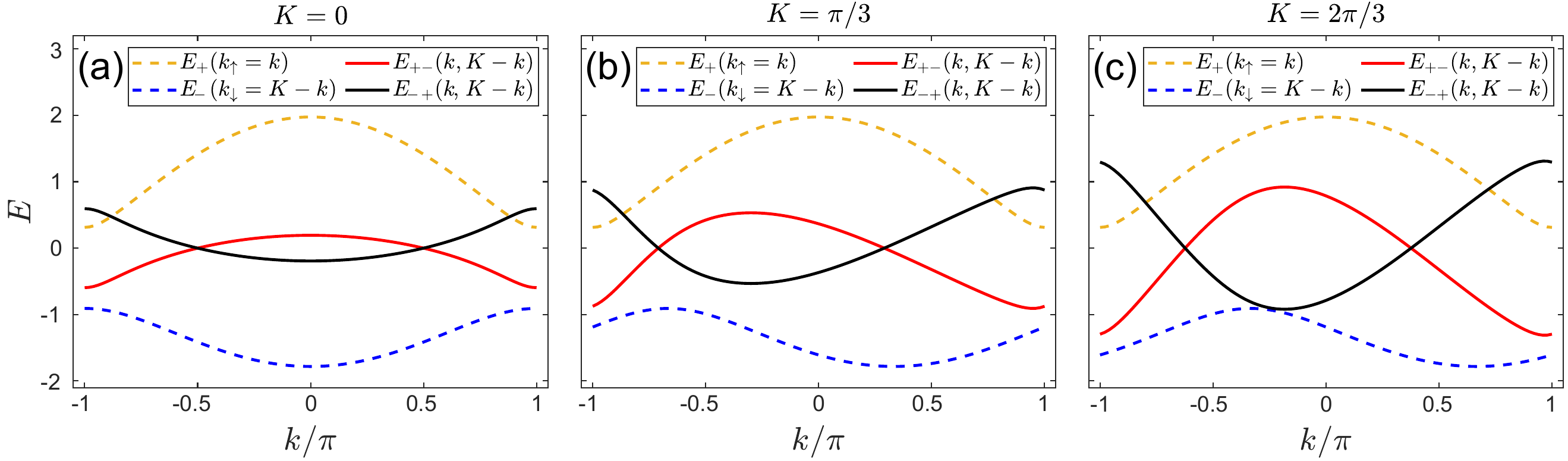}
   \caption{\textbf{The SSH-eigenenergies in several subspaces of $K$.} 
(a) The single-particle energies $E_{+}(k_{\uparrow}=k)$ and $E_{-}(k_{\downarrow}=K-k)$, and $E_{\pm\mp}(k,K-k)$ defined in Eq.~\eqref{SMDGF1} in $K=0$ subspace. (b) and (c) are similar to (a), but for $K=\pi/3$ and $K=2\pi/3$ subspaces, respectively.
It is seen that the fluctuation of $E_{\pm\mp}(k,K-k)$ is the smallest when $K=0$, where the extrema of $E_{+}(k_{\uparrow}=k)$ and $E_{-}(k_{\downarrow}=K-k)$ occur at the same $k$. 
In all panels, $\theta_{\uparrow}=0.2\pi$ and $\theta_{\downarrow}=0.1\pi$, with $\theta_{\sigma}={\rm arg}(u_{\sigma}+iv_{\sigma})$ and $|u_{\sigma}+iv_{\sigma}|=\sqrt{2}$.
Note that the chosen parameters satisfy $u_\sigma v_\sigma>0$ for both pseudospin-up and pseudospin-down particles. If one (and only one) of them has $u_\sigma v_\sigma<0$, the subspace with the minimal fluctuation of $E_{\pm\mp}(k,K-k)$ is $K=\pi$.
}
\label{figS3}
\end{figure}

\begin{figure}
    \centering
    \includegraphics[width=0.95\linewidth]{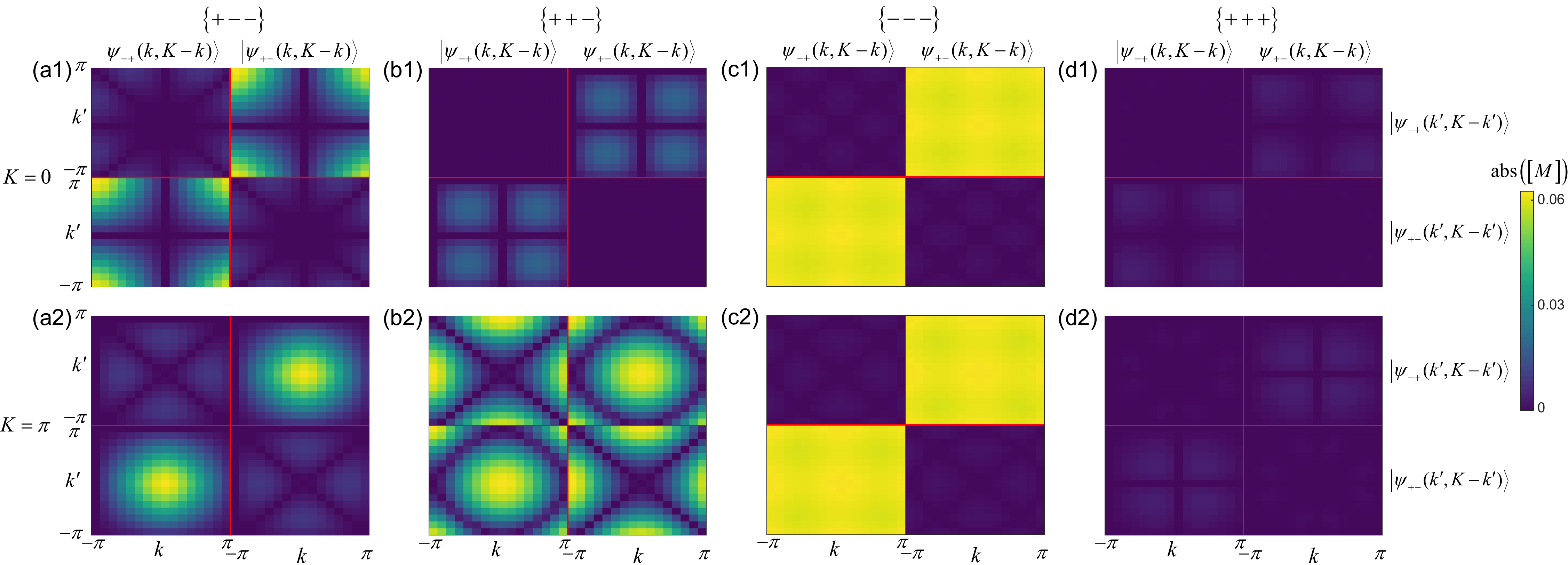}
   \caption{\textbf{The representation of DGF and inter-species band inversion.} 
(a1) to (d1) The absolute value of elements of $[M]$ calculated using Eqs.~\eqref{SMDGF1} and \eqref{DGF3} in the $K = 0$ subspace with different parameters. The parameters are chosen as follows: 
(a1) $\theta_{\uparrow}=0.1\pi$ and $\theta_{\downarrow}=0.4\pi$, 
(b1) $\theta_{\uparrow}=0.4\pi$ and $\theta_{\downarrow}=0.6\pi$, 
(c1) $\theta_{\uparrow}=0.1\pi$ and $\theta_{\downarrow}=0.9\pi$, and 
(d1) $\theta_{\uparrow}=0.05\pi$ and $\theta_{\downarrow}=0.15\pi$, with $|u_{\sigma}+iv_{\sigma}|=\sqrt{2}$. 
%The parameter used in (a1-b1) [(c1-d1)] are the same as those used in Fig. \ref{figSBBS} (Fig. 3 of the main text).} 
The parameter used in (a) and (b) [(c) and (d)] are the same as those used in Fig. \ref{figSBBS}(a) and (b) (Fig. 3(a) and (b) of the main text), respectively. 
The associated topological invariants $\{I_{00},I_{\pi\pi},I_{0\pi}\}$ are given above each panel. 
(a2) to (d2) are the same as (a1) to (d1), but for the $K=\pi$ subspace.
In all panels, $t=0.5$ is set, and the total number of sites is $L = 32$.}
\label{figS4}
\end{figure}

Specifically, we obtain the effect of DGF acting on eigenbasis as
\begin{eqnarray}\label{effDG1} 
H_{\rm DGF}|\psi_{\alpha\beta}(k_{\uparrow},k_{\downarrow})\rangle=&&\frac{t}{\sqrt{L}}\sum_j^{L/2}{\big \{}e^{i[\phi_{\uparrow}(k_{\uparrow})/2+k_{\uparrow}j]}|2j-1\rangle_{\uparrow}\otimes|\tilde{\varphi}^{(\uparrow,j)}_{\beta}(k_{\downarrow})\rangle-\alpha e^{i[-\phi_{\uparrow}(k_{\uparrow})/2+k_{\uparrow}j]}|2j\rangle_{\uparrow}\otimes|\tilde{\varphi}^{(\downarrow,j)}_{\beta}(k_{\downarrow})\rangle{\big \}}\nonumber\\
&&+\frac{t}{\sqrt{L}}\sum_j^{L/2}{\big \{}e^{i[\phi_{\downarrow}(k_{\downarrow})/2+k_{\downarrow}j]}|\tilde{\varphi}^{(\uparrow,j)}_{\alpha}(k_{\uparrow})\rangle\otimes|2j-1\rangle_{\downarrow}-\beta e^{i[-\phi_{\downarrow}(k_{\downarrow})/2+k_{\downarrow}j]}|\tilde{\varphi}^{(\downarrow,j)}_{\alpha}(k_{\uparrow})\rangle\otimes|2j\rangle_{\downarrow}{\big \}},
\end{eqnarray} 
where 
\begin{eqnarray}\label{tarkDG1} 
|\tilde{\varphi}^{(\uparrow,j)}_{\alpha}(k_{\uparrow})\rangle=\frac{e^{ik_{\uparrow}j}}{\sqrt{L}}[\alpha e^{-i\phi_{\uparrow}(k_{\uparrow})/2}|2j-1\rangle_{\uparrow}-e^{i\phi_{\uparrow}(k_{\uparrow})/2}|2j\rangle_{\uparrow}]=\frac{\alpha e^{ik_{\uparrow}j}\sqrt{2}}{\sqrt{L}}[|\mu^{(\uparrow)}_{-\alpha}(k_{\uparrow})\rangle]^*\otimes|j\rangle_{\uparrow}^{\rm cell},\nonumber\\
|\tilde{\varphi}^{(\downarrow,j)}_{\beta}(k_{\downarrow})\rangle=\frac{e^{ik_{\downarrow}j}}{\sqrt{L}}[\beta e^{-i\phi_{\downarrow}(k_{\downarrow})/2}|2j-1\rangle_{\downarrow}-e^{i\phi_{\downarrow}(k_{\downarrow})/2}|2j\rangle_{\downarrow}]=\frac{\beta e^{ik_{\downarrow}j}\sqrt{2}}{\sqrt{L}}[|\mu^{(\downarrow)}_{-\beta}(k_{\downarrow})\rangle]^*\otimes|j\rangle_{\downarrow}^{\rm cell}.
\end{eqnarray}
Here $|2j-1\rangle$ and $|2j\rangle$ are the basis of the two sublattices in the $j$th unit cell.
Thereby,
\begin{eqnarray}\label{effDG2} 
H_{\rm DGF}|\psi_{\alpha\beta}(k_{\uparrow},k_{\downarrow})\rangle=\frac{2t}{{L}}\sum_j^{L/2}e^{i(k_{\uparrow}+k_{\downarrow})j}\{\beta[|\mu^{(\uparrow)}_{-\alpha}(k_{\uparrow})\rangle\otimes|j\rangle_{\uparrow}^{\rm cell}]\otimes[|\mu^{(\downarrow)}_{-\beta}(k_{\downarrow})\rangle\otimes|j\rangle_{\downarrow}^{\rm cell}]^*+\alpha[|\mu^{(\uparrow)}_{-\alpha}(k_{\uparrow})\rangle\otimes|j\rangle_{\uparrow}^{\rm cell}]^*\otimes[|\mu^{(\downarrow)}_{-\beta}(k_{\downarrow})\rangle\otimes|j\rangle_{\downarrow}^{\rm cell}]\}.\nonumber\\
\end{eqnarray}
As a result, the elements of $[M]$ is given by
\begin{eqnarray}\label{effDG3}
[M]^{(\alpha'\beta'),(\alpha\beta)}_{(k'_{\uparrow},k'_{\downarrow}),(k_{\uparrow},k_{\downarrow})}=&&\frac{4t}{L^2}\sum_{j}^{L/2}e^{i(k_{\uparrow}+k_{\downarrow}-k'_{\uparrow}-k'_{\downarrow})j}\beta\langle\mu^{(\uparrow)}_{\alpha'}(k'_{\uparrow})|\mu^{(\uparrow)}_{-\alpha}(k_{\uparrow})\rangle\langle\mu^{(\downarrow)}_{\beta'}(k'_{\downarrow})|[|\mu^{(\downarrow)}_{-\beta}(k_{\downarrow})\rangle]^*\nonumber\\
&&+\frac{4t}{{L^2}}\sum_{j}^{L/2}e^{i(k_{\uparrow}+k_{\downarrow}-k'_{\uparrow}-k'_{\downarrow})j}\alpha\langle\mu^{(\uparrow)}_{\alpha'}(k'_{\uparrow})|[|\mu^{(\uparrow)}_{-\alpha}(k_{\uparrow})\rangle]^*\langle\mu^{(\downarrow)}_{\beta'}(k'_{\downarrow})|\mu^{(\downarrow)}_{-\beta}(k_{\downarrow})\rangle\nonumber\\
=&&\delta_{k'_{\uparrow}+k'_{\downarrow},k_{\uparrow}+k_{\downarrow}}\frac{2t}{{L}}\beta\langle\mu^{(\uparrow)}_{\alpha'}(k'_{\uparrow})|\mu^{(\uparrow)}_{-\alpha}(k_{\uparrow})\rangle\langle\mu^{(\downarrow)}_{\beta'}(k'_{\downarrow})|[|\mu^{(\downarrow)}_{-\beta}(k_{\downarrow})\rangle]^*\nonumber\\
&&+\delta_{k'_{\uparrow}+k'_{\downarrow},k_{\uparrow}+k_{\downarrow}}\frac{2t}{{L}}\alpha\langle\mu^{(\uparrow)}_{\alpha'}(k'_{\uparrow})|[|\mu^{(\uparrow)}_{-\alpha}(k_{\uparrow})\rangle]^*\langle\mu^{(\downarrow)}_{\beta'}(k'_{\downarrow})|\mu^{(\downarrow)}_{-\beta}(k_{\downarrow})\rangle,
\end{eqnarray}
where %the Kronecker delta that the
$\delta_{k'_{\uparrow}+k'_{\downarrow},k_{\uparrow}+k_{\downarrow}}$
indicates that the total momentum $K=k_\uparrow+k_\downarrow$ is conserved. 
Additionally, we focus only on the elements of $[M]$ matrix on eigenbasis with $\alpha=-\beta$,
namely, with the two particles occupying opposite single-particle energy bands.
This is because in this scenario, the total energy of two particles without DGF, $E_{\alpha\beta}(k_{\uparrow},k_{\downarrow})$, is close to ${\rm Re}E=0$, which gives bulk bound states in our numerical results in Figs. 3(a) of the main text and Fig. \ref{figSBBS}.
Substituting the form of the periodic part of the Bloch state, the matrix elements of $[M]$ in $k_{\uparrow}+k_{\downarrow}=K$ subspace are provided as:
\begin{align}
%[M]^{(\pm\mp),(\pm\mp)}_{(k'_{\uparrow},k'_{\downarrow}),(k_{\uparrow},k_{\downarrow})}=\frac{\pm4t}{{N}}\{&\cos\frac{\phi_{\uparrow}(k_{\uparrow})}{2}\sin\frac{\phi_{\uparrow}(k'_{\uparrow})}{2}\sin\frac{\phi_{\downarrow}(k_{\downarrow})}{2}\cos\frac{\phi_{\downarrow}(k'_{\downarrow})}{2}
%\nonumber\\-&\sin\frac{\phi_{\uparrow}(k_{\uparrow})}{2}\cos\frac{\phi_{\uparrow}(k'_{\uparrow})}{2}\cos\frac{\phi_{\downarrow}(k_{\downarrow})}{2}\sin\frac{\phi_{\downarrow}(k'_{\downarrow})}{2}\},\nonumber\\
%[M]^{(\pm\mp),(\mp\pm)}_{(k',K-k'),(k,K-k)}=\frac{\pm4t}{{N}}\{&\sin\frac{\phi_{\uparrow}(k)}{2}\sin\frac{\phi_{\uparrow}(k')}{2}\cos\frac{\phi_{\downarrow}(K-k)}{2}\cos\frac{\phi_{\downarrow}(K-k')}{2}\nonumber\\
%-&\cos\frac{\phi_{\uparrow}(k)}{2}\cos\frac{\phi_{\uparrow}(k')}{2}\sin\frac{\phi_{\downarrow}(K-k)}{2}\sin\frac{\phi_{\downarrow}(K-k')}{2}\}.\nonumber\\
[M]^{(\pm\mp),(\pm\mp)}_{(k'_{\uparrow},k'_{\downarrow}),(k_{\uparrow},k_{\downarrow})}=&\frac{\pm2t}{L}\left\{\sin\frac{\phi_{\uparrow}(k_{\uparrow})+\phi_{\uparrow}(k'_{\uparrow})}{2}\sin\frac{\phi_{\downarrow}(k_{\downarrow})-\phi_{\downarrow}(k'_{\downarrow})}{2}
-\sin\frac{\phi_{\uparrow}(k_{\uparrow})-\phi_{\uparrow}(k'_{\uparrow})}{2}\sin\frac{\phi_{\downarrow}(k_{\downarrow})+\phi_{\downarrow}(k'_{\downarrow})}{2}\right\},\nonumber\\
=&\frac{\pm4t}{L}\left\{\cos\frac{\phi_{\uparrow}(k_{\uparrow})}{2}\sin\frac{\phi_{\uparrow}(k'_{\uparrow})}{2}\sin\frac{\phi_{\downarrow}(k_{\downarrow})}{2}\cos\frac{\phi_{\downarrow}(k'_{\downarrow})}{2}-\sin\frac{\phi_{\uparrow}(k_{\uparrow})}{2}\cos\frac{\phi_{\uparrow}(k'_{\uparrow})}{2}\cos\frac{\phi_{\downarrow}(k_{\downarrow})}{2}\sin\frac{\phi_{\downarrow}(k'_{\downarrow})}{2}\right\},\nonumber\\
[M]^{(\pm\mp),(\mp\pm)}_{(k'_{\uparrow},k'_{\downarrow}),(k_{\uparrow},k_{\downarrow})}=&\frac{\pm2t}{L}\left\{\cos\frac{\phi_{\uparrow}(k_{\uparrow})-\phi_{\uparrow}(k'_{\uparrow})}{2}\cos\frac{\phi_{\downarrow}(k_{\downarrow})+\phi_{\downarrow}(k'_{\downarrow})}{2}
-\cos\frac{\phi_{\uparrow}(k_{\uparrow})+\phi_{\uparrow}(k'_{\uparrow})}{2}\cos\frac{\phi_{\downarrow}(k_{\downarrow})-\phi_{\downarrow}(k'_{\downarrow})}{2}\right\},\nonumber\\
=&\frac{\pm4t}{L}\left\{\sin\frac{\phi_{\uparrow}(k_{\uparrow})}{2}\sin\frac{\phi_{\uparrow}(k'_{\uparrow})}{2}\cos\frac{\phi_{\downarrow}(k_{\downarrow})}{2}\cos\frac{\phi_{\downarrow}(k'_{\downarrow})}{2}-\cos\frac{\phi_{\uparrow}(k_{\uparrow})}{2}\cos\frac{\phi_{\uparrow}(k'_{\uparrow})}{2}\sin\frac{\phi_{\downarrow}(k_{\downarrow})}{2}\sin\frac{\phi_{\downarrow}(k'_{\downarrow})}{2}\right\}.
\label{DGF3}
\end{align}
Due to the  inversion symmetry of the SSH model $H_\sigma$, its single-particle eigenenery $E_{\pm}(k_{\sigma})$ takes extrema at $k_\sigma=0$ or $k_\sigma=\pi$, being either maximum or minimum depending on the the sign of $u_\sigma$ (since $v_\sigma$ is chosen to be non-negative).
Thus the eigenenergies $E_{\pm\mp}(k,K-k)$ at $K=0$ or $\pi$ undergo less significant changes when varying $k$~(e.g., see Fig. \ref{figS3}), acting approximately as an identity matrix. Consequently, $[M]$ plays a more significant role in these subspaces.
Within these two subspaces, the possible extrema of matrix elements $[M]^{(\alpha',-\alpha'),(\alpha,-\alpha)}_{(k',K-k'),(k,K-k)}$ are located at the symmetric momenta $k,k'\in\{0,\pi\}$ (where the derivative of $[M]^{(\alpha',-\alpha'),(\alpha,-\alpha)}_{(k',K-k'),(k,K-k)}$ over either $k$ or $k'$ is zero), with $\phi_{\sigma}$ being either $0$ or $\pi$.
Furthermore,  when $\phi_{\uparrow}(k)=\phi_{\downarrow}(K-k)$ and $\phi_{\uparrow}(k')=\phi_{\downarrow}(K'-k')$ at $k,k'\in\{0,\pi\}$,
these possible extrema of matrix elements turn out to be zero, making all other elements vanishing provided there is no other extrema.
Numerically, we observe that as long as the above condition of $\phi_\sigma$ is met,
matrix elements of $[M]$ always approach zero even when extrema appears at momenta other than $k,k'\in\{0,\pi\}$. An example is shown in Fig. \ref{figS4}(b1), where the matrix elements vanish when $k,k'\in\{0,\pi\}$, yet possess small values at other momenta, indicating the existence of other extrema.

Based on above analyses, we find that the DGF exerts its influence only when $\phi_{\uparrow}\neq\phi_{\downarrow}$ at least at 
one set of symmetric momenta $k,k'\in\{0,\pi\}$.
By definition, at $k_\sigma\in\{0,\pi\}$, $\phi_{\sigma}(k_{\sigma})={\rm arg}(u_{\sigma}+v_{\sigma}e^{-ik_{\sigma}})=0$ ($\pi$) means that the 
single-particle eigenstate polarized positively along $\sigma_x$ in the pseudospin space of sublattices has a positive (negative) eigenenergy.
Therefore, $\phi_{\uparrow}\neq\phi_{\downarrow}$ at a set of $k,k'\in\{0,\pi\}$ means that inter-species band inversion occurs for the two particles at $k$ and $k'$, respectively.
These features of the system can be characterized by topological invariants
$I_{k_\uparrow k_\downarrow}$ with $k_\uparrow=k$ and $k_\downarrow=K-k$ (each being $0$ or $\pi$),
as demonstrated by the results in Fig. 3(c) of the main text.

%Especially, the matrix defined in this way plays a more significant role in two higher-symmetric subspaces, namely $K=0$ and $K=\pi$.
%In these subspaces, due to the chiral and inversion symmetries of the SSH model, the eigenenergies $E_{\pm\mp}(k,K-k)$ undergo less significant changes. 
%More importantly, the possible extrema of matrix elements of $[M]$ within these two subspaces are located at the symmetric momenta \LLH{$k=k_0\in\{0,\pi\}$} \red{[explanation]}, where the value of $\phi_{\sigma}$ is either $0$ or $\pi$.    
%Interestingly, these matrix elements turn out to be zero when $\phi_{\uparrow}=\phi_{\downarrow}$, 
%and all matrix elements of $[M]$ approach zero in this subspace \LLH{when this condition is met} for all symmetric momenta.

To have a clearer demonstration of this relationship between $[M]$ and $I_{k_\uparrow k\downarrow}$, we display the absolute value of elements of $[M]$ with different parameters in Fig.~\ref{figS4}. As demonstrated, the elements of $[M]$ assume considerable values in a subspace when the associated topological invariants have non-trivial values $I_{k_\uparrow k\downarrow}=-1$. Specifically, for the cases shown in Figs.~\ref{figS4}(a1-a2) and \ref{figS4}(c1-c2), especially the latter, lots of elements take notable values in both the $K=0$ and $K=\pi$ subspaces because inter-species band inversion occurs in these two subspaces. In contrast, both topological invariants are trivial for the case in Figs.~\ref{figS4}(d1-d2); thus, all elements in these two subspaces are nearly zero. 
More intriguingly, the inter-species band inversion occurs only in the $K=\pi$ subspace for the case shown in Figs.~\ref{figS4}(b1-b2). As a result, the elements are relatively significant in the $K=\pi$ subspace, whereas they are relatively insignificant in the $K=0$ subspace. This clear contrast further confirms the relationship between the degree of the leading effect of DGF and inter-species band inversion.
Combining these results with the appearance of bulk bound states shown in Fig. 3 of the main text and Fig.~\ref{figSBBS}, 
%where we present the spectra and the appearance of bulk bound states under the conditions with parameters used in Fig.~\ref{figS4}(a1)-(a2) and (b1)-(b2), 
we can draw the conclusion: the DGF plays a vital role and induces non-accidental two-particle bulk bound states only when inter-species band inversion occurs (with the corresponding $I_{k_\uparrow k\downarrow}=-1$ for at least one set of high-symmetric momenta).

%The conservation of the total momentum $K=k_\uparrow+k_\downarrow$~\cite{PhysRevA.95.063630,PhysRevA.101.023620,PhysRevB.107.125161,PhysRevLett.133.140202}
%(also see Supplementary Material~\cite{suppmat}),
%further allows us to treat the subspaces with $K=0$ and $K=\pi$ separately.
%Additionally, we focus only on the elements of $[M]$ matrix on eigenbasis with $\alpha=-\beta$,
%namely, with the two particles occupying the lower and higher single-particle energy bands, respectively.
%This is because the total energy in this scenario is close to ${\rm Re}E=0$, which gives  $\mathcal{PT}$-broken bulk bound states in our numerical results in Fig. \ref{fig2}(a-c).

%Finally, with detailed derivation given in the Supplemental Materials~\cite{suppmat},
%we find that the concerned DGF matrix elements of $[M]$ take non-vanishing values in the subspace $K=0$ ($K=\pi$) only when $I_{00}$ and/or $I_{\pi\pi}$ ($I_{0\pi}$ and/or $I_{\pi0}$) takes a nontrivial value $-1$. 
%Note that in the absence of the DGF, the system is simply the direct sum of two single-particle Hamiltonians, which may only accidentally host some unstable two-particle bound states.
%Therefore we reach a conclusion that non-accidental two-particle bulk bound states may emerge only when inter-species band inversion occurs at least at one set of single-particle high-symmetric points (with the corresponding $I_{kk'}=-1$).
%
%
%Fig.S4 the obvolute value of element of $[M]$ for each case.
%
%The integral form of Eq.(7)

\subsection{Eigenenergies of bulk bound states with symmetric parameters}\label{appC2}
In this subsection we derive the energies of bulk bound states with symmetric parameters, to see the degeneracy of eigenenergies in the absence of DGF, and how their complex values emerge due to DGF.

We use the representation derived in the previous subsection to obtain eigenenergies of bulk bound states. Specifically, we focus on symmetric parameters, such as $\{|u_{\uparrow}|=|u_{\downarrow}|,|v_{\uparrow}|=|v_{\downarrow}|\}$ or $\{|u_{\uparrow}|=|v_{\downarrow}|,|v_{\uparrow}|=|u_{\downarrow}|\}$, which correspond to the cases shown in Fig.3(a) of the main text and Fig.~\ref{figSBBS}. In these cases, the SSH eigenenergies $E_{\pm\mp}(k,K-k)$ defined in Eq.~\eqref{SMDGF1} become fully degenerate at zero energy in the $K=0$ ($K=\pi$) subspace, when ${\rm sgn}(u_{\uparrow}v_{\uparrow})={\rm sgn}(u_{\downarrow}v_{\downarrow})$ [${\rm sgn}(u_{\uparrow}v_{\uparrow})=-{\rm sgn}(u_{\downarrow}v_{\downarrow})$]. As demonstrated below, the DGF induces pronounced bulk bound states in subspaces where $E_{\pm\mp}(k,K-k)=0$ and the inter-species band inversion occurs. In contrast, in trivial cases with $I_{kk'}=1$, the large degeneracy $E_{\pm\mp}(k,K-k)=0$ may also lead to accidental bulk bound states irrelevant to DGF but they are not robust against disorder, as displayed in Fig. 3(f) of the main text. To proceed, we temporarily neglect the dissipation term $-i\sum_{j=1}^{L/2}\gamma_\sigma n_{\sigma,2j}$. In the above mentioned subspaces with symmetric parameters (so that $E_{\pm\mp}(k,K-k)=0$), the Hamiltonian $H$ in Eq. (1) of the main text is dominated by the DGF, which is represented as a real antisymmetric $L\times L$ matrix $[M]$.

However, we find that all the $3$rd-order minors (the associated matrix is denoted as $[M_3]$) of $[M]$ is zero. 
The proof is as follows.
Firstly, if all three columns of $[M_3]$ correspond to states $|\psi_{\alpha\beta}(k,K-k)\rangle$ (see Eq. \eqref{SMDGF1} and recall that $k_\uparrow+k_\downarrow=K$) with the same $\alpha$,
%states from the same block of $[M]$, i.e., $\alpha_1=\alpha_2=\alpha_3$ in Eqs.~\eqref{SMDGF1} and \eqref{DGF3}, with momenta $k_1$, $k_2$ and $k_3$, 
these columns will be linearly correlated.
That is, 
\begin{eqnarray}\label{K0A} 
[M_3]\vec{A} = 0,~~\vec{A}=[A_1,A_2,A_3]^T,~~A_i=\sum_{j = 1}^3\sum_{l = 1}^3\varepsilon_{ijl}\sin\frac{\phi_{\uparrow}(k_j)}{2}\cos\frac{\phi_{\downarrow}(K-k_j)}{2}\cos\frac{\phi_{\uparrow}(k_l)}{2}\sin\frac{\phi_{\downarrow}(K-k_l)}{2}
\end{eqnarray}
where $\varepsilon_{ijl}$ is the Levi-Civita symbol, and $k_1,k_2,k_3$ the three momenta associated with the corresponding states.
And if only one column of $[M_3]$ is associated with a state with its $\alpha$ different from the other two (supposing this state is placed as the last column and with a momentum $k=k_3$), we also have
\begin{eqnarray}\label{K0B} 
[M_3]\vec{B} = 0,~~~~\vec{B}=\vec{A}[{\phi_{\uparrow}(k_3)\rightarrow}\pi-\phi_{\uparrow}(k_3),{\phi_{\downarrow}(k_3)\rightarrow}\pi-\phi_{\downarrow}(k_3)].
\end{eqnarray}
Therefore, we can conclude that $\det([M_3]) = 0$. This result indicates that the rank of the matrix $[M]$ satisfies $1\leq\mathrm{Rank}([M])<3$ (if $[M]$ is nonzero), thus allowing only one or two nonzero eigenvalues.
However, since $[M]$ is a real antisymmetric matrix, nonzero eigenvalues must come in pairs with opposite imaginary values.

Thereby, the characteristic polynomial of $[M]$, given by $\det([M]-\lambda I) =0$, takes the form:  
\begin{eqnarray}\label{poly}
\lambda^L + a_{2}\lambda^{L - 2}=0,\quad a_{2}=-\sum_{i}\sum_{j>i}[M]_{ij}[M]_{ji}=\frac{1}{2}\sum_{i}\sum_j [M]^2_{ij},
\end{eqnarray}  
indicating that $[M]$ has $L-2$ zero eigenvalues and $2$ pure imaginary eigenvalues $\pm i\sqrt{a_{2}}$. These pure imaginary eigenvalues correspond to the energies of the bulk bound states.

%please check all results for $K=\pi$ space. Comparing to PT_ManyPa_Lei version, some equations are changed, because \phi_{\downarrow}=\arg(|u_{\downarrow}|+|v_{\downarrow}|e^{-ik_{\downarrow}}) in PT_ManyPa_Lei while \phi_{\downarrow}=\arg(u_{\downarrow}+v_{\downarrow}e^{-ik_{\downarrow}}) here.

Based on the above analysis, in cases with symmetric parameters (so that $E_{\pm\mp}(k,K-k)=0$ for $K=0$ or $\pi$), as long as $I_{k,K-k}=-1$ for $k\in\{0,\pi\}$ (so that $[M]$ does not vanish),
%where symmetric parameters \LLH{lead to} $E_{\pm\mp}(k,K-k)=0$ and either $I_{0K}=-1$, $I_{\pi(K-\pi)}=-1$, or both, 
the DGF will dominate in the corresponding $K$ subspace, and may induce topologically-correlated bulk bound states.
Accompanying Eq.~\eqref{DGF3}, we can obtain the two imaginary energies as $E_{\rm BBS}=\pm i\sqrt{a_{2}}$, where
\begin{eqnarray}\label{Eanal0}
a_{2}=\frac{4t^{2}}{L^{2}}\sum_{k}\sum_{k'}\left\{\left(1 - \cos[\phi_{\uparrow}(k)]\cos[\phi_{\downarrow}(K-k)]\right)\left(1 - \cos[\phi_{\uparrow}(k')]\cos[\phi_{\downarrow}(K-k')]\right)\right.\nonumber\\
\left.-\sin[\phi_{\uparrow}(k)]\sin[\phi_{\downarrow}(K-k)]\sin[\phi_{\uparrow}(k')]\sin[\phi_{\downarrow}(K-k')]\right\},
\end{eqnarray}
which match well with the eigenenergies with maximal and minimal imaginary values in our numerical results in Fig. 3 of the main text and Fig.~\ref{figSBBS} (upon an imaginary energy shift $-i\gamma_\uparrow /2$).
When $L\rightarrow\infty$, it can be written in an integral form
\begin{eqnarray}\label{Eanalin0}
a_{2}=\frac{t^{2}}{4\pi^{2}}\int_{-\pi}^{\pi}dk\int_{-\pi}^{\pi}dk'\left\{\left(1 - \cos[\phi_{\uparrow}(k)]\cos[\phi_{\downarrow}(K-k)]\right)\left(1 - \cos[\phi_{\uparrow}(k')]\cos[\phi_{\downarrow}(K-k')]\right)\right.\nonumber\\
\left.-\sin[\phi_{\uparrow}(k)]\sin[\phi_{\downarrow}(K-k)]\sin[\phi_{\uparrow}(k')]\sin[\phi_{\downarrow}(K-k')]\right\}.
\end{eqnarray}
The analytical expressions for the energies of bulk bound states obtained here, accompanied by their relationship with $[M]$ and inter-species band inversion revealed in Subsec.\ref{appC1}, validate that the formation of bulk bound states is attributed to the DGF and inter-species topology.

In addition, eigenvalues of $[M]$ can also be obtained for several flat band limits. Below we shall consider four examples with the same topological invariants as in Fig. \ref{figS4}, but different explicit values of parameters.

\begin{itemize} 
%Moreover, the outcomes for several flat band limit cases can be computed easily.
\item The first example is the trivial flat band limit without inter-species inversion, where 
$v_{\uparrow}=v_{\downarrow}=0$ and $u_{\uparrow}=u_{\downarrow}$.
%either $v_{\uparrow}=v_{\downarrow}=0<u_{\uparrow}=u_{\downarrow}$ or $u_{\uparrow}=u_{\downarrow}<v_{\uparrow}=v_{\downarrow}=0$.
In this case, we have $\phi_{\uparrow}(k)=\phi_{\downarrow}(k)={\rm arg}(u_{\downarrow}/|u_{\downarrow}|)$, which leads to $E_{\rm BBS}=a_{2}=0$. This result is in consistent with the absence of topologically-correlated bulk bound states; in fact, the $[M]$ is a zero matrix in these cases, meaning that the system is simply described by the direct sum of the two single-particle Hamiltonian. This example corresponds to the same topological phase as that in Fig. \ref{figS4}(d), where all topological invariants are trivial, and $[M]$ approximately vanishes for both $K=0$ and $K=\pi$.

\item Another example is when $v_{\uparrow}=v_{\downarrow}=0$ and $u_{\uparrow}=-u_{\downarrow}\neq0$, where all invariants $I_{kk'}$ at symmetric momenta are nontrivial, indicating that bulk bound states can be induced in both of the two subspaces ($K=0$ or $\pi$). Specifically, $\phi_{\uparrow}(k)={\rm arg}(u_{\uparrow}/|u_{\uparrow}|)$ and $\phi_{\downarrow}(k)=\pi-\phi_{\uparrow}(k)$ in this case. The integral yields $a_{2}=\frac{4t^{2}}{4\pi^{2}}\int_{-\pi}^{\pi}dk\int_{-\pi}^{\pi}dk'=4t^{2}$, thus resulting in $E_{\rm BBS}=\pm i2t$.
In fact, in this case, for both species of particles, the eigenstates of the SSH model $H_{\sigma}$ are intra-cell disconnected dimers, and above result can also be easily obtained from the real space Hamiltonian.
This example corresponds to the same topological phase as the one in Fig. \ref{figS4}(c), where all topological invariants are nontrivial, and $[M]$ does not vanish for both $K=0$ and $K=\pi$.

\item Similarly, for the case of $u_{\uparrow}=v_{\downarrow}=0<v_{\uparrow}=|u_{\downarrow}|$, one of the invariants $I_{kk'}$ in each subspace is nontrivial, indicating that bulk bound states can also be induced in both of the two subspaces. 
In this case, $\phi_{\uparrow}(k)=-k$ and $\phi_{\downarrow}(k)={\rm arg}(u_{\downarrow}/|u_{\downarrow}|)$. The integral gives $a_{2}=\frac{t^{2}}{4\pi^{2}}\int_{-\pi}^{\pi}dk\int_{-\pi}^{\pi}dk'=t^{2}$, and $E_{\rm BBS}=\pm it$.
This example corresponds to the same topological phase as the one in Fig. \ref{figS4}(a), with $\{I_{00},I_{\pi\pi},I_{0\pi}\}=\{+--\}$, and $[M]$ does not vanish for both $K=0$ and $K=\pi$.

\item Finally, when $u_{\uparrow}=u_{\downarrow}=0<v_{\uparrow}=v_{\downarrow}$, inter-species band inversion occurs only in the $K=\pi$ subspace, with $I_{0\pi}=I_{\pi0}=-1$. 
Since $\phi_{\uparrow}(k)=\phi_{\downarrow}(k)=-k$, the integral gives $a_{2}=\frac{2t^{2}}{4\pi^{2}}\int_{-\pi}^{\pi}dk\int_{-\pi}^{\pi}dk' = 2t^{2}$ and $E_{\rm BBS}=\pm i\sqrt{2}t$ for $K=\pi$, while $E_{\rm BBS}=a_{2}=0$ for $K=0$.
This example corresponds to the same topological phase as the one in in Fig. \ref{figS4}(b), with $\{I_{00},I_{\pi\pi},I_{0\pi}\}=\{++-\}$, and $[M]$ does not vanish only for $K=\pi$.
\end{itemize} 

Finally, we further incorporate the dissipation term into our consideration. 
However, it merely modifies these bulk bound states mildly and shifts their energies close to the value of $E_{\rm BBS}=\pm i\sqrt{a_2}-i\gamma_{\uparrow}/2$.

%Taking these cases shown in Figs.2(a)-(c) of the main text as examples, the analytical value of $E^{(+)}_{\rm b}-i\gamma_{\uparrow}/2$ are $0.1608$, $0.7232$, and $0.4571$ which agree well with the numerical values $0.1756$, $0.7344$ and $0.4867$ corresponding to the largest imaginary energies in these plots.
%In addition, this formula can be transformed into integral form under the thermodynamic limit, as shown in the supplementary material.

\subsection{Accidental bulk bound states in Fig. 3(e) of the main text}\label{appC3}
In Fig. 3(e) of the main text, we also observe states with large inter-species entanglement entropy $S_m$ along the diagonal regime, $\theta_\uparrow=\theta_\downarrow$ with  $\theta_\sigma={\rm arg}(u_\sigma+iv_\sigma)$, even in the $\{+++\}$ phase without inter-species band inversion.
Note that we have also chosen $|u_\uparrow+iv_\uparrow|=|u_\downarrow+iv_\downarrow|$ to obtain the results there. Thus, the above diagonal condition yields $u_\uparrow=u_\downarrow=u$ and $v_\uparrow=v_\downarrow=v$.
In the high-symmetric subspace with $K=0$,
the energy without DGF becomes $E_{\alpha\beta}(k,K-k)\in\{0, \pm 2\sqrt{u^2+v^2+2uv\cos k}\}$,
and the effect of DGF vanishes in the $\{+++\}$ phase. 
In other words, such a parameter regime hosts a strong degeneracy at $E=0$ for eigenstates at different $k$ with $K=0$ and $\alpha=-\beta$.
The superposition of these degenerate states may give rise to some accidental bulk bound states with large $S_m$,
just as that the nontrivial bulk bound states are given by superpositions of Bloch states in high-symmetric subspaces (which are eigenstates of $[M]$ matrix, as discussed in the last subsection).

We also note that we have restricted our discussion in the parameter regime with $v_{\sigma}\geq0$ (see the beginning of Sec. \ref{appC}), and Fig. 3 of the main text has only demonstrated results with $\theta_\sigma\in [0,\pi]$.
Otherwise, the strong degeneracy at $E=0$ occurs in the other symmetric subspace $K=\pi$, with $\theta_\sigma\in [\pi,2\pi]$.

\section{Supplementary Note 4: the $\mathcal{PT}$ phase transition of the system}
\MOD{In this section, we discuss the $\mathcal{PT}$-breaking conditions for different inter-species topological states and non-topological states in our system. 
Explicitly, the eigenstates can be categorized into three groups: (i) edge (anti-)confined states, (ii) bulk bound states, and (iii) other bulk eigenstates. Among them, group (i) emerges only under OBCs, while groups (ii) and (iii) exist under both OBCs and PBCs. However, since the later two do not show NHSE, their eigenenergies remain nearly identical under different boundary conditions. Thus, we will discuss the PT-broken/unbroken conditions under OBCs for (i), and PBCs for (ii) and (iii).}

\subsection{$\mathcal{PT}$-breaking of edge (anti-)confined states}
\MOD{Edge (anti-)confined states represent a topological boundary effect under OBCs, possessing spatial-inversion asymmetric distributions across the system. 
As a consequence, they inevitably break $\mathcal{PT}$ symmetry and acquire imaginary energy components due to the on-site loss term. In other words, the presence of these states always signals a $\mathcal{PT}$-broken phase under OBCs, which corresponds to all phases in our phase diagram except $\{---\}$, $\{+++\}$, and $\{++-\}$. Additionally, in the $\{++-\}$ phase, both species exhibit nontrivial topological properties at the single-particle level, giving rise to the coincidental edge (anti-)confined states discussed in Sec.~\ref{appA}. These states also display edge localization and violate $\mathcal{PT}$ symmetry. To summarize, $\mathcal{PT}$ breaking associated with edge (anti-)confined states takes place when $|u_{\uparrow}| < |v_{\uparrow}|$ and/or $|u_{\downarrow}| < |v_{\downarrow}|$. }

\MOD{Note that for these states, the DGF acts as non-reciprocal hopping terms inducing skin-like localization, yet generally does not generate additional imaginary energies—analogous to the NHSE in the non-Hermitian SSH model. The imaginary energies of edge (anti-)confined states come almost entirely from the on-site loss term, which is applied exclusively to even sites. Specifically, in this work, we consider a loss term only for the pseudospin-up particle, with $\gamma_{\uparrow} \neq 0$ and $\gamma_{\downarrow} = 0$. When $|u_{\uparrow}| < |v_{\uparrow}|$, the pseudospin-up particle is topologically nontrivial at the single-particle level, thereby exhibiting topological polarization at either odd or even sites. Consequently, these states possess imaginary energies ${\rm Im}E = -\gamma_{\uparrow}$ or ${\rm Im}E = 0$, which significantly deviate from the $\mathcal{PT}$-symmetric energy ${\rm Im}E= -\gamma_{\uparrow}/2$. However, if $|u_{\uparrow}| > |v_{\uparrow}|$ and $|u_{\downarrow}| < |v_{\downarrow}|$, the pseudospin-up particle is topologically trivial at the single-particle level, and its spatial-inversion asymmetric distributions arise from DGF-mediated skin-like localization. Such localization lacks strong polarization on even/odd sites. Thus, despite their asymmetric distributions already breaking $\mathcal{PT}$ symmetry, edge (anti-)confined states in this scenario exhibit negligible deviations from the $\mathcal{PT}$-symmetric imaginary energy.}

\subsection{$\mathcal{PT}$-breaking of bulk bound states}
\MOD{For bulk bound states, we introduce ansatzes of these states in different phases, utilizing the symmetric parameters (particularly in the flat band limit) presented in Subsec.~\ref{appC2}. It should be noted that, due to the relatively homogenous distribution of bulk bound states between even and odd sites, the on-site loss $\gamma_{\sigma}$ added exclusively to even sites (with opposite signs on even and odd sites after a global energy shift) generally does not induce $\mathcal{PT}$ breaking for these states. For simplicity, we therefore set $\gamma_{\sigma}=0$ in the subsequent discussion of bulk bound states.}

\begin{figure}[H]
    \centering
    \includegraphics[width=0.8\linewidth]{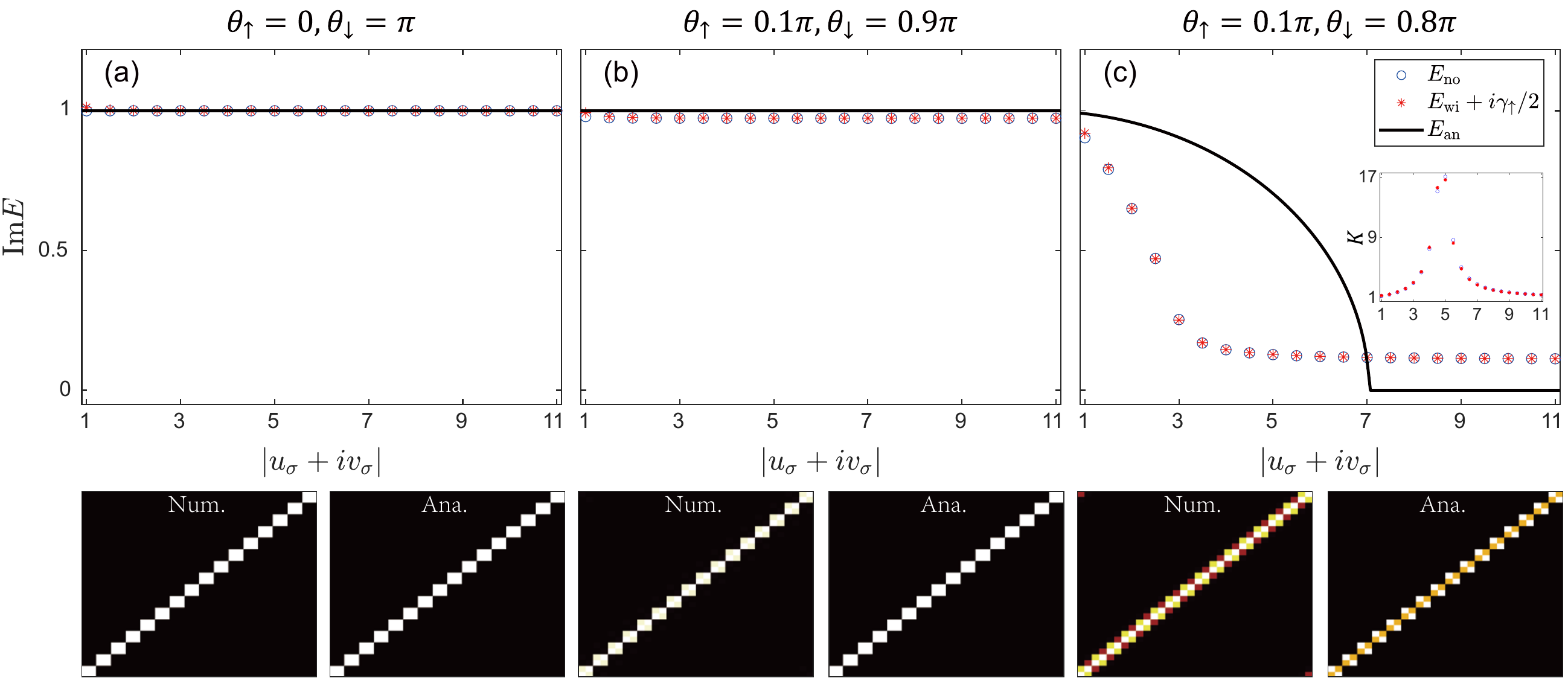}
   \caption{\MOD{\textbf{The comparison of the $\mathcal{PT}$ phase transition between analytical and numerical calculations for the $\{---\}$ phase.}
The upper column show the maximum imaginary energy versus $|u_{\sigma}+iv_{\sigma}|$ obtained from three different methods:
the effective Hamiltonian $H_{\rm eff}^{\{---\}}$ in Eq.~\eqref{effmmm}, the Hamiltonian $H$ in Eq.~(1) of the main text without dissipation, and the Hamiltonian $H$ in Eq.~(1) of the main text with dissipation $\gamma_{\uparrow}=0.5$. These are denoted as $E_{\rm an}$, $E_{\rm no}$, and $E_{\rm wi}$, and plotted as black lines, blue circles, and red asterisks, respectively. For better comparison, $E_{\rm wi}$ has been shifted by $i\gamma_{\uparrow}/2$. In the lower column, the normalized two-particle correction of bulk bound states with $|u_{\sigma}+iv_{\sigma}|=1.5$ is provided, obtained from both the effective Hamiltonian $H_{\rm eff}^{\{---\}}$ (labeled "Ana.") and the Hamiltonian $H$ without dissipation (labeled "Num."). The parameters are set as $t=0.5$ for all panels, with $(\theta_{\uparrow}=0,\theta_{\downarrow}=\pi)$ for (a), $(\theta_{\uparrow}=0.1\pi,\theta_{\downarrow}=0.9\pi)$ for (b), and $(\theta_{\uparrow}=0.1\pi,\theta_{\downarrow}=0.8\pi)$ for (c). 
%The reasons for the discrepancy between the analytical and numerical results in (c), i.e., the asymmetric parameters situation, are discussed in Fig.\ref{PTMMMdev}. 
\LZT{In the inset of (c), we also plot the Petermann factor defined in Eq.~\eqref{pe_fac}  for the state with maximal imaginary eigenenergy, which takes values significantly greater than $1$ near the exceptional point.}
The chosen system size is $L=32$, with $L^2=1024$ the Hamiltonian dimension.}}
\label{PTMMM}
\end{figure}

\MOD{Firstly, we introduce two-mode ansatz solution for the $\{---\}$ phase:
\begin{eqnarray}\label{trymmm}
|\psi_{\pm}\rangle=\frac{\sqrt{2}}{\sqrt{L}}\sum_{j=1}^{L/2}|a^{\uparrow}_{j,\pm}\rangle\otimes|a^{\downarrow}_{j,\pm}\rangle,
\end{eqnarray}
with $|a^{\sigma}_{j,\pm}\rangle=\frac{1}{\sqrt{2}}[|2j-1\rangle_{\sigma}\pm|2j\rangle_{\sigma}]$.
The Hamiltonian's representation within the space of these two states can be derived as follows: $\langle\psi_\pm|H_{\rm DGF}|\psi_\pm\rangle=\langle\psi_\mp|\sum_\sigma H_\sigma|\psi_\pm\rangle=0$, $\langle\psi_\mp|H_{\rm DGF}|\psi_\pm\rangle=\pm2t$, and $\langle\psi_\pm|\sum_\sigma H_\sigma|\psi_\pm\rangle=\pm\sum_{\sigma}u_{\sigma}$. From these results, we obtain an effective two-level $\mathcal{PT}$-symmetric Hamiltonian:
\begin{eqnarray}\label{effmmm}
H_{\rm eff}^{\{---\}}=\left(\sum_{\sigma}u_{\sigma}\right)\tau_z+2it\tau_x,
\end{eqnarray}
with $\tau_{x,z}$ being Pauli matrix acting on $|\psi_{\pm}\rangle$ space. Accordingly, the $\mathcal{PT}$ unbroken (broken) regimes can be identified as $|\sum_{\sigma}u_{\sigma}|>t$ ($|\sum_{\sigma}u_{\sigma}|<t$). To validate this result, we consider three sets of $\theta_{\sigma}$ parameters: those corresponding to the flat band limit ($\theta_{\uparrow}=0$, $\theta_{\downarrow}=\pi$), general symmetric parameters ($\theta_{\uparrow}=\pi-\theta_{\downarrow}$), and asymmetric parameters. For each set, we vary the magnitude $|u_{\sigma}+iv_{\sigma}|$, calculate the maximum imaginary part of the eigenvalues in the absence of dissipation under PBC (denoted as $E_{\rm no}$, represented by blue circles), and compare it with the prediction from the above effective model (black line), as shown in Fig.\ref{PTMMM}. We have also calculated the maximum imaginary part of the eigenvalues, $E_{\rm wi}$, for the original system with the on-site loss term $\gamma_\uparrow$. Fig.\ref{PTMMM} shows that with an energy shift, $E_{\rm wi} + i\gamma_\uparrow/2 \approx E_{\rm no}$, confirming that the loss term can be neglected. In the asymmetric parameters regime [Fig.\ref{PTMMM}(c)], the imaginary energy vanishes once $|u_\sigma + v_\sigma|$ exceeds a critical value, signifying a transition from $\mathcal{PT}$-broken to $\mathcal{PT}$-unbroken phases. However, our analytical results differ from the numerical simulations. This arises because the system is highly parameter-sensitive, and the ansatz we adopt becomes inaccurate when moving away from cases with symmetric parameters [Figs.\ref{PTMMM}(a) and (b)], as will be discussed in detail later.
Additionally, for each parameter regime, the analytically derived two-particle correlation function of the bulk bound state and its numerically obtained counterpart are presented below the energy spectrum. This validates the reliability of our ansatz.}

\begin{figure}
    \centering
    \includegraphics[width=0.8\linewidth]{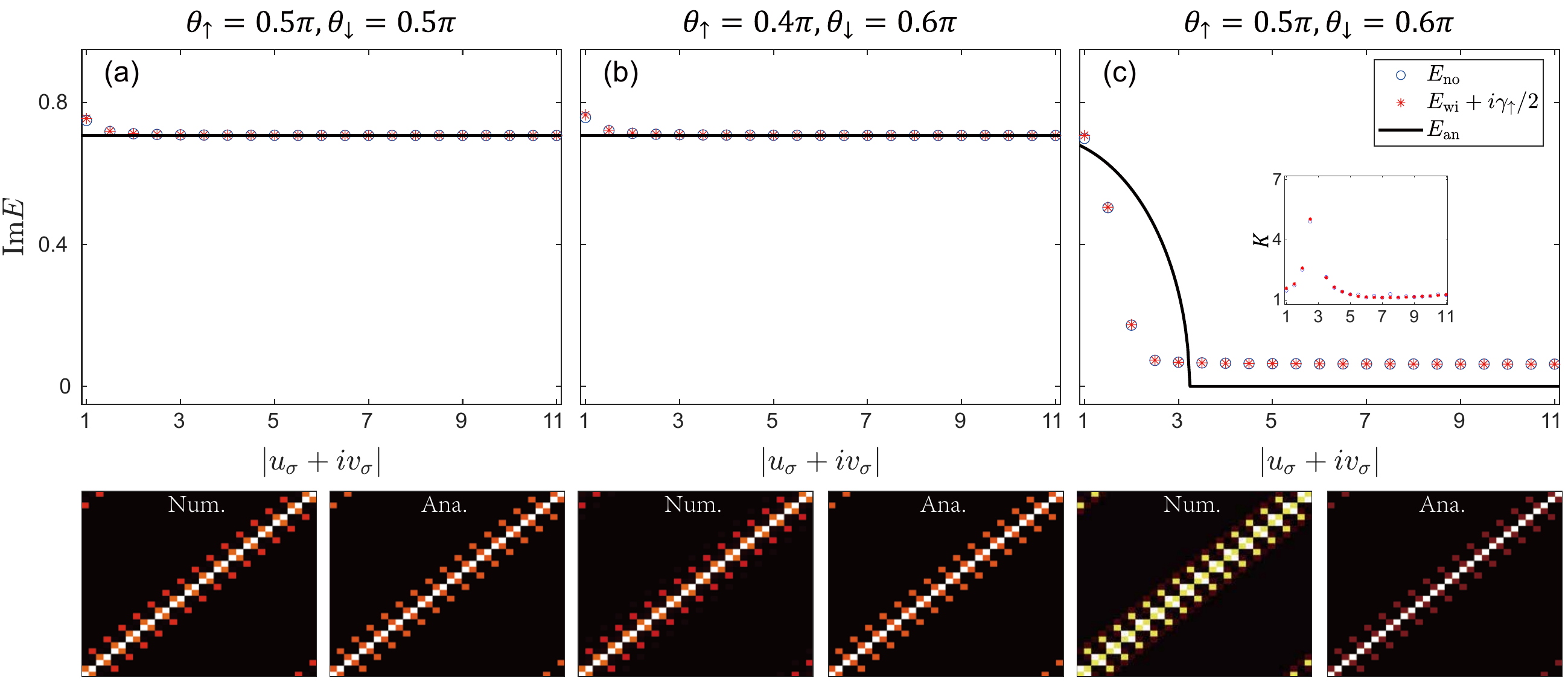}
   \caption{\MOD{\textbf{The comparison of the $\mathcal{PT}$ phase transition between analytical and numerical processes for the $\{++-\}$ phase.} The upper column show the maximum imaginary energy versus $|u_{\sigma}+iv_{\sigma}|$ obtained from three different methods: the effective Hamiltonian $H_{\rm eff}^{\{++-\}}$ in Eq.~\eqref{effppm}, the Hamiltonian $H$ in Eq.~(1) of the main text without dissipation, and the Hamiltonian $H$ in Eq.~(1) of the main text with dissipation $\gamma_{\uparrow}=0.5$. These are denoted as $E_{\rm an}$, $E_{\rm no}$, and $E_{\rm wi}$, and plotted as black lines, blue circles, and red asterisks, respectively. For better comparison, $E_{\rm wi}$ has been shifted by $i\gamma_{\uparrow}/2$. In the lower column, the normalized two-particle correction of bulk bound states with $|u_{\sigma}+iv_{\sigma}|=1.5$ is provided, obtained from both the effective Hamiltonian $H_{\rm eff}^{\{++-\}}$ (labeled "Ana.") and the Hamiltonian $H$ without dissipation (labeled "Num."). The parameters are set as $t=0.5$ for all panels, with $(\theta_{\uparrow}=0.5\pi,\theta_{\downarrow}=0.5\pi)$ for (a), $(\theta_{\uparrow}=0.4\pi,\theta_{\downarrow}=0.5\pi)$ for (b), and $(\theta_{\uparrow}=0.5\pi,\theta_{\downarrow}=0.6\pi)$ for (c). \LZT{In the inset of (c), we also plot the Petermann factor defined in Eq.~\eqref{pe_fac} for the state with maximal imaginary eigenenergy, which takes values significantly greater than $1$ near the exceptional point.} The chosen system size is $L=32$, with $L^2=1024$ the Hamiltonian dimension.}}
\label{PTPPM}
\end{figure}

\MOD{Similarly, in the $\{++-\}$ phase, the ansatz takes the form 
\begin{equation}\label{tryppm}
\begin{split}
|\psi_{\pm}\rangle &= \frac{1}{2\sqrt{L}} \sum_{j=1}^{L/2} \left[ \sqrt{2} \left( |2j-1\rangle_{\uparrow} \otimes |2j-1\rangle_{\downarrow} + |2j\rangle_{\uparrow} \otimes |2j\rangle_{\downarrow} \right) \right. \\
& \left. \pm \left( |2j-1\rangle_{\uparrow} \otimes |2j\rangle_{\downarrow} + |2j\rangle_{\uparrow} \otimes |2j-1\rangle_{\downarrow} + |2j\rangle_{\uparrow} \otimes |2j+3\rangle_{\downarrow} + |2j+3\rangle_{\uparrow} \otimes |2j\rangle_{\downarrow} \right) \right].
\end{split}
\end{equation}
It satisfies $\langle\psi_{\pm}|H_{\rm DGF}|\psi_{\pm}\rangle=\langle \psi_{\mp}|\sum_{\sigma}H_{\sigma}|\psi_{\pm}\rangle=0$, $\langle\psi_{\mp}|H_{\rm DGF}|\psi_{\pm}\rangle=\pm\sqrt{2}t$, and $\langle \psi_{\pm}|\sum_{\sigma}H_{\sigma}|\psi_{\pm}\rangle = \pm\frac{1}{\sqrt{2}}\sum_{\sigma}u_{\sigma}$. The corresponding effective effective Hamiltonian is given by
\begin{eqnarray}\label{effppm}
H_{\rm eff}^{\{++-\}}=\frac{1}{\sqrt{2}}\left(\sum_{\sigma}u_{\sigma}\right)\tau_z+\sqrt{2}it\tau_x,
\end{eqnarray}
Clearly, the boundary between the $\mathcal{PT}$-broken and $\mathcal{PT}$-unbroken phases is also given by $|\sum_{\sigma}u_{\sigma}|=2t$. The comparison between the numerical and analytical results is given in Fig.~\ref{PTPPM}.}

\begin{figure}
    \centering
    \includegraphics[width=0.8\linewidth]{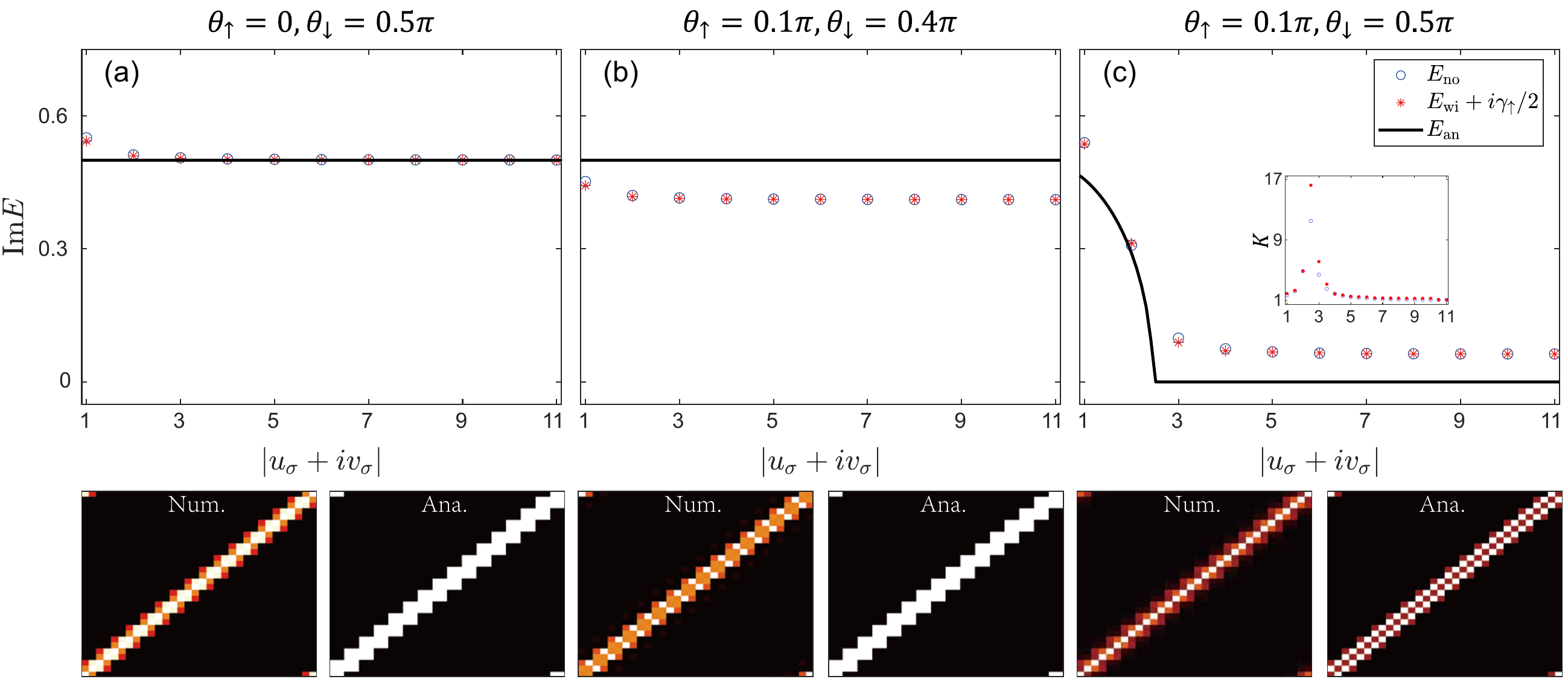}
   \caption{\MOD{\textbf{The comparison of the $\mathcal{PT}$ phase transition between analytical and numerical processes for the $\{+--\}$ phase.} The upper column show the maximum imaginary energy versus $|u_{\sigma}+iv_{\sigma}|$ obtained from three different methods: the effective Hamiltonian $H_{\rm eff}^{\{+--\}}$ in Eq.~\eqref{effpmm}, the Hamiltonian $H$ in Eq.~(1) of the main text without dissipation, and the Hamiltonian $H$ in Eq.~(1) of the main text with dissipation $\gamma_{\uparrow}=0.5$. These are denoted as $E_{\rm an}$, $E_{\rm no}$, and $E_{\rm wi}$, and plotted as black lines, blue circles, and red asterisks, respectively. For better comparison, $E_{\rm wi}$ has been shifted by $i\gamma_{\uparrow}/2$. In the lower column, the normalized two-particle correction of bulk bound states with $|u_{\sigma}+iv_{\sigma}|=1.5$ is provided, obtained from both the effective Hamiltonian $H_{\rm eff}^{\{+--\}}$ (labeled "Ana.") and the Hamiltonian $H$ without dissipation (labeled "Num."). The parameters are set as $t=0.5$ for all panels, with $(\theta_{\uparrow}=0,\theta_{\downarrow}=0.5\pi)$ for (a), $(\theta_{\uparrow}=0.1\pi,\theta_{\downarrow}=0.4\pi)$ for (b), and $(\theta_{\uparrow}=0.1\pi,\theta_{\downarrow}=0.5\pi)$ for (c). \LZT{In the inset of (c), we also plot the Petermann factor defined in Eq.~\eqref{pe_fac}  for the state with maximal imaginary eigenenergy, which takes values significantly greater than $1$ near the exceptional point.} The chosen system size is $L=32$, with $L^2=1024$ the Hamiltonian dimension.}}
\label{PTPMM}
\end{figure}

\MOD{Finally, we consider the $\{+--\}$ phase, whose results can be generalized to the $\{-+- \}$, $\{-++\}$, and $\{+-+\}$ phases via specific parameter symmetry transformations. In particular, the ansatz for the bulk bound state in the $\{+--\}$ phase is given by
\begin{eqnarray}\label{trypmm}
|\psi_{\pm}\rangle = \frac{1}{\sqrt{2L}} \sum_{j=1}^{L/2} |a_{j,\pm}^{\uparrow}\rangle \otimes \left( |2j-2\rangle_{\downarrow} \mp |2j-1\rangle_{\downarrow} - |2j\rangle_{\downarrow} \pm |2j+1\rangle_{\downarrow} \right),
\end{eqnarray}
which yields $\langle \psi_{\pm} | H_{\rm DGF} | \psi_{\pm} \rangle = \langle \psi_{\mp} | \sum_{\sigma} H_{\sigma} | \psi_{\pm} \rangle = 0$, $\langle \psi_{\mp} | H_{\rm DGF} | \psi_{\pm} \rangle = \pm t$, and $\langle \psi_{\pm} | \sum_{\sigma} H_{\sigma} | \psi_{\pm} \rangle = \pm \left( u_{\uparrow} - \frac{v_{\uparrow}}{2} + \frac{u_{\downarrow}}{2} - v_{\downarrow} \right)$. The corresponding effective Hamiltonian reads
\begin{eqnarray}\label{effpmm}
H_{\rm eff}^{\{+--\}} = \left( u_{\uparrow} - \frac{v_{\uparrow}}{2} + \frac{u_{\downarrow}}{2} - v_{\downarrow} \right) \tau_z + i t \tau_y.
\end{eqnarray}
The boundary between $\mathcal{PT}$-broken and $\mathcal{PT}$-unbroken phases is similarly given by $\left| u_{\uparrow} - \frac{v_{\uparrow}}{2} + \frac{u_{\downarrow}}{2} - v_{\downarrow} \right| = t$. A comparison between the numerical and analytical results is presented in Fig.~\ref{PTPMM}. 
}

\MOD{
For the $\{-++\}$ phase, which shares the same topological character as the $\{+--\}$ phase (with only the pseudospin-down particle being topologically nontrivial) but has $u_{\uparrow}$ taking a negative value, the trial states take the form
\begin{eqnarray}\label{trymmp}
|\psi_{\pm}\rangle = \frac{1}{\sqrt{2L}} \sum_{j=1}^{L/2} |a_{j,\pm}^{\uparrow}\rangle \otimes \left( |2j-2\rangle_{\downarrow} \pm |2j-1\rangle_{\downarrow} - |2j\rangle_{\downarrow} \mp |2j+1\rangle_{\downarrow} \right).
\end{eqnarray}
The corresponding $\mathcal{PT}$ broken-unbroken boundary is given by $|u_{\uparrow} - v_{\uparrow}/2 - u_{\downarrow}/2 + v_{\downarrow}| = t$. In summary, the $\mathcal{PT}$ broken-unbroken boundary for $\{\pm\mp\mp\}$ phases is $|u_{\uparrow} - v_{\uparrow}/2 \pm (u_{\downarrow}/2 - v_{\downarrow})| = t$. Lastly, the $\{+-+\}$ ($\{-+- \}$) phase can be mapped to the $\{+--\}$ ($\{-++\}$) phase via the transformation $\theta_{\uparrow} \leftrightarrow \theta_{\downarrow}$. With similar derivation, the $\mathcal{PT}$ broken-unbroken boundary for $\{\pm\mp\pm\}$ phases can be obtained as $|u_{\downarrow} - v_{\downarrow}/2 \pm (u_{\uparrow}/2 - v_{\uparrow})| = t$.
}

\begin{figure}
    \centering
    \includegraphics[width=0.4\linewidth]{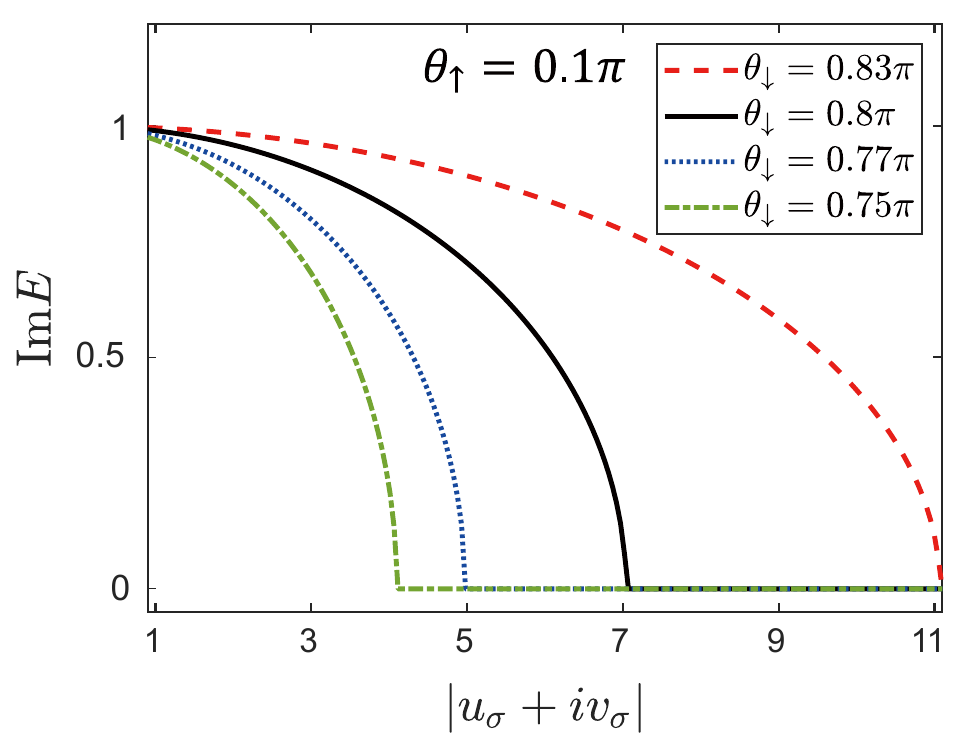}
   \caption{\MOD{\textbf{The high sensitivity of $\mathcal{PT}$ phase transition point for the $\{+--\}$ phase.} The maximum imaginary part of energy versus $|u_{\sigma}+iv_{\sigma}|$, obtained from the effective Hamiltonian $H_{\rm eff}^{\{---\}}$ in Eq.~\eqref{effmmm}. Here the parameters $t=0.5$ and $\theta_{\uparrow}=0.1\pi$ are fixed, while different lines correspond to different values of $\theta_{\downarrow}$ ($0.83\pi$, $0.8\pi$, $0.77\pi$, and $0.75\pi$). As can be seen, a small change in $\theta_{\downarrow}$ leads to a significant variation in both the imaginary part of the energy and the $\mathcal{PT}$ phase transition point. Note that the results for $\theta_{\downarrow} = 0.8\pi$ have also appeared in Fig.~\ref{PTMMM}(c).}}
\label{PTMMMdev}
\end{figure}

\MOD{In the three scenarios discussed above, numerical results reveal that bulk bound states consistently exhibit $\mathcal{PT}$ symmetry breaking under symmetric parameter conditions: $\sum_{\sigma} \theta_{\sigma} = \pi$ for the first two scenarios, and $\sum_{\sigma} \theta_{\sigma} = \pi/2$ or $3\pi/2$ for the last scenario. This behavior aligns with the fact that the effective Hamiltonian 
contains only the term of $i\tau_y$ under these parameter settings. 
However, the ansatz wave function is obtained under symmetric parameters, and deviates from the actual one in the asymmetric scenario.
Thus, the effective Hamiltonian becomes less accurate, and discrepancies emerge between the $\mathcal{PT}$ phase transition points predicted by the effective Hamiltonian and those obtained numerically.
Furthermore, the eigenvalues of the non-Hermitian Hamiltonian are highly sensitive to these parameter variations. For instance, in the $\{---\}$ phase, the effective Hamiltonian $H_{\rm eff}^{\{---\}}$ depends on the parameter $u_{\sigma}$, which varies significantly within this phase (with $|\partial_{\theta_{\sigma}} u_{\sigma}|$ being relatively large). In Fig. \ref{PTMMMdev}, we fix $\theta_{\uparrow}$ and slightly vary $\theta_{\downarrow}$, while illustrating the imaginary part of the eigenvalues derived from the effective Hamiltonian.
It is evident that even a slight change in $\theta_{\downarrow}$ leads to a significant discrepancy in the $\mathcal{PT}$ transition point, confirming the high sensitivity of our system.
}

\LZT{To further characterize the process of the $\mathcal{PT}$-symmetric phase transition and the emergence of the exceptional point, we also define and compute the Petermann factor—mathematically equivalent to the condition number~\cite{PhysRevResearch.5.033042,PhysRevResearch.6.013044}—of the state with the largest imaginary part, which is defined as:
\begin{eqnarray}\label{pe_fac}
K= \frac{{}_L\langle \psi | \psi \rangle_L~ {}_R\langle \psi | \psi \rangle_R}{\left| {}_L\langle \psi | \psi \rangle_R \right|^2}
\end{eqnarray}
where $| \psi \rangle_R$ ($| \psi \rangle_L$) denotes the right (left) eigenstate of the Hamiltonian. The Petermann factor is significantly larger than $1$ in the vicinity of the exceptional point. As illustrated in the insets of Figs. \ref{PTMMM}(c), \ref{PTPPM}(c), and \ref{PTPMM}(c), the emergence of the Exceptional Point (EP) is unambiguously marked by the transition point at which the imaginary part of this state vanishes (i.e., changes from non-zero to zero).}

\subsection{$\mathcal{PT}$-breaking of other bulk eigenstates}
\MOD{Finally, other bulk eigenstates can also break $\mathcal{PT}$ symmetry. They consists of other non-topological eigenstates, which generally correspond to single-particle bulk states of the two species and are less influenced by the DGF. This is because a strong DGF effect already induces the bulk bound states discussed earlier. Therefore, the $\mathcal{PT}$-breaking condition for these non-topological states can be approximately described by the single-particle Hamiltonian with the DGF neglected. Within our considered parameter regime—where the loss term is applied exclusively to the pseudospin-up particle—the $\mathcal{PT}$-breaking condition is given by $\gamma_{\uparrow}^2 > 4(u_{\uparrow}^2 + v_{\uparrow}^2 - 2|u_{\uparrow}v_{\uparrow}|)$, derived from the single-particle Bloch Hamiltonian for the pseudospin-up particle. This condition determines the gray regions in the phase diagrams presented in Figs. 3 and 4 of the main text.
}

\MOD{In summary, the $\mathcal{PT}$-broken condition in our system can be categorized into three sources: 
\begin{itemize}
\item (i) $|u_{\uparrow}| < |v_{\uparrow}|$ or $|u_{\downarrow}| < |v_{\downarrow}|$, ensuring that $\mathcal{PT}$-broken edge (anti-)confined states emerge. 
\item (ii) For bulk bound states to break $\mathcal{PT}$ symmetry: $|\sum_{\sigma} u_{\sigma}| < 2t$ for the $\{---\}$ and $\{++-\}$ phases; $|u_{\uparrow} - v_{\uparrow}/2 \pm (u_{\downarrow}/2 - v_{\downarrow})| < t$ for the $\{\pm\mp\mp\}$ phases; and $|u_{\downarrow} - v_{\downarrow}/2 \pm (u_{\uparrow}/2 - v_{\uparrow})| < t$ for the $\{\pm\mp\pm\}$ phases. 
\item(iii) $\gamma_{\uparrow}^2 > 4(u_{\uparrow}^2 + v_{\uparrow}^2 - 2|u_{\uparrow}v_{\uparrow}|)$, so that the non-topological states breaks $\mathcal{PT}$ symmetry.
\end{itemize}
Note that these parameter constraints are derived with certain approximations and may become less accurate when the underlying approximations break down (e.g., when deviating from symmetric parameters, as discussed for bulk bound states). Additionally, states from source (i) arise exclusively under OBCs, whereas those from sources (ii) and (iii) exist under both OBCs and PBCs. The latter do not exhibit the NHSE, and their eigenenergies remain nearly identical across different boundary conditions.
}
  
\section{Supplementary Note 5: numerical instability of diagonalizing non-Hermitian matrices}\label{appCO}
Non-Hermitian systems with non-reciprocal hopping exhibit spectral instabilities in some cases, especially when their parameters are near exceptional points where Hamiltonians appear similar to a Jordan block~\cite{RevModPhys.93.015005,2408.01265}. 
In this section, we discuss the numerical stability issue.
To provide a more general scheme for identifying the numerical instability/stability of non-Hermitian matrices $H$, we note that numerical instability can be attributed to the non-normality of a matrix~\cite{g1cw-tk7f,1bvp-p2cz}, which can be characterized by a condition number defined as~\cite{g1cw-tk7f}:
\begin{eqnarray}\label{CO}
{\rm cond}(V)=||V||\cdot||V^{-1}||.
\end{eqnarray}
Here $V$ is an invertible matrix diagonaling the matrix $H$, $H=V\Lambda V^{-1}$, $\Lambda$ a diagonal matrix, and $||A||=s_{\rm max}(A)$ is the $2$-norm of matrix $A$ with $s_{\rm max}(A)$ being its largest singular value. 
The condition number satisfies ${\rm cond}(V)=1$ for a normal matrix, and becomes extremely large for a highly non-normal matrix.

\begin{figure}
    \centering
    \includegraphics[width=0.95\linewidth]{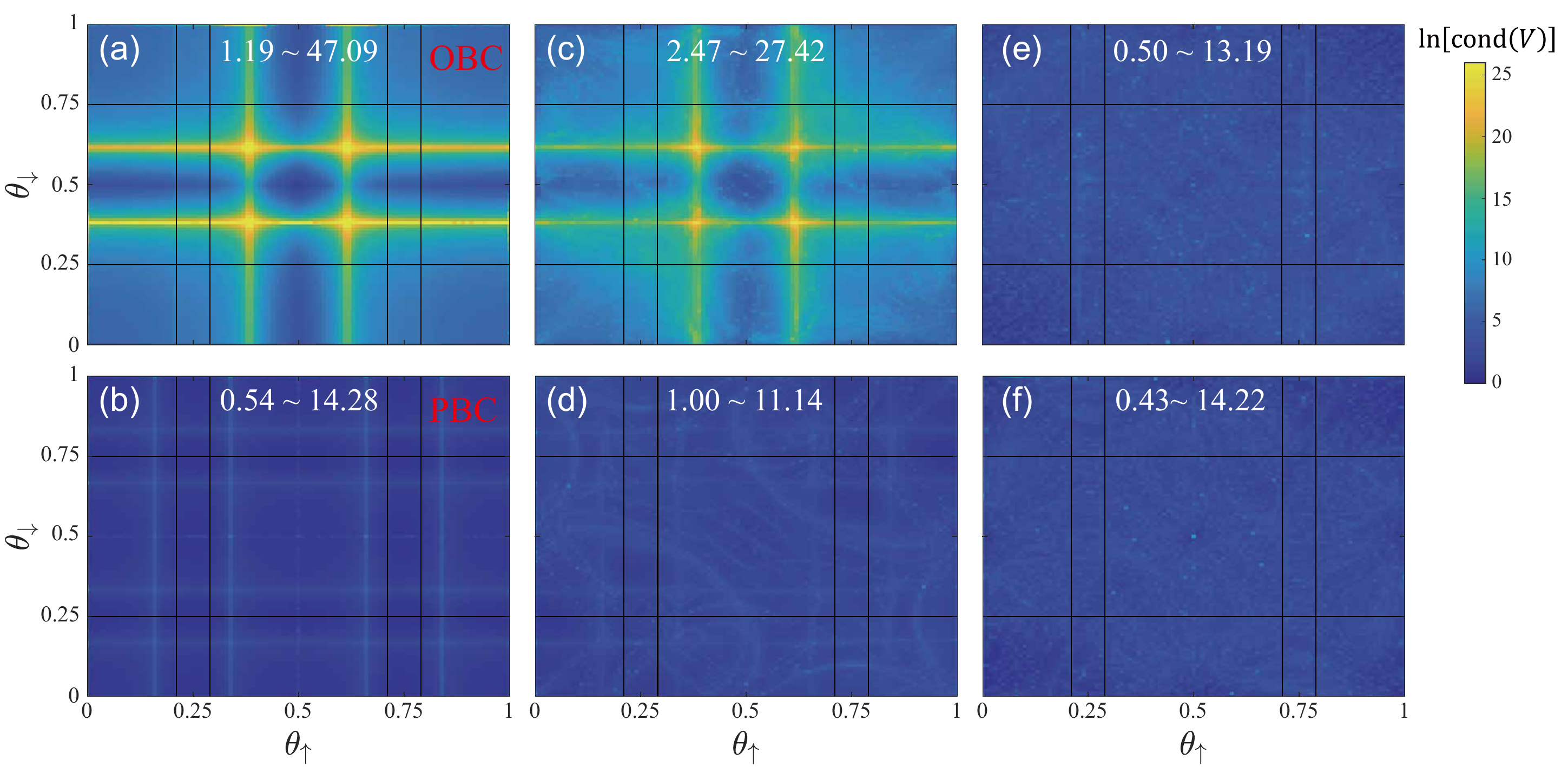}
   \caption{\MOD{\textbf{The comparison of condition numbers for our system and similar ones with non-reciprocal hopping.} (a) and (b) The logarithms of the condition numbers, as defined in Eq.\eqref{CO} and spanned over the $\theta_{\uparrow}$-$\theta_{\downarrow}$ parameter space, under OBC and PBC, respectively. Here, the intra-cell non-reciprocal hopping term $H_{\rm NRH}$ [defined in Eq.~\eqref{NRH}] is included instead of DGF $H_{\rm DGF}$. (c) and (d) [(e) and (f)] are similar to (a) and (b), but includes both $H_{\rm NRH}$ and $H_{\rm DGF}$ ($H_{\rm DGF}$ instead of $H_{\rm NRH}$). The range of ${\rm ln}[{\rm cond}(V)]$ is given in each panel. Other parameters are fixed with $|u_{\sigma}+iv_{\sigma}|=\sqrt{2}$ and $\gamma_\uparrow=t=0.5$. We note that the situations discussed in (e) and (f), which exhibit relatively smaller condition numbers under both OBC and PBC, are exactly the ones studied in this work. The chosen system size is $L=16$, with $L^2=256$ the Hamiltonian dimension.}}
\label{cond}
\end{figure}

\MOD{To more intuitively demonstrate the numerical stability of the system under our study, we additionally introduce an intra-cell non-reciprocal hopping term:
\begin{eqnarray}\label{NRH}
H_{\rm NRH}=\sum_{\sigma}\sum_{j=1}^{L/2}ta^\dagger_{\sigma,2j-1}a_{\sigma,2j}-h.c.,
\end{eqnarray}
with same strength of DGF. 
In Fig.~\ref{cond}, we compare the values of ${\rm cond}(V)$ for three scenarios of our model:}
\begin{itemize} 
%Moreover, the outcomes for several flat flat band limit cases can be computed easily.
\item[(i)] \MOD{without the DGF $H_{\rm DGF}$ but with an intra-cell non-reciprocal hopping term $H_{\rm NRH}$. 
In this scenario,
the system is equivalent to two single-particle chains, and $H_{\rm NRH}$ may induce strong instability by driving the system closer to exceptional points. 
As illustrated in Figs.~\ref{cond}(a) and (b), the condition number cond$(V)$ of our system reaches an extremely large value of approximately ${\rm cond}(V)\approx e^{47}$ under OBCs, whereas it is significantly smaller under PBCs. This finding aligns with the single-particle framework: a system under PBCs can be characterized by Bloch wavefunctions and are thus numerically stable, while the NHSE that gives rise to numerical instability only manifests under OBCs.}

\item[(ii)] \MOD{with both the DGF $H_{\rm DGF}$ and the extra non-reciprocal term $H_{\rm NRH}$. Notably, the condition number under OBCs becomes relatively smaller (${\rm cond}(V)\approx e^{27}$), as shown in Fig.~\ref{cond}(c). 
This result implies that the DGF reduces the instability of the system, possibly due to that it induces different non-reciprocity to states with different particle distributions, thereby driving the system away from possible exceptional points. In other words, the DGF actually enhances the stability of our system.}

\item[(iii)] \MOD{with only the DGF $H_{\rm DGF}$, which is the case we study in our paper. We find that, as shown in Figs. \ref{cond}(e) and (f), the non-normality induced by the DGF corresponds to a maximal condition number of approximately ${\rm cond}(V)\approx e^{13}$ and ${\rm cond}(V)\approx e^{14}$ under OBCs and PBCs, respectively, and only at some discrete points. This indicates that the system exhibits only weak instability, which occurs only at a few discrete parameter points, regardless of whether the boundary conditions are OBCs or PBCs.}
\end{itemize}

\MOD{As a summary of this section, we verified the computational stability of the studied system by calculating the condition number. This ensures that the results are not affected by the machine precision in numerical calculations, and there is no need to resort to infinite precision to circumvent such an impact~\cite{2408.01265}.}

\begin{figure}
    \centering
    \includegraphics[width=0.5\linewidth]{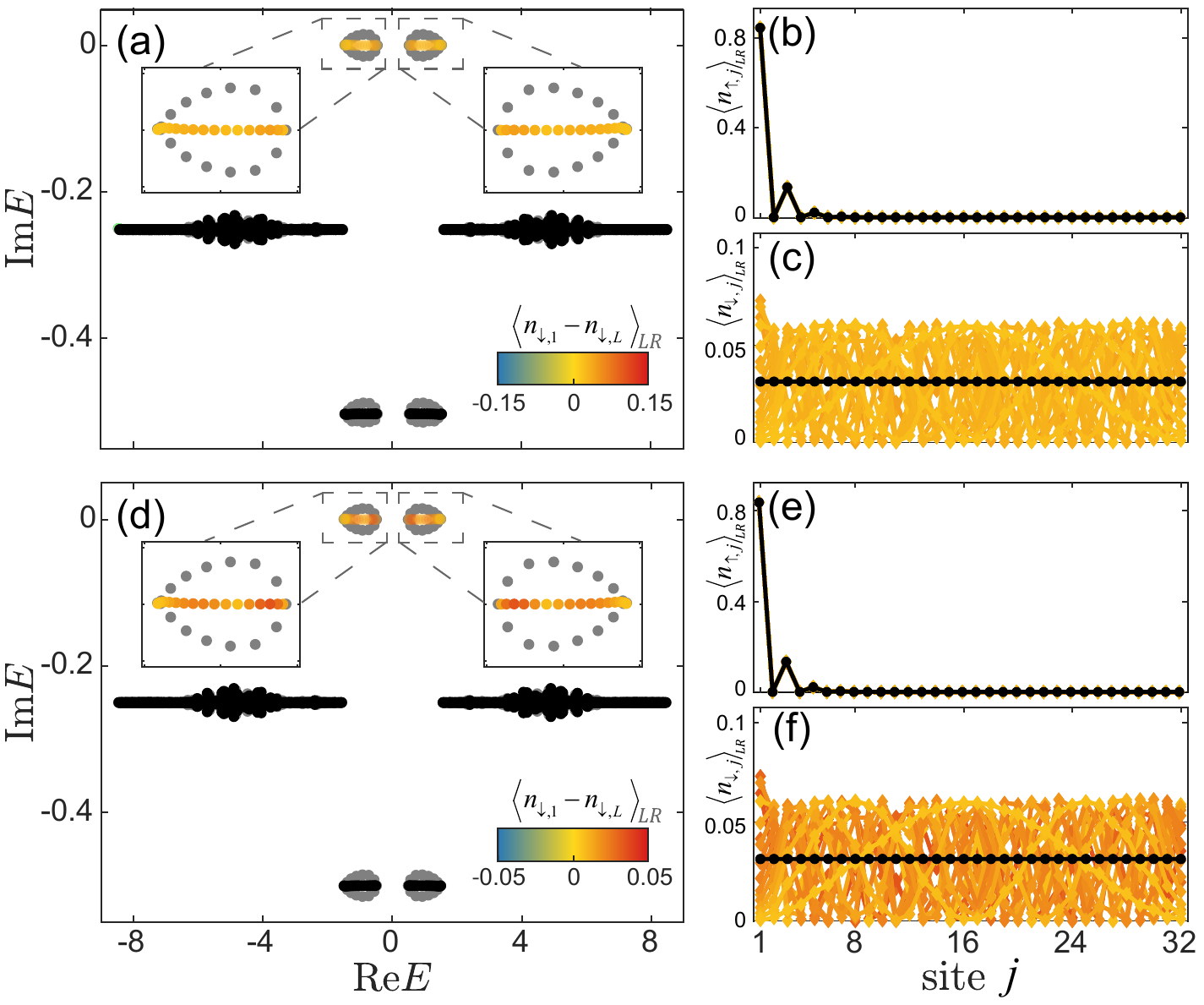}
   \caption{\LZT{\textbf{Edge confined and anti-confined states under BO basis.} (a) OBC spectrum of the Hamiltonian (black and colored dots) in Eq. (1) of the main text, where the eigestates with ${\rm Im}E\approx0$ are marked by the edge-density imbalance of the pseudospin-down particle under BO basis. Gray dots are the eigenenergies with PBCs taken only for the pseudospin-down particle. (b) and (c) distributions of pseudospin-up and -down particles under BO basis, respectively, for OBC eigenstates in the enlarged insets in (a) marked by the same colors. Their average distributions are shown in black in each panel. Parameters in (a) to (c) are $v_{\uparrow}=5$, $v_{\downarrow}=0.5$, $\gamma_{\uparrow}=t=0.5$, $u_{\uparrow}=2$, and $u_{\downarrow}=1$. (d) to (f) the same as (a) to (c), but with $u_{\uparrow}=-2$ and $u_{\downarrow}=-1$. The pseudospin-down particle do not exhibit any noticeable inhomogeneous distribution at the edges.
The chosen system size is $L=32$, with $L^2=1024$ the Hamiltonian dimension.}}
\label{EdgBO}
\end{figure}

\begin{figure}[H]
    \centering
    \includegraphics[width=0.3\linewidth]{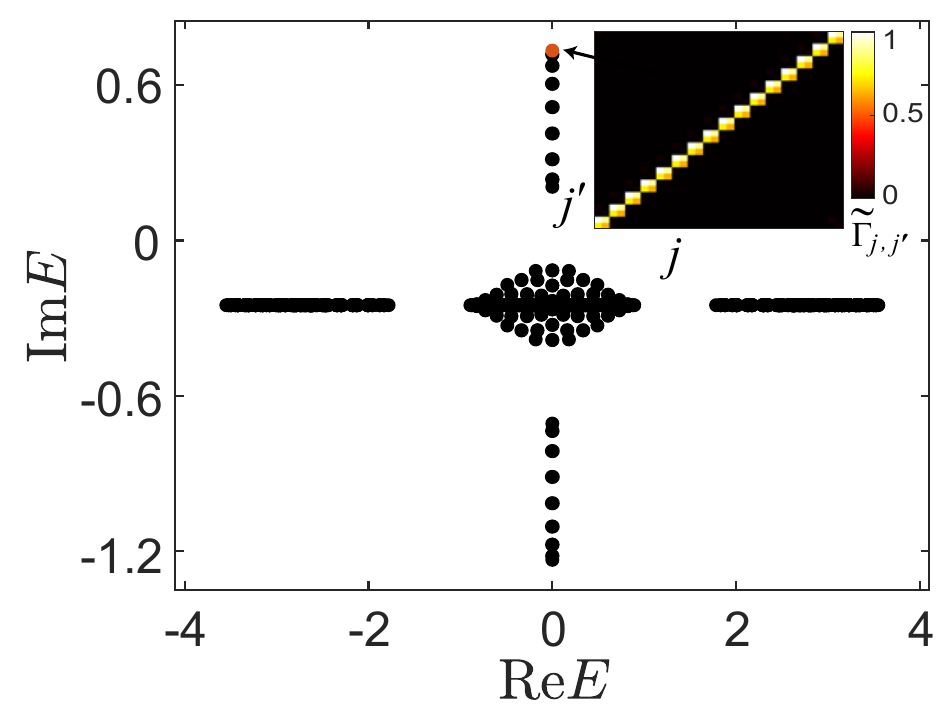}
   \caption{\LZT{\textbf{Bulk bound states under BO basis.} The PBC spectrum of the Hamiltonian in Eq. (1) of the main text, 
with $\theta_{\uparrow}=0.1\pi$, $\theta_{\downarrow}=0.9\pi$, $\theta_{\sigma}={\rm arg}(u_{\sigma}+iv_{\sigma})$, $|u_{\sigma}+iv_{\sigma}|=\sqrt{2}$, and $\gamma_\uparrow=t=0.5$,  i.e., the same as Fig.3(a) of the main text.
The normalized two-particle correction under BO basis [Eq.~\eqref{bulBO}] of the state with the highest imaginary energy (marked by arrows) is shown in the inset. The chosen system size is $L=32$, with $L^2=1024$ the Hamiltonian dimension.}}
\label{BulBO}
\end{figure}

\LZT{\section{Supplementary Note 6: The edge (anti-)confined states and bulk bound states under bi-orthogonal basis}\label{appBO}
	In non-Hermitian systems, eigenstates generally do not satisfy the orthonormal condition, but rather the bi-orthonormal condition (BO) between left and right eigenvectors~\cite{Brody_2014,PhysRevLett.121.026808}. To give a complete understanding of the inter-species topological states in our model, we provide additional results for edge (anti-)confined states and bulk bound states under BO basis in this section. Explicitly, we introduce the particle distribution under BO basis as
\begin{eqnarray}\label{edgBO}
\langle n_{\sigma,j}\rangle_{LR}={}_L\langle\psi_m|n_{\sigma,j}|\psi_m\rangle_R,
\end{eqnarray}
where $H|\psi_m\rangle_R=E_m|\psi_m\rangle_R$ and $H^{\dagger}|\psi_m\rangle_L=E^*_m|\psi_m\rangle_L$ give the right and left eigenstates, respectively, satisfying  ${}_L\langle\psi_m|\psi_m'\rangle_R=\delta_{mm'}$. As depicted in Fig.~\ref{EdgBO}, the pseudospin-up particle retains its topological localization property, whereas the pseudospin-down particle shows an extended distribution in the BO basis. In other words, both the edge-confined and anti-confined distribution characteristics disappear in the BO basis, which suggests that their origin—aside from the DGF-driven mechanism—is similar to that of NHSE. %That is, the presence of the NHSE can break the conventional bulk-boundary correspondence through the use of left/right eigenvectors\cite{PhysRevLett.116.133903,PhysRevLett.121.086803,RevModPhys.93.015005}.
%In contrast, using the BO basis it has been shown that the conventional bulk-boundary correspondence properly predicts the topological phase boundaries\cite{PhysRevLett.121.026808}.
In contrast, bulk bound states exhibit an extended distribution and are therefore expected to persist in the BO basis as well. To verify this, we compute the two-particle correlation function in the BO basis, which is defined as
\begin{eqnarray}\label{bulBO}
\widetilde{\Gamma}_{j,j'}=\Gamma_{j,j'}/{\rm Max}(\Gamma_{j,j'}),~~~\Gamma_{j,j'}={}_L\langle n_{\uparrow,j}n_{\downarrow,j'}\rangle_R,
\end{eqnarray}
using the same parameters as those in Fig. 3(a) of the main text. As illustrated in Fig. \ref{BulBO}, the key characteristics of the bulk bound states remain intact.}

\section{Supplementary Note 7: realization of our model using cold atoms}\label{appE}

In this section, we outline schemes for realizing the Hamiltonian in Eq. (1) of the main text. 
%As already mentioned in the main text, the pseudospin-dependent SSH model can be implemented in cold atom systems. Here, we focus on the scheme for realizing the DGF. 
Specifically, in Subsec.~\ref{appE1}, we demonstrate how to induce intra-cell hopping with non-reciprocal phases in the SSH model via a three-step Floquet protocol. Building upon these implementations, we construct the DGF by tuning the strength of non-reciprocal hopping and inter-species interactions, as detailed in Subsec.~\ref{appE2}.
Alternatively, the resultant model in Subsec.~\ref{appE1} may also be implemented in the Lindblad master equation framework, as presented in Subsec.~\ref{appE3}. 
%\MOD{In fact, considering the diverse experimental techniques available in cold-atom experiments and other platforms—along with the realization of nonreciprocal hopping~\cite{PhysRevLett.129.070401} and DGFs (through Floquet approaches)~\cite{PhysRevLett.121.030402,Gorg2019,Schweizer2019} in atomic experiments—our work can be further refined by incorporating these advancements.}

\subsection{A three-step modulation process to realize non-reciprocity hopping}\label{appE1}
We consider a three-step Floquet model with a frequency $\Omega=2\pi/T$, described by the following Hamiltonian,
\begin{eqnarray}\label{SMTDHam0}
H(t)&=&H_0+\sum_{\sigma}V_{\sigma}(t),\\
H_0&=&\sum_{\sigma=\uparrow,\downarrow}\left(H_{\sigma}-i\sum_{j=1}^{L/2}\gamma'_{\sigma}n_{\sigma,2j}\right),~~
V_{\sigma}(t)=
\begin{cases}
V_1=\sum^{L/2}_{j}\Delta_{\sigma}(ia_{\sigma,2j-1}^{\dagger}a_{\sigma,2j}+H.c.), & 0\leq t< T/3 \nonumber\\
V_2=-i\sum^{L/2}_{j}\mu_{\sigma}n_{\sigma,2j}, & T/3\leq t< 2T/3 \nonumber\\
V_3=-\sum^{L/2}_{j}\Delta_{\sigma}(ia_{\sigma,2j-1}^{\dagger}a_{\sigma,2j}+H.c.). & 2T/3\leq t<T 
\end{cases}.\nonumber
\end{eqnarray} 
$H_\sigma$ is the pseudospin-dependent SSH model, which can be realized by  ``tune-out" wavelengths with suitable polarization for Bose–Einstein condensates (BEC) systems such as $^{87}$Rb atoms~\cite{Wen:21,Meng2023} in dimerized optical lattices~\cite{Atala2013}.
$\gamma'_\sigma$ and $\mu_\sigma$ in $V_2$ are static and periodically-modulated components of pseudospin-dependent loss, which can be realized by applying a resonant optical beam~\cite{Li2019,PhysRevLett.129.070401,Zhao2025}.
Additionally, a periodically-modulated intra-cell hopping $\pm i\Delta_\sigma$ in $V_{1}$ and $V_3$ with a phase difference $\pm\pi/2$ with respect to the natural tunneling can be implemented through Raman-assisted tunneling~\cite{PhysRevLett.107.255301,PhysRevLett.111.185301,PhysRevLett.111.185302,Jotzu2014,Aidelsburger2015,Kennedy2015,doi:10.1126/science.1259052}.

In the high frequency limit, we can obtain an effective Hamiltonian up to the order of $1/\Omega$ via perturbative expansion~\cite{PhysRevX.4.031027}.
This Hamiltonian consists of the SSH model and non-Hermitian intra-cell hopping with a non-reciprocity phase:
\begin{eqnarray}
H=\sum_{\sigma=\uparrow,\downarrow}\left[H_{\sigma}-i\sum_{j=1}^{L/2}\gamma_{\sigma}n_{\sigma,2j}+i\chi_{\sigma}\sum_{j}^{L/2}(a_{\sigma,2j-1}^{\dagger}a_{\sigma,2j}+H.c.)\right].
\label{SMTDHam0}
\end{eqnarray}
Here $\gamma_{\sigma}\equiv\gamma'_{\sigma}-\mu_{\sigma}/3$ and $\chi_{\sigma}\equiv\frac{2\pi}{27\Omega}\Delta_{\sigma}\mu_{\sigma}$.
Note that in order for $\gamma_\sigma$ (the parameter in our DGF Hamiltonian in Eq. (1) of the main text) to choose arbitrary non-negative value, $\gamma'_{\sigma}$ may need to take negative value that represents gain instead of loss.
To ensure no pure gain in our system, we can introduce an extra background loss term $-i\sum_{j = 1}^{L}\kappa_{\sigma}n_{\sigma,j}$ 
and require $\gamma'_{\sigma}+\kappa_{\sigma}\geq0$.

In the following analysis, we choose  $\chi_{\sigma}$ to be pseudospin-independent, i.e. $\chi_{\uparrow}=\chi_{\downarrow}\equiv\chi$.
Then we obtain a SSH model with non-Hermitian intra-cell hopping, which provides a basis to realize DGF.

\subsection{Inducing DGF through a second Floquet modulation}\label{appE2}
Next, we consider another periodic modulation to the system, $\chi\rightarrow\chi\cos(\omega t)$, with an extra modulated inter-species interaction $V\sin(\omega t)\sum_{j=1}^{L}n_{\uparrow,j}n_{\downarrow,j}$ induced by Feshbach resonance~\cite{RevModPhys.82.1225}.
Assuming $\omega\ll\Omega$, we can obtain an effective Hamiltonian up to the order of $1/\omega$ by means of the high frequency expansion~\cite{PhysRevA.68.013820,Eckardt_2015}, 
%After obtaining the SSH model with non-Hermitian intra-cell hopping in Eq.~\eqref{SMTDHam0}, we further modulate the strength of non-Hermitian hopping, specifically $\chi\rightarrow\chi\cos(\omega t)$. Additionally, we introduce a modulated inter-species interaction $V\sin(\omega t)\sum_{j=1}^{N}n_{\uparrow,j}n_{\downarrow,j}$ through Feshbach resonance~\cite{RevModPhys.82.1225}. These modulations occur on another time-scale $1/\omega$ with $\omega\ll\Omega$. By means of the high frequency expansion~\cite{PhysRevA.68.013820,Eckardt_2015}, we obtain the effective Hamiltonian up to the order of $1/\omega$:
\begin{eqnarray}
H_{\rm eff}&=&\sum_{\sigma=\uparrow,\downarrow}\left(H_{\sigma}+H_{\rm DGF}-i\sum_{j=1}^{L/2}\gamma_\sigma n_{\sigma,2j}\right)\nonumber\\
H_{\rm DGF}&=&\frac{\chi V}{4\omega}\sum_{\sigma\neq\bar{\sigma}}\sum_{j=1}^{L/2}\left[\left(n_{\bar{\sigma},2j-1}-n_{\bar{\sigma},2j}\right)a^\dagger_{\sigma,2j-1}a_{\sigma,2j}\right]-h.c.,
%H_{\rm AH}&=&\sum_{j}t\left(n_{\downarrow,2j-1}-n_{\downarrow,2j}\right)a^\dagger_{\uparrow,2j-1}a_{\uparrow,2j}-h.c.-i\gamma_\uparrow n_{\uparrow,2j}+\uparrow\leftrightarrow\downarrow
\label{SMSSHDGFeff}
\end{eqnarray}
which has exactly the same form as the Hamiltonian in Eq.(1) of the main text.

\MOD{The Floquet process often induces heating effects, and here we briefly discuss their impact on the observation of ISTPs. Firstly, it should be noted that our proposed scheme is based on the high-frequency regime and its effective Hamiltonian formulation. Additionally, according to a recent study on Floquet heating~\cite{PhysRevA.100.033406}, the heating rate $P$ due to two-body collisions in different frequency regimes follows a general semiclassical expression $P\propto\rho\sigma vE_0$, implying that $P$ can be reduced at lower atomic densities $\rho$. The core of our theoretical work focuses on few-body physics, with key results observable in the low-density regime. Furthermore, the modified density of states in low dimensions can significantly suppress the heating rate. Specifically, 1D system we focus on, the heating formula in the high-frequency regime takes the form: $P\propto n_{1D} (\hbar/m)^2\sqrt{m/\hbar\omega}\sigma/l_0^4E_0$.
This indicates that the heating rate $P$ is suppressed as the modulation frequency $\omega$ increases. Moreover, our work focuses on scenarios where natural tunneling is weak (on the same order as the dynamic gauge field), which requires a deep lattice potential corresponding to a large band gap between the $s$ and $p$ bands. This can also suppress interband transitions induced by heating. 
}

\MOD{Moreover, we perform a rough estimation for strength of DGF by referencing relevant $^{87}\text{Rb}$ experiments and theoretical works closely tied to such atomic experiments. According to a recent study~\cite{PhysRevA.111.013319}, for a deep lattice depth ($30E_r$, where $E_r$ denotes the recoil energy), the natural tunneling strength is $t_0 = 70{\rm Hz}$, while the laser-induced hopping can reach $t_{{\rm SOC}} = 700\,{\rm Hz} = 10t_0$. The interaction strength can be as large as $U = 1.5{\rm kHz} \approx 20t_0$, which can be further enhanced via Feshbach resonance. Additionally, atomic dissipation can be effectively induced and controlled using an additional laser field, as demonstrated in the $^{87}{\rm Rb}$ experiment that realized the NHSE~\cite{PhysRevLett.129.070401}, where the dissipation strength is $\gamma = 1.3{\rm kHz}\approx20t_0$.
Assuming two control frequencies $\Omega = 10\omega = 100t_0$, the strength of the dynamic gauge field is estimated as $t = \frac{2\pi t_{\text{SOC}}\gamma U}{108\omega\Omega}\approx0.23t_0$, which is slightly smaller than the value primarily used in the main text. Nevertheless, this magnitude is still sufficient to observe edge (anti-)confined states and bulk bound states, particularly considering that both interactions and dissipation can be further increased.
 }

\subsection{Effective Hamiltonian of the Lindblad master equation}\label{appE3}

Besides the three-step modulation with a frenquency $\Omega$, %we show the form of Hamiltonian in
Eq.~\eqref{SMTDHam0} can also be realized as the effective Hamiltonian of the Lindblad master equation.
In this way, only a  single-period modulation with a frequency $\omega$ is required to realize the final Hamiltonian with DGF.

Specifically, we further consider two excited states $|e_{\uparrow}\rangle$ and $|e_{\downarrow}\rangle$, which experience an opposite Stark shift to the one for $|\uparrow\rangle$ and $|\downarrow\rangle$ (the ground states of the two species of particles). Due to the Stark shift, the lattice potential minima of $|e_{\sigma}\rangle$ will locate in the middles of the minima of $|\sigma\rangle$, leading to a dimerized lattice for excited and grond states of each species.%among the ones for ground states. 
That is to say, the $(2j-1)$-site of $|e_{\sigma}\rangle$ will locate between the $(2j-1)$- and $(2j)$-site of $|\sigma\rangle$. The excited state is expected to have a dissipation with rate $g_{\sigma}$.
We further introduce lasers to couple the pseudospin states $|\sigma\rangle$ at the $(2j)$-th site and the $(2j-1)$-th site to the state $|e_{\sigma}\rangle$ at the $(2j-1)$-th site. The corresponding Rabi frequencies are $\Omega_{\sigma,1}$ and $-\Omega_{\sigma,2}$, respectively, and their atom-light detuning is $\Delta_{\sigma}$.
%The sign difference between two Rabi frequencies is from the phases accumulation of laser from $2j-1$-site to $2j$-site of $|\sigma\rangle$~\cite{PhysRevX.8.031079}.  
These processes are described by the Lindblad master equation:
\begin{eqnarray}
\dot{\rho}_t=-i[H_t,\rho_t]+\sum_{j = 1}^{L}D[L_{\sigma,j}]\rho_t,
\label{SMMaster}
\end{eqnarray}
where \(D[L_{\sigma,j}]\rho_t = L_{\sigma,j}\rho_tL_{\sigma,j}^{\dag}-\frac{1}{2}\left\{L_{\sigma,j}^{\dag}L_{\sigma,j},\rho_t\right\}\) and \(L_{\sigma,j}=g_{\sigma}e_{\sigma,j}\).
The Hamiltonian is:
\begin{eqnarray}
H_t=\sum_{\sigma}\left\{H_{\sigma}+\sum_{j = 1}^{L/2}\left(\Omega_{\sigma,1} a^\dagger_{\sigma,2j - 1}e_{\sigma,2j - 1}-\Omega_{\sigma,2}a^\dagger_{\sigma,2j}e_{\sigma,2j - 1}+H.c.\right)
-\Delta_{\sigma}\sum_{j = 1}^{L}e^\dagger_{\sigma,j}e_{\sigma,j}\right\}.
\label{SMTotHam}
\end{eqnarray}
In the regime of ${\rm max}(\Delta_{\sigma},g_{\sigma})\gg{\rm max}(\Omega_{\sigma,1},\Omega_{\sigma,2})$, the excited states $|e_{\sigma}\rangle$ can be adiabatically eliminated, resulting in an effective Lindblad master equation:
\begin{eqnarray}
\dot{\rho}_a&=-i\left[\sum_{\sigma}H_\sigma,\rho_a\right]+\sum_{\sigma}\sum_{j}D[\tilde{L}_{\sigma,j}]\rho_a,\\
\tilde{L}_{\sigma,j}&=\frac{\sqrt{g_{\sigma}}}{\sqrt{g^2_{\sigma}+4\Delta^2_{\sigma}}}(\Omega_{\sigma,1}a_{\sigma,2j - 1}-\Omega_{\sigma,2}a_{\sigma,2j}),
\label{SMEFFMaster}
\end{eqnarray}
where $\tilde{L}_{\sigma,j}$ is the effective Lindblad operator.
\MOD{Thus, an effective non-Hermitian Hamiltonian can be obtained by neglecting the quantum jumps}:
\begin{eqnarray}
H_{\rm eff}&=&\sum_{\sigma}(H_{\sigma}-i\sum_{j}L^{\dagger}_{\sigma,j}L_{\sigma,j})\nonumber\\
&=&\sum_{\sigma}\left\{H_{\sigma}+\sum_{j}^{L/2}[-i(\kappa_{\sigma,o}a^\dagger_{\sigma,2j-1}a_{\sigma,2j-1}
+\kappa_{\sigma,e}a^\dagger_{\sigma,2j}a_{\sigma,2j})+i\chi_{\sigma}(a^\dagger_{\sigma,2j-1}a_{\sigma,2j}+H.c.)]\right\},
\label{SMPos}
\end{eqnarray}
where $\kappa_{\sigma,o}=\frac{\Omega^2_{\sigma,1}{g_{\sigma}}}{{g^2_{\sigma}+4\Delta^2_{\sigma}}}$, $\kappa_{\sigma,e}=\frac{\Omega^2_{\sigma,2}{g_{\sigma}}}{{g^2_{\sigma}+4\Delta^2_{\sigma}}}$, and $\chi_{\sigma}=\frac{\Omega_{\sigma,1}\Omega_{\sigma,2}{g_{\sigma}}}{{g^2_{\sigma}+4\Delta^2_{\sigma}}}$. 
Requiring $|\Omega_{\downarrow,1}|=|\Omega_{\uparrow,1}|\leq|\Omega_{\sigma,2}|$ and ignoring the background particle loss $\kappa_{\sigma,o}$,
%After ignoring the background particle loss , 
we can see this effective Hamiltonian has the same form as the one in Eq.~\eqref{SMTDHam0}.

\MOD{
To justify the effective non-Hermitian Hamiltonian,  
we note that in certain open systems where dissipation corresponds to atomic loss from the main system to additional energy levels (also termed the environment), the quantum jump term only induces decoherence between the environment and the system of interest. Consequently, the main system can be described very accurately by an effective non-Hermitian Hamiltonian. This explains why, in studies investigating properties of Hamiltonian exceptional points~\cite{Li2019,Zhang2025} and realizing NHSE~\cite{PhysRevLett.129.070401} based on quantum open systems, the observed phenomena are in excellent agreement with those described by the effective non-Hermitian Hamiltonian. Furthermore, in some of these experimental works—for example, Ref.~\cite{Li2019} which discusses the influence of periodic parameter modulation on $\mathcal{PT}$ phase transitions, and Ref.~\cite{Zhang2025} which measures dynamics around exceptional points—the phenomena can all be predicted by the Floquet non-Hermitian Hamiltonian. This demonstrates that neglecting the quantum jump term is compatible with the Floquet process. 
%Our proposed experimental scheme primarily draws on these aforementioned works, where the challenging postselection approach is unnecessary.
%Additionally, it is worth noting that, in contrast to such systems, open systems involving particle decay between energy levels within the main system (e.g., experiments on quantum heat engines\cite{Zhang2022,PhysRevLett.130.110402}) cannot directly neglect the quantum jump term. For instance, the aforementioned studies consistently focus on Liouvillian exceptional points rather than Hamiltonian exceptional points.
}

\MOD{Finally, we note that both the initial Floquet process and the effective Hamiltonian of the master equation in the experimental schemes we mentioned serve to generate non-reciprocal hopping. Thanks to the extensive discussions on NHSE in recent years, numerous experimental studies on generating such nonreciprocity have been reported, as exemplified by~\cite{PhysRevLett.129.070401,Zhao2025} Our work and the proposed experimental schemes can be viewed as building upon these well-established experimental efforts, with the incorporation of an additional Floquet process.
}

\section{Supplementary Note 8: inter-species topological phases with more particles}\label{appD}
In the main text, we mainly discuss the situation with particle numbers $N_{\uparrow}=N_{\downarrow}=1$,
namely one pseudospin-$\uparrow$ particle and one pseudospin-$\downarrow$ particle. 
In this section, we extend the results to a situation with more particles ($N_{\uparrow}=N_{\downarrow}=2$) and show that the ISTPs holding different topological states still exist.
\MOD{Specifically, in Secs.~\ref{appD1} and \ref{appD2}, we assume both species are bosons and fermions, respectively.
In Sec.~\ref{appD3}, we study the fermion-boson mixture scenario.}

\begin{figure}
    \centering
    \includegraphics[width=0.5\linewidth]{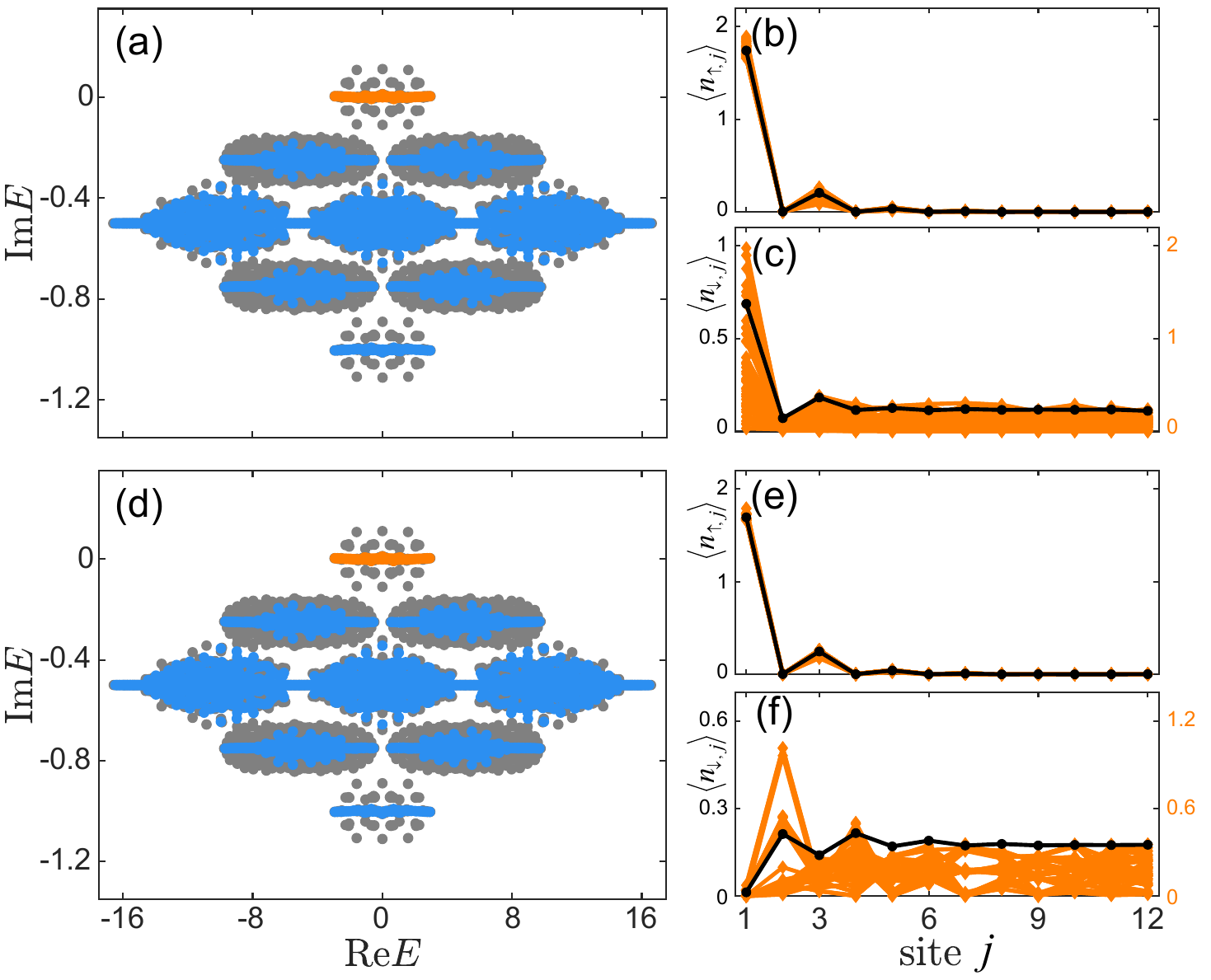}
   \caption{\textbf{Edge confined and anti-confined states for extrinsic ISTPs with $N_{\uparrow} = N_{\downarrow} = 2$ \MOD{where all particles are bosons}.} (a) OBC spectrum of Hamiltonian in Eq.(1) of the main text, represented by blue and orange dots. 
   The gray dots denote the eigenenergies obtained under PBCs for the pseudospin-down particles only. 
   %Here, spin-$\uparrow$ particle is always under OBCs, and spin-$\downarrow$ particle takes OBCs (PBCs) for colored and black (gray) circles. 
   (b) and (c) show the distributions of pseudospin-up and pseudospin-down particles, respectively, for the eigenstates marked by orange color in (a) (78 in total). 
   Orange lines and dots represent the distribution of each single eigenstate, and black line and dots denote their average.
   %The orange lines and diamonds represent the distributions of single eigenstates with $\mathrm{Im}E\approx 0$ [corresponding to the orange dots in (a)], along with their average values (black line and circles).
   The parameters in (a) to (c) are $v_{\uparrow}=5$, $v_{\downarrow}=0.5$, $\gamma_{\uparrow}=t = 0.5$, $u_{\uparrow}=2$, and $u_{\downarrow}=1$.
   (d) to (f) the same as (a) to (c), but with $u_{\uparrow}=-2$ and $u_{\downarrow}=-1$. \MOD{In (c) and (f), for a better demonstration, distributions of single eigenstates and their average correspond to different y-axis coordinates on the right and left sides of the figure, respectively.} In all panels, the total number of each specie of particles is $N_{\uparrow}=N_{\downarrow}=2$, and the number of sites is $L=12$, \MOD{which gives a Hamiltonian matrix of dimension $(L^2+L)^2/4=6084$.}}
\label{figS5}
\end{figure}

\subsection{The bosonic scenario}\label{appD1}
\MOD{In this subsection, we investigate the ISTPs involving more particles ($N_{\uparrow} = N_{\downarrow} = 2$) where the two species of particles are bosonic.} 
We first investigate the appearance and properties of the edge \MOD{confined} and anti-\MOD{confined} states for the extrinsic ISTPs. 
As shown in Figs.~\ref{figS5}(a) and (c), where pseudospin-up (pseudospin-down) particles are in the topologically nontrivial (trivial) region, more branches of eigenvalues appear, with the imaginary parts of eigenenergies diverging from the value $-\gamma_{\uparrow}N_{\uparrow}/2$.
However, there are still some states with imaginary energies close to zero, similar to those in Fig. 2 of the main text. 
The distributions of pseudospin-up particles for these states are shown in Figs.~\ref{figS5}(b) and (e), which exhibit a strong localization at the left edge.
When $u_{\downarrow}>0$, as shown in Fig.~\ref{figS5}(c), the distribution of pseudospin-down particles bulges at the left edge, and shows a uniform value away from it.
In contrast, Fig.~\ref{figS5}(e) confirms the emergence of anti-\MOD{confined} states when $u_{\downarrow}<0$, with a dip in their distribution at the left edge of the system.
These observations indicate that the edge \MOD{confined} and anti-\MOD{confined} states here origin from the same mechanism as that with $N_\uparrow=N_\downarrow=1$.
Furthermore, these edge \MOD{confined} and anti-\MOD{confined} states have the largest imaginary energies among all eigenstates, thus they shall also dominate the evolution process.
%Thereby, in the situation with more particles, we confirm that the edge bound and anti-bound states preserve their main distribution features and can still dominate the evolution process owing to their having the largest imaginary energies.

\begin{figure}
    \centering
    \includegraphics[width=0.8\linewidth]{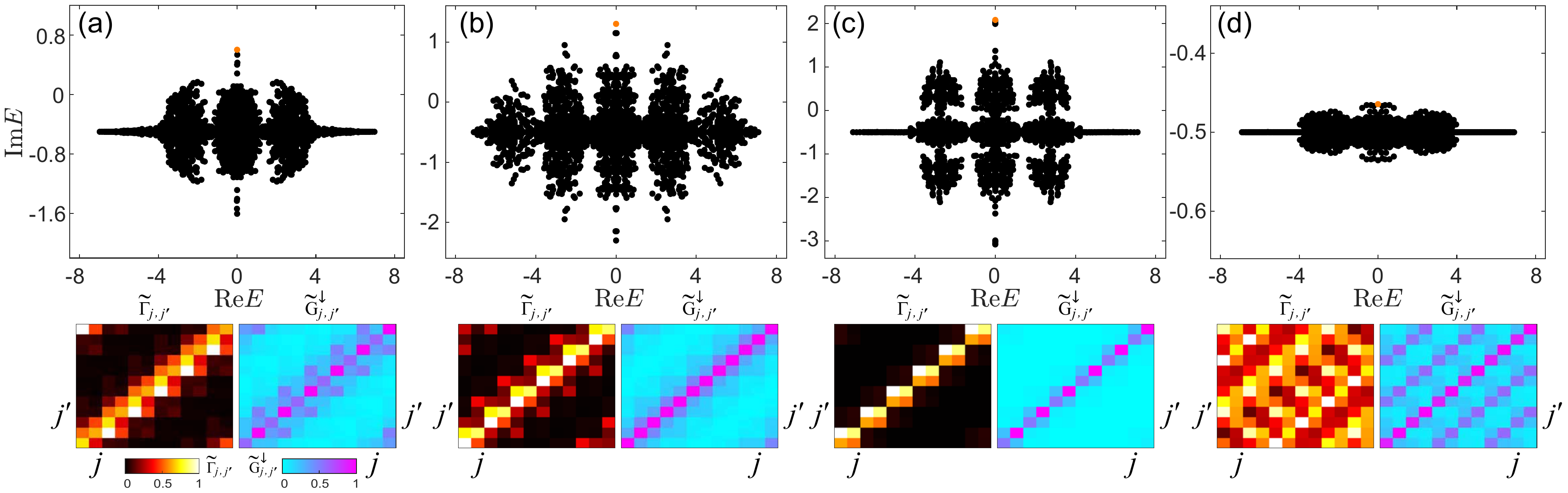}
   \caption{\textbf{Bulk bound states for intrinsic ISTPs with $N_{\uparrow}=N_{\downarrow}=2$ \MOD{where all particles are bosons}.} (a) to (d) The PBC spectra of the Hamiltonian in Eq.(1) with different parameters. Below each spectrum, we display the normalized inter-species two-particle correction $\widetilde{\Gamma}_{j,j'}$ [defined in Eq.(4) of the main text], and the intra-species ones for pseudospin-down particles $\widetilde{G}^{\downarrow}_{j,j'}$ [defined in Eq.~\eqref{TwocorrDown}], of the eigenstate with the highest imaginary energy (represented by the orange dots in the spectra). The parameters are chosen as follows: (a) $\theta_{\uparrow}=0.1\pi$ and $\theta_{\downarrow}=0.4\pi$, (b) $\theta_{\uparrow}=0.4\pi$ and $\theta_{\downarrow}=0.6\pi$, (c) $\theta_{\uparrow}=0.1\pi$ and $\theta_{\downarrow}=0.9\pi$, and (d) $\theta_{\uparrow}=0.05\pi$ and $\theta_{\downarrow}=0.15\pi$, with $|u_{\sigma}+iv_{\sigma}|=\sqrt{2}$ and $\gamma_{\uparrow}=t = 0.5$. 
   The parameter used in (a) and (b) [(c) and (d)] are the same as those used in Fig. \ref{figSBBS}(a) and (b) [Fig.3(a) and (b) of the main text].
In all panels, the total number of each species of particles is $N_{\uparrow}=N_{\downarrow}=2$, and the number of sites is $L=12$, \MOD{which gives a Hamiltonian matrix of dimension $(L^2+L)^2/4=6084$.}}
\label{figS6}
\end{figure}

Next, we focus on intrinsic ISTPs with bulk bound states induced by inter-species band inversion, with $N_\uparrow=N_\downarrow=2$. 
In Figs.~\ref{figS6}(a), (b), and (c), we display the PBC spectra in the situations with nontrivial inter-species band inversion, while Fig.~\ref{figS6}(d) corresponds to the trivial situation.
As shown here, in nontrivial situations, some eigenstates are separated from the energy clusters (black) and have relatively larger imaginary energies (as marked by orange color in the figures).
In the trivial situation, the eigenstate with the largest imaginary energy is also marked orange, which merges into the energy cluster around ${\rm Im}E=-\gamma_{\uparrow}N_{\uparrow}/2$.
%In the trivial situation, the imaginary part of the spectrum shows only a slight divergence from the value $-\gamma_{\uparrow}M_{\uparrow}/2$.
We also calculate the normalized inter-species two-particle correction $\widetilde{\Gamma}_{j,j'}$, which is defined in Eq.(4) of the main text, for the states with the largest imaginary energies.
The results are shown beneath the spectra, which confirm that the bulk bound states
%, in which two species of particles are bound together and extend throughout the bulk, 
also appear in the nontrivial cases and vanish in the trivial case, in the situation with more particles.
\MOD{Furthermore, we define a two-particle correlation function between the same species of particles, $G^{\sigma}_{j,j'}$, and its normalized form $\widetilde{G}^{\sigma}_{j,j'}$, as
\begin{eqnarray}
G^{\sigma}_{j,j'}=\langle a^{\dag}_{\sigma,j}a^{\dag}_{\sigma,j'}a_{\sigma,j'}a_{\sigma,j}\rangle, ~~~ \widetilde{G}^{\sigma}_{j,j'}=G^{\sigma}_{j,j'}/\mathrm{Max}(G^{\sigma}_{j,j'}).
\label{TwocorrDown}
\end{eqnarray}
The results for $\widetilde{G}^{\downarrow}_{j,j'}$ are displayed below the spectra, confirming that particles of the same species are also forced to bind together in the case of bulk bound states.
Note that $\widetilde{G}^{\uparrow}_{j,j'}$ exhibits similar features and thus is not shown in the figure.
} %This observation is beyond the situations with $N_{\uparrow}=N_{\downarrow}=1$.

\begin{figure}
    \centering
    \includegraphics[width=0.5\linewidth]{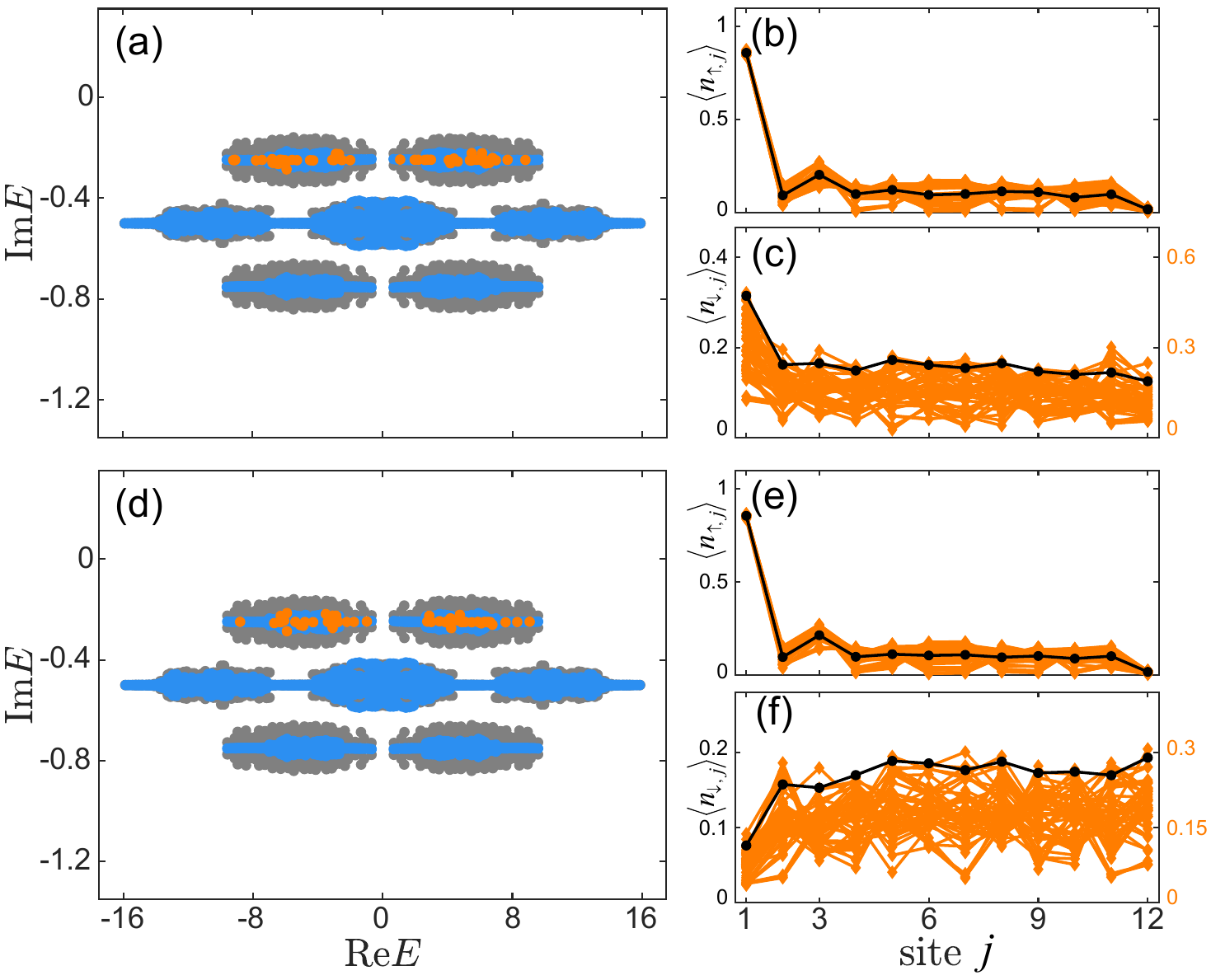}
   \caption{\MOD{\textbf{Edge confined and anti-confined states for extrinsic ISTPs with $N_{\uparrow}=N_{\downarrow}=2$, where all particles are fermions.} (a) OBC spectrum of Hamiltonian in Eq.(1) of the main text, represented by blue and orange dots. 
   The gray dots denote the eigenenergies obtained under PBCs for the pseudospin-down particles only. 
   %Here, spin-$\uparrow$ particle is always under OBCs, and spin-$\downarrow$ particle takes OBCs (PBCs) for colored and black (gray) circles. 
   (b) and (c) show the distributions of pseudospin-up and pseudospin-down particles, respectively, for the eigenstates marked by orange color in (a) (50 randomly chosen ones among 660 in the energy cluster near ${\rm Im}E\approx -\gamma_\uparrow/2=-0.25$). 
   Orange lines and dots represent the distribution of each single eigenstate, and black line and dots denote their average.
   %The orange lines and diamonds represent the distributions of single eigenstates with $\mathrm{Im}E\approx 0$ [corresponding to the orange dots in (a)], along with their average values (black line and circles).
   The parameters in (a) to (c) are $v_{\uparrow}=5$, $v_{\downarrow}=0.5$, $\gamma_{\uparrow}=t = 0.5$, $u_{\uparrow}=2$, and $u_{\downarrow}=1$.
   (d) to (f) the same as (a) to (c), but with $u_{\uparrow}=-2$ and $u_{\downarrow}=-1$. In (c) and (f), for a better demonstration, distributions of single eigenstates and their average correspond to different y-axis coordinates on the right and left sides of the figure, respectively. In all panels, the total number of each specie of particles is $N_{\uparrow}=N_{\downarrow}=2$, and the number of sites is $L=12$, which gives a Hamiltonian matrix of dimension $(L^2-L)^2/4=4356$.}}
\label{figS5bothFer}
\end{figure}

\subsection{The fermionic scenario}\label{appD2}
\MOD{In this subsection, we convert all species of particles into fermions.
Using the same parameters and particle number as in Fig.~\ref{figS5}, we plot the spectra of extrinsic ISTPs in Fig.~\ref{figS5bothFer} for this scenario. In this scenario, since pseudospin-up particles are fermions, their distinct occupation patterns give rise to three clusters of eigenstates with differing imaginary energies. Specifically: (i) The imaginary energy ${\rm Im}E$ is centered at $-\gamma_{\uparrow}/2$ when one fermion occupies the left-localized single-particle topological edge state and the other occupies a single-particle bulk state without even/odd site polarization; (ii) ${\rm Im}E$ centers at $-3\gamma_{\uparrow}/2$ when one fermion occupies the right-localized single-particle topological edge state and the other occupies a single-particle bulk state; and (iii) ${\rm Im}E$ centers at $-\gamma_{\uparrow}$ when the two fermions occupy the left- and right-localized single-particle topological edge states respectively, or both occupy single-particle bulk states. Among these, clusters (i) and (ii) correspond to the inter-species edge (anti-)confined states. Figs.~\ref{figS5bothFer}(b-c) and (e-f) display the distributions of $50$ randomly selected eigenstates ($660$ in total) belonging to cluster (i), where the characteristics of edge (anti-)confined states remain identifiable.} 

\begin{figure}
    \centering
    \includegraphics[width=0.8\linewidth]{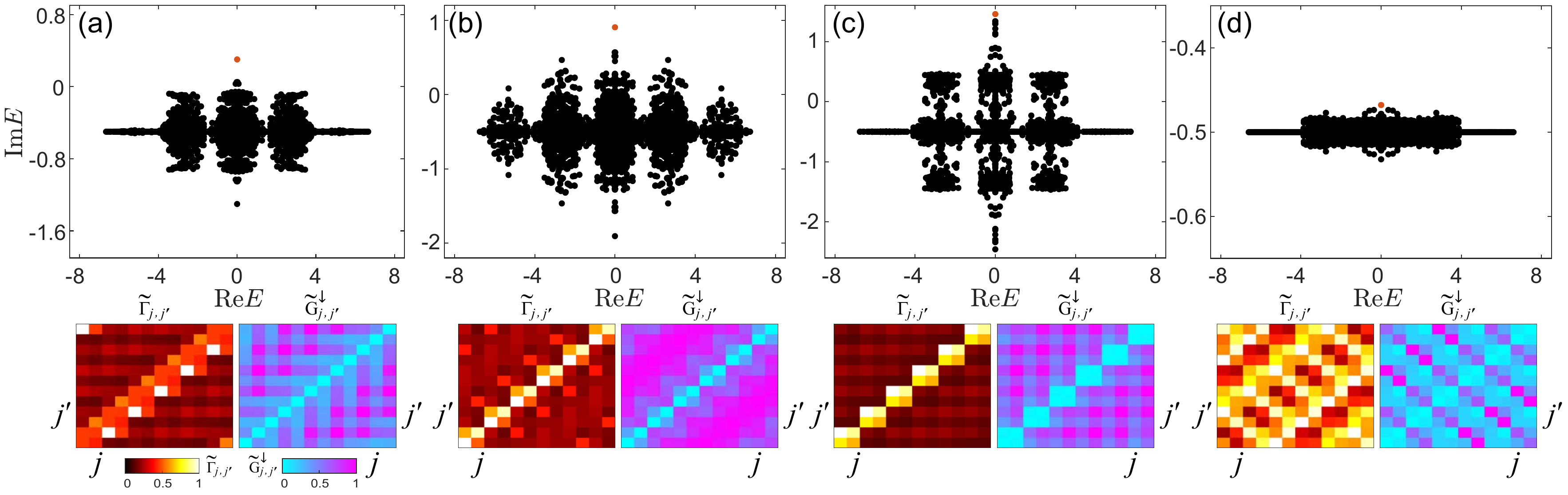}
   \caption{\MOD{\textbf{Bulk bound states for intrinsic ISTPs with $N_{\uparrow}=N_{\downarrow}=2$ where all particles are fermions.} (a) to (d) The PBC spectra of the Hamiltonian in Eq.(1) with different parameters. Below each spectrum, we display the normalized inter-species two-particle correction $\widetilde{\Gamma}_{j,j'}$ [defined in Eq.(4) of the main text], and the intra-species ones for pseudospin-down particles $\widetilde{G}^{\downarrow}_{j,j'}$ [defined in Eq.~\eqref{TwocorrDown}], of the eigenstate with the highest imaginary energy (represented by the orange dots in the spectra). The parameters are chosen as follows: (a) $\theta_{\uparrow}=0.1\pi$ and $\theta_{\downarrow}=0.4\pi$, (b) $\theta_{\uparrow}=0.4\pi$ and $\theta_{\downarrow}=0.6\pi$, (c) $\theta_{\uparrow}=0.1\pi$ and $\theta_{\downarrow}=0.9\pi$, and (d) $\theta_{\uparrow}=0.05\pi$ and $\theta_{\downarrow}=0.15\pi$, with $|u_{\sigma}+iv_{\sigma}|=\sqrt{2}$ and $\gamma_{\uparrow}=t = 0.5$. 
   The parameter used in (a) and (b) [(c) and (d)] are the same as those used in Fig. \ref{figSBBS}(a) and (b) [Fig.3(a) and (b) of the main text].
In all panels, the total number of each species of particles is $N_{\uparrow}=N_{\downarrow}=2$, and the number of sites is $L=12$, \MOD{which gives a Hamiltonian matrix of dimension $(L^2-L)^2/4=4356$.}}}
\label{figS6bothFer}
\end{figure}

\MOD{For the intrinsic ISTP with bulk bound states, results for fermions with the same parameters and particle number as in Fig.~\ref{figS6} are shown in Fig.~\ref{figS6bothFer}. In Figs.~\ref{figS6bothFer}(a-c), the inter-species two-particle correction $\widetilde{\Gamma}_{j,j'}$ still exhibits significant diagonal and near-diagonal components, manifesting the characteristics of bulk bound states. In contrast, the intra-species two-particle correlation function $\widetilde{G}^{\sigma}_{j,j'}$ shows vanishing diagonal components due to the Pauli exclusion principle. For simplicity, we only present $\widetilde{G}^{\downarrow}_{j,j'}$ here, noting that $\widetilde{G}^{\uparrow}_{j,j'}$ displays similar features.}

\subsection{The fermion-boson mixture scenario}\label{appD3} 
\MOD{In this subsection, we provide further discussion and demonstration for different fermion-boson mixtures in our model.}

\begin{figure}[H]
    \centering
    \includegraphics[width=0.5\linewidth]{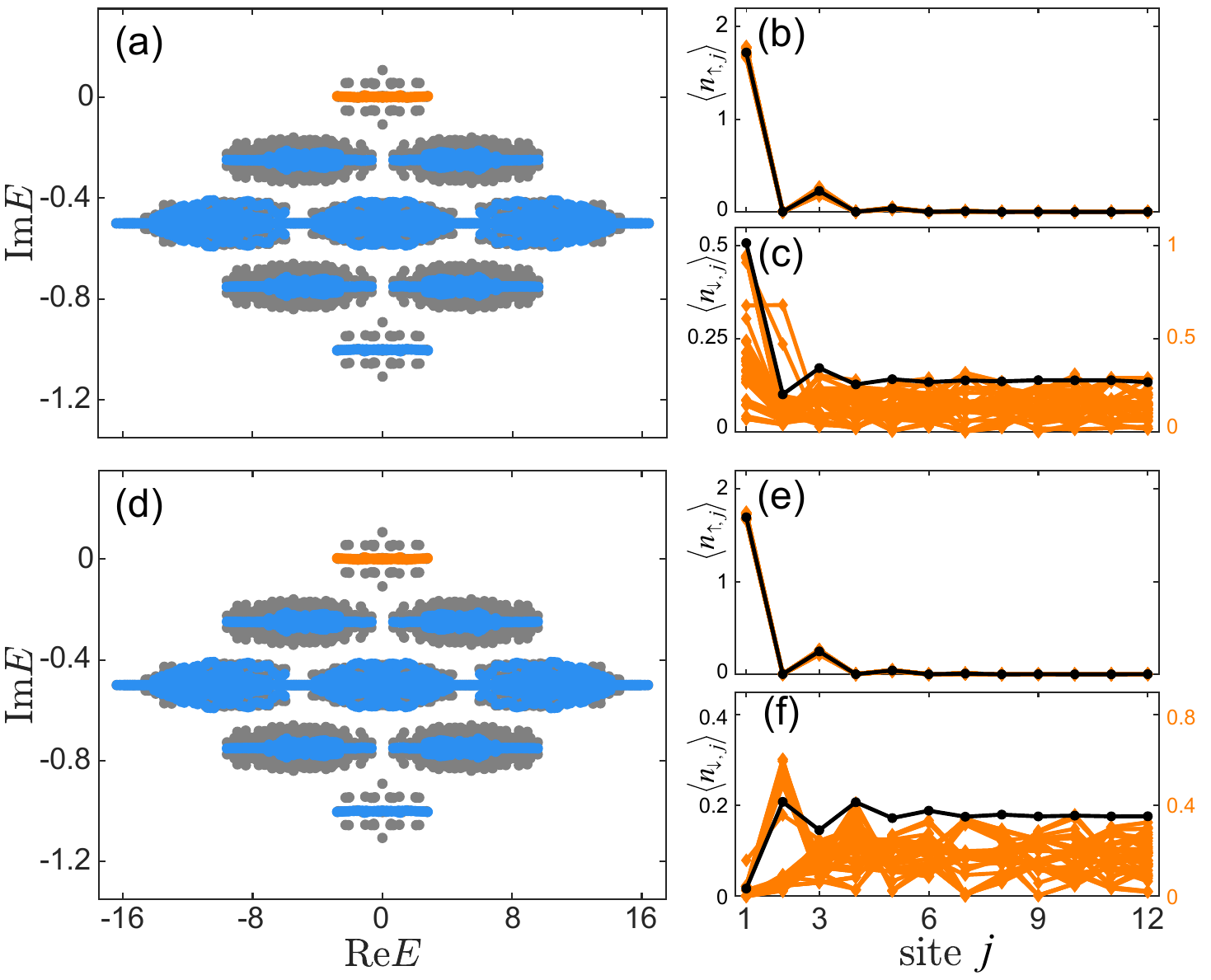}
   \caption{\MOD{\textbf{Edge confined and anti-confined states for extrinsic ISTPs with $N_{\uparrow}=N_{\downarrow}=2$, where pseudospin-up (down) particles are bosons (fermions).} (a) OBC spectrum of Hamiltonian in Eq.(1) of the main text, represented by blue and orange dots. 
   The gray dots denote the eigenenergies obtained under PBCs for the pseudospin-down particles only. 
   %Here, spin-$\uparrow$ particle is always under OBCs, and spin-$\downarrow$ particle takes OBCs (PBCs) for colored and black (gray) circles. 
   (b) and (c) show the distributions of pseudospin-up and pseudospin-down particles, respectively, for the eigenstates marked by orange color in (a) (78 in total). 
   Orange lines and dots represent the distribution of each single eigenstate, and black line and dots denote their average.
   %The orange lines and diamonds represent the distributions of single eigenstates with $\mathrm{Im}E\approx 0$ [corresponding to the orange dots in (a)], along with their average values (black line and circles).
   The parameters in (a) to (c) are $v_{\uparrow}=5$, $v_{\downarrow}=0.5$, $\gamma_{\uparrow}=t = 0.5$, $u_{\uparrow}=2$, and $u_{\downarrow}=1$.
   (d) to (f) the same as (a) to (c), but with $u_{\uparrow}=-2$ and $u_{\downarrow}=-1$. In (c) and (f), for a better demonstration, distributions of single eigenstates and their average correspond to different y-axis coordinates on the right and left sides of the figure, respectively. In all panels, the total number of each specie of particles is $N_{\uparrow}=N_{\downarrow}=2$, and the number of sites is $L=12$, which gives a Hamiltonian matrix of dimension $(L^2+L)(L^2-L)/4=5148$.}}
\label{figS5downFer}
\end{figure}

\MOD{Firstly, we consider a scenario where pseudospin-up particles are bosons, pseudospin-down particles are fermions, and there are two particles for each species. 
As shown in 
%We use the same parameter settings as in Fig.~\ref{figS5}, and present the results for the extrinsic ISTPs in this scenario in 
Fig.~\ref{figS5downFer}, 
it is apparent that the edge (anti-)confined states still arise, 
with similar features in imaginary energies as in the bosonic scenario (since the pseudospin-up particles with on-site loss are bosons).
%and display analogous features under this scenario.
} 

\begin{figure}[H]
    \centering
    \includegraphics[width=0.8\linewidth]{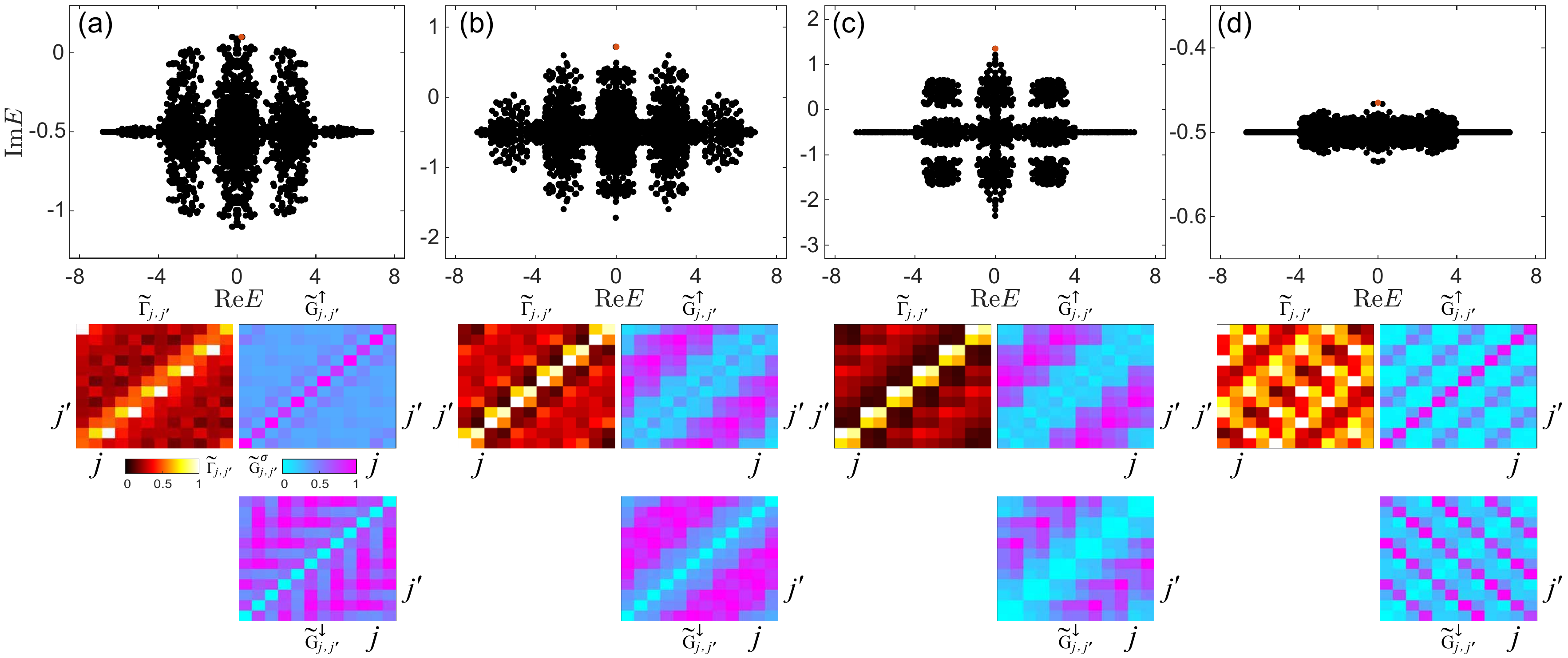}
   \caption{\MOD{\textbf{Bulk bound states for intrinsic ISTPs with $N_{\uparrow}=N_{\downarrow}=2$, where pseudospin-up (down) particles are bosons (fermions).} (a) to (d) The PBC spectra of the Hamiltonian in Eq.(1) with different parameters. Below each spectrum, we display the normalized inter-species two-particle correction $\widetilde{\Gamma}_{j,j'}$ [defined in Eq.(4) of the main text], and the intra-species ones for pseudospin-up particles $\widetilde{G}^{\uparrow}_{j,j'}$ as well as pseudospin-down particles $\widetilde{G}^{\downarrow}_{j,j'}$ [defined in Eq.~\eqref{TwocorrDown}], of the eigenstate with the highest imaginary energy (represented by the orange dots in the spectra). The parameters are chosen as follows: (a) $\theta_{\uparrow}=0.1\pi$ and $\theta_{\downarrow}=0.4\pi$, (b) $\theta_{\uparrow}=0.4\pi$ and $\theta_{\downarrow}=0.6\pi$, (c) $\theta_{\uparrow}=0.1\pi$ and $\theta_{\downarrow}=0.9\pi$, and (d) $\theta_{\uparrow}=0.05\pi$ and $\theta_{\downarrow}=0.15\pi$, with $|u_{\sigma}+iv_{\sigma}|=\sqrt{2}$ and $\gamma_{\uparrow}=t = 0.5$. 
   The parameter used in (a) and (b) [(c) and (d)] are the same as those used in Fig. \ref{figSBBS}(a) and (b) [Fig.3(a) and (b) of the main text].
In all panels, the total number of each species of particles is $N_{\uparrow}=N_{\downarrow}=2$, and the number of sites is $L=12$, which gives a Hamiltonian matrix of dimension $(L^2+L)(L^2-L)/4=5148$.}}
\label{figS6downFer}
\end{figure}

\MOD{%Similarly, when using parameters identical to those in Fig.~\ref{figS6}, 
On the other hand,
bulk bound states also emerge for intrinsic ISTPs, but with slightly different details. In the panels beneath each spectrum of Fig.~\ref{figS6downFer}, we present the normalized inter-species two-particle correction $\widetilde{\Gamma}_{j,j'}$, along with the intra-species two-particle correlation functions $\widetilde{G}^{\uparrow}_{j,j'}$ and $\widetilde{G}^{\downarrow}_{j,j'}$, corresponding to the eigenstate with the largest imaginary energy. It is observed that $\widetilde{\Gamma}_{j,j'}$ attains its maximum when $j \approx j'$ in Figs.~\ref{figS6downFer}(a) to (c), while taking relatively uniform values in panel Fig.~\ref{figS6downFer}(d). This indicates that the two species are bound together only in the former cases, which possess non-trivial inter-species topology. In contrast, due to the Pauli exclusion principle, the two-particle correlation function of fermions (pseudospin-down), $\widetilde{G}^{\downarrow}_{j,j'}$, vanishes along the diagonal. On the other hand, the two-particle correlation function of bosons (pseudospin-up), $\widetilde{G}^{\uparrow}_{j,j'}$, exhibits different properties across the three inter-species topologically non-trivial cases. In Fig.~\ref{figS6downFer}(a), $\widetilde{G}^{\uparrow}_{j,j'}$ takes larger values along the diagonal, suggesting that the two bosons are also bound together. However, in Figs.~\ref{figS6downFer}(b) and (c), the bosons (pseudospin-up) display an unbounded feature similar to that of the fermions (pseudospin-down). These observations are consistent with the gradual increase of the diagonal components of the inter-species correlation $\widetilde{\Gamma}_{j,j'}$ from Figs.~\ref{figS6downFer}(a) to (c), 
which leads to increasingly similar behaviors between the two species.
}


\begin{figure}[H]
    \centering
    \includegraphics[width=0.5\linewidth]{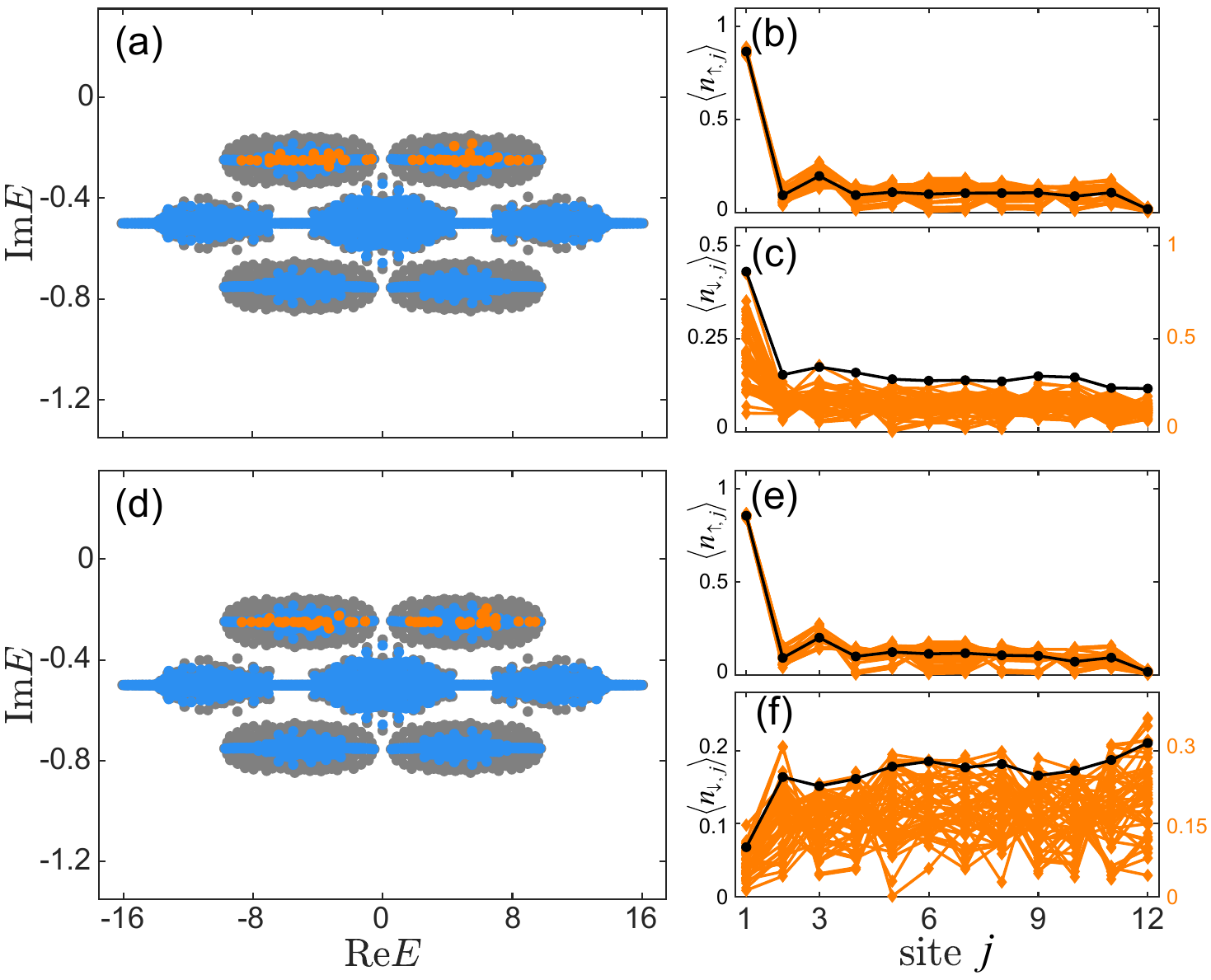}
   \caption{\MOD{\textbf{Edge confined and anti-confined states for extrinsic ISTPs with $N_{\uparrow}=N_{\downarrow}=2$, where pseudospin-up (down) particles are fermions (bosons).} 
   (a) OBC spectrum (blue and orange dots.) of Hamiltonian in Eq.(1) of the main text. 
   The gray dots denote the eigenenergies obtained under PBCs for the pseudospin-down particles only. 
show the distributions of pseudospin-up and pseudospin-down particles, respectively, for the eigenstates marked by orange color in (a)
(50 randomly chosen ones among 780 in the energy cluster near ${\rm Im}E\approx -\gamma_\uparrow/2=-0.25$).
   Orange color represents the distribution of each single eigenstate, and black color denotes their average.
   %The orange lines and diamonds represent the distributions of single eigenstates with $\mathrm{Im}E\approx 0$ [corresponding to the orange dots in (a)], along with their average values (black line and circles).
   The parameters in (a) to (c) are $v_{\uparrow}=5$, $v_{\downarrow}=0.5$, $\gamma_{\uparrow}=t = 0.5$, $u_{\uparrow}=2$, and $u_{\downarrow}=1$.
   (d) to (f) the same as (a) to (c), but with $u_{\uparrow}=-2$ and $u_{\downarrow}=-1$. In (c) and (f), distributions of single eigenstates and their average correspond to different y-axis coordinates on the right and left sides of the figure, respectively. The number of sites is $L=12$, which gives a Hamiltonian matrix of dimension $(L^2+L)(L^2-L)/4=5148$.}}
\label{figS5upFer}
\end{figure}

\begin{figure}[H]
    \centering
    \includegraphics[width=0.8\linewidth]{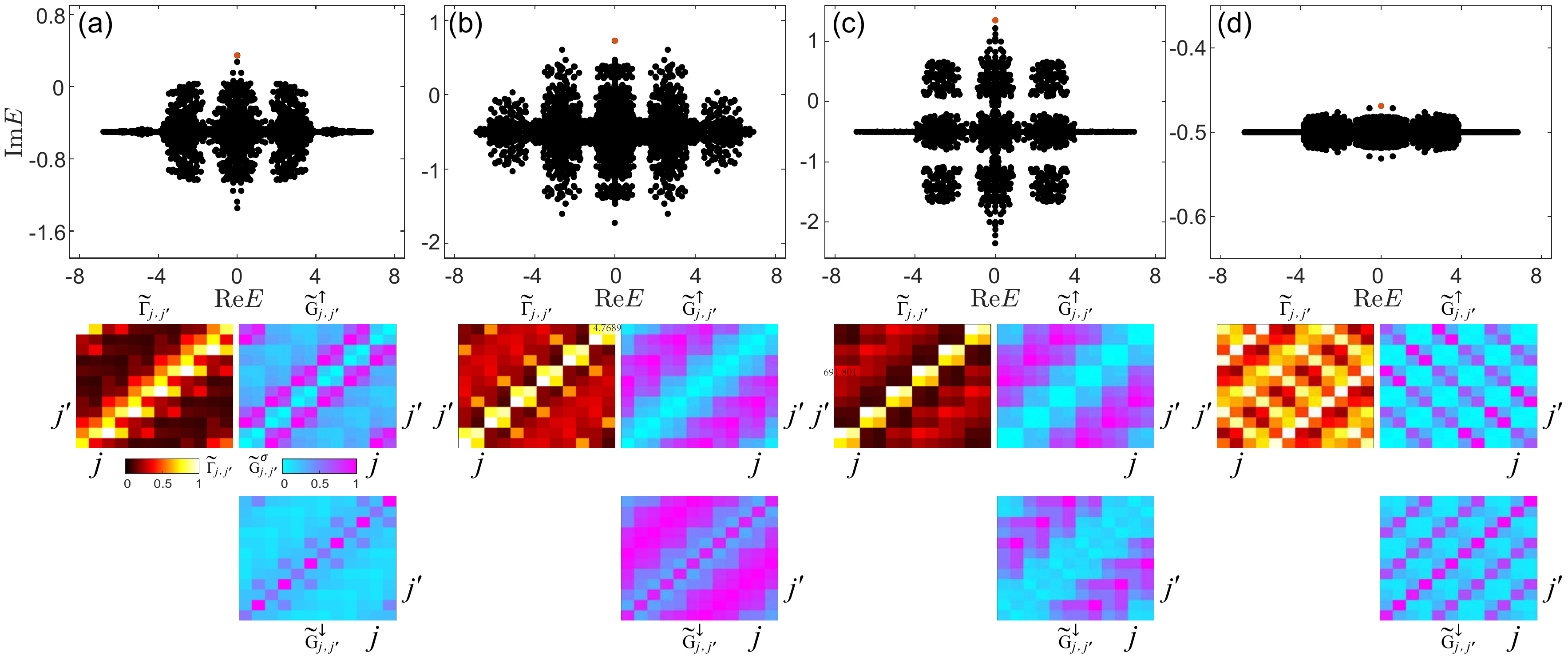}
   \caption{\MOD{\textbf{Bulk bound states for intrinsic ISTPs with $N_{\uparrow}=N_{\downarrow}=2$, where pseudospin-up (down) particles are fermionic (bosonic).} (a) to (d) The PBC spectra of the Hamiltonian in Eq.(1) of the main text with different parameters. Below each spectrum, we display the normalized inter-species two-particle correction $\widetilde{\Gamma}_{j,j'}$ [defined in Eq.(4) of the main text], and the intra-species ones $\widetilde{G}^{\uparrow}_{j,j'}$ and $\widetilde{G}^{\downarrow}_{j,j'}$ [defined in Eq.~\eqref{TwocorrDown}], of the eigenstate with the highest imaginary energy (represented by the orange dots in the spectra). The parameters are chosen as follows: (a) $\theta_{\uparrow}=0.1\pi$ and $\theta_{\downarrow}=0.4\pi$, (b) $\theta_{\uparrow}=0.4\pi$ and $\theta_{\downarrow}=0.6\pi$, (c) $\theta_{\uparrow}=0.1\pi$ and $\theta_{\downarrow}=0.9\pi$, and (d) $\theta_{\uparrow}=0.05\pi$ and $\theta_{\downarrow}=0.15\pi$, with $|u_{\sigma}+iv_{\sigma}|=\sqrt{2}$ and $\gamma_{\uparrow}=t = 0.5$. 
   The parameter used in (a) and (b) [(c) and (d)] are the same as those used in Fig. \ref{figSBBS}(a) and (b) [Fig.3(a) and (b) of the main text].
The number of sites is $L=12$, which gives a Hamiltonian matrix of dimension $(L^2+L)(L^2-L)/4=5148$.}}
\label{figS6upFer}
\end{figure}

\MOD{Next, we present the results with pseudospin-up (-down) particles being fermions (bosons). 
%The edge (anti-)confined states are illustrated in Fig.~\ref{figS5upFer}.
Similar to the scenario with fermions only, three clusters of eigenstates with distinct imaginary energies [Figs.\ref{figS5upFer}(a) and (d)] emerge in this scenario because pseudospin-up particles (with loss) are fermions. 
Edge (anti-)confined states can be identified in Figs.~\ref{figS5upFer}(b-c) and (e-f),
where we show the distributions of $50$ randomly selected eigenstates (from a total of $780$) belonging to the cluster with ${\rm Im}E \approx -\gamma_{\uparrow}/2$.
%In Figs.~\ref{figS5upFer}(b-c) and (e-f), we show the distributions of $50$ randomly selected eigenstates (from a total of $780$) belonging to the cluster with ${\rm Im}E \approx -\gamma_{\uparrow}/2$ in the spectra, where the characteristics of edge (anti-)confined states can still be identified.
In Fig.~\ref{figS6upFer}, we demonstrate the emergence of bulk bound states for intrinsic ISTPs under this scenario. The same characteristics of bulk bound states are also observed, consistent with the previous case with bosonic (fermionic) pseudospin-up (-down) particles.}

\end{document}